\documentclass{aa}
\usepackage{natbib, graphicx}
\usepackage{amssymb, amsmath}
\usepackage{lscape}
\usepackage[varg]{txfonts}
\usepackage{color}
\begin{document}

\title{Supernovae and their host galaxies}
\subtitle{I. The SDSS DR8 database and statistics\thanks{Full Table~\ref{bigbigtable}
is only available at the CDS via anonymous ftp to
\texttt{cdsarc.u-strasbg.fr (130.79.128.5)}
or via \texttt{http://cdsarc.u-strasbg.fr/cgi-bin/qcat?J/A+A/544/A81}}}
\titlerunning{Supernovae and their host galaxies. I.}

\author{A.~A.~Hakobyan\inst{1,2,3}
\and V.~Zh.~Adibekyan\inst{4}
\and L.~S.~Aramyan\inst{1,2,3}
\and A.~R.~Petrosian\inst{1,3}
\and J.~M.~Gomes\inst{4}
\and \\ G.~A.~Mamon\inst{5}
\and D.~Kunth\inst{5}
\and M.~Turatto\inst{6}
}
\institute{
Byurakan Astrophysical Observatory, 0213 Byurakan, Aragatsotn province, Armenia
\\ \email{hakobyan@bao.sci.am}
\and
Department of General Physics and Astrophysics, Yerevan State University,
1 Alex Manoogian, 0025 Yerevan, Armenia
\and
Isaac Newton Institute of Chile, Armenian Branch, 0213 Byurakan, Aragatsotn province, Armenia
\and
Centro de Astrof\'{i}sica da Universidade do Porto, Rua das Estrelas, 4150-762 Porto, Portugal
\and
Institut d'Astrophysique de Paris (UMR 7095: CNRS \& UPMC), 98bis Bd Arago, 75014 Paris, France
\and
INAF - Osservatorio Astronomico di Padova, Vicolo dell'Osservatorio 5, 35122 Padova, Italy
}

\date{Received 4 May 2012 / Accepted 20 June 2012}

\abstract
{In this first paper of a series, we report the creation of large and
well-defined database that combines extensive new measurements and
a literature search of 3876 supernovae (SNe) and their 3679 host galaxies
located in the sky area covered by the
Sloan Digital Sky Survey (SDSS) Data Release 8 (DR8).}
{This database should be much larger than previous ones, and should contain a
homogenous set of global parameters of SN hosts, including morphological
classifications and measures of nuclear activity.}
{The measurements of apparent magnitudes, diameters $(D_{25})$, axial ratios $(b/a)$,
and position angles (PA) of SN host galaxies were made using the images
extracted from the SDSS $g$-band. For each host galaxy, we analyzed RGB
images of the SDSS to
accurately measure the position of its nucleus to provide the SDSS name.
With these images, we also provide the host galaxy's morphological type, and
note if it has a bar, a disturbed disk, and whether it is part of an
interacting or merging system.
In addition, the SDSS nuclear spectra were analyzed to diagnose
the central power source of the galaxies. Special attention was
paid to collect accurate data on the spectroscopic classes, coordinates,
offsets of SNe, and heliocentric redshifts of the host galaxies.}
{Identification of the host galaxy sample
is 91\% complete (with 3536 SNe in 3340 hosts), of which the SDSS names of
$\sim$1100 anonymous hosts are listed for the first time.
The morphological classification is available for 2104 host galaxies,
including 73 (56) hosts in interacting (merging) systems.
The \emph{total sample} of host galaxies collects heliocentric redshifts
for 3317 ($\sim$90\%) galaxies.
The $g$-band magnitudes, $D_{25}$, $b/a$, and PA are available for
2030 hosts of the morphologically classified sample of galaxies.
Nuclear activity measures are provided for 1189 host galaxies.
We analyze and discuss many selection effects and biases that
can significantly affect any future analysis of our sample.}
{The creation of this large database will help to better understand
how the different types of SNe are
correlated with the properties of the nuclei and global physical
parameters of the host galaxies, and minimize
possible selection effects and errors that often arise when data
are selected from different sources and catalogs.}


\maketitle

\section{Introduction}

A crucial aspect of many recent studies
of extragalactic supernovae (SNe) is to establish the
links between the nature of SN progenitors and stellar
populations of their host galaxies.
The most direct method for realizing this
is through their identification on pre-SN images.
However, the number of such SNe is small and is limited to the nearby
core-collapse (CC) events \citep[e.g.,][]{2009ARA&A..47...63S}.
This limitation and small-number statistics are the main reasons to investigate
the properties of SN progenitors through indirect methods.
In this context, the properties of SN host galaxies, such as the morphology, color,
nuclear activity, star formation
rate, metallicity, stellar population, age etc. provide strong
clues to the understanding of the progenitors
\citep[e.g.,][]{1995A&A...297...49P,1997Ap.....40..296K,2002MNRAS.331L..25B,
2003A&A...406..259P,2005A&A...433..807M,2008ApJ...673..999P,
2009A&A...503..137B,2010ApJ...721..777A,2010ApJ...724..502H,2011arXiv1110.1377K}.
In addition, valuable information of the nature of progenitors can be
obtained through the study of the spatial distribution of SNe
\citep[e.g.,][]{2008MNRAS.388L..74F,2009A&A...508.1259H,2010MNRAS.405.2529W}
and environments
\citep[e.g.,][]{2008MNRAS.390.1527A,2011ApJ...731L...4M,2011A&A...530A..95L}.

Over the past decade, many studies have investigated the nature
of SN progenitors in the nearby Universe via local and global properties of
their host galaxies.
For example, \citet{2008ApJ...673..999P} investigated how the different types of SNe
are correlated with the metallicity of their host galaxy.
They showed strong evidence that SNe~Ibc\footnote{{\tiny We use SNe~Ibc to mainly denote the Ib,
Ic and mixed Ib/c SNe types whose specific Ib or Ic classification is uncertain.}}
occur in higher metallicity hosts
than SNe~II, while there is no such effect for SNe~Ia relative to SNe~II.
\citet{2008MNRAS.388L..74F} studied the radial distribution of SNe~Ia in morphologically
selected early-type host galaxies from the SDSS, and found that
there is no statistically significant difference between
the radial distribution of SNe~Ia and the light profile of
their early-type host galaxies.
\citet{2009A&A...508.1259H} compared the radial distribution of CC SNe within their spiral
hosts with the distributions of stars and ionized gas in spiral disks.
They concluded that the normalized radial distribution of all CC SNe is
consistent with an exponential law, the scale length of the distribution
of SNe~II appears to be significantly larger than that of the stellar disks
of their host galaxies, while SNe~Ibc
have a significantly smaller scale length than SNe~II
\citep[see also][]{2009MNRAS.399..559A}.
The scale length of the radial distribution of CC~SNe
shows no significant correlation with
the host galaxy morphological type, or the presence of bar structure.

Several authors have studied the radial distributions of SNe
of different types in large numbers of galaxies
\citep[e.g.,][]{1975PASJ...27..411I,1992A&A...264..428B,1997AJ....113..197V,
2008MNRAS.388L..74F,2009A&A...508.1259H}, but
none of these studies attempted to categorize the hosts according
to their activity level. However, other authors have shown that the SNe
distributions in galaxies with various activity levels might be different
\citep[e.g.,][]{1990A&A...239...63P,2005AJ....129.1369P,
2008Ap.....51...69H,2010MNRAS.405.2529W,2012A&A...540L...5H}.
For example, \citet{2010MNRAS.405.2529W} directly measured number
and surface density distributions of SNe~II in their hosts,
and indicated that SNe~II detected in star-forming galaxies follow
an exponential law, in contrast, the distribution of SNe~II detected in
Active Galactic Nuclei (AGN) hosts
significantly deviates from an exponential law.
\citet{2005AJ....129.1369P} studying
a sample of CC SNe in galaxies hosting AGN found that
SNe~Ibc in active/star-forming galaxies are more
centrally concentrated than are the SNe~II,
but given the small sample,
this difference was not statistically significant.
The results of \citeauthor{2005AJ....129.1369P}
were confirmed with larger samples of CC~SNe
by \citet{2008Ap.....51...69H}.

The locations of SN explosions in multiple galaxy systems
have also been studied. In interacting galaxies,
CC~SNe are not preferentially located towards
the companion galaxy
\citep[e.g.,][]{2001MNRAS.328.1181N}.
Similarly, the azimuthal distributions inside the
host members of galaxy groups are consistent with being
isotropic \citep{2001MNRAS.328.1181N}.
\citet{2010ApJ...724..502H} found that
SNe~Ia are more likely to occur in isolated
galaxies without close neighbors.

However, many similar studies
\citep[e.g.,][]{2009MNRAS.399..559A,2009A&A...508.1259H,2010MNRAS.405.2529W}
presented above are suffered from poor statistics,
as well as strong biases on the SNe and their host galaxies sample.
Often, they were random selections of nearby SNe and their hosts
from the Asiago Supernova
Catalogue\footnote{\texttt{{\tiny http://web.oapd.inaf.it/supern/cat/}}}
\citep[ASC,][]{1999A&AS..139..531B} or the Sternberg Astronomical Institute
(SAI) Supernova
Catalogue\footnote{\texttt{{\tiny http://www.sai.msu.su/sn/sncat/}}}
\citep[SSC,][]{2004AstL...30..729T} or the official list of all
the discovered SNe on the Central
Bureau for Astronomical Telegrams (CBAT)
website\footnote{\texttt{{\tiny http://www.cbat.eps.harvard.edu/lists/Supernovae.html}}}.
Most recently, \citet{2012A&A...538A.120L} published
a unified SN catalog for 5526 extragalactic SN that were discovered
up to 2010 December~31.
The unified catalog mostly combines ASC, SSC, and data from CBAT in a consistent way,
and adopts all of the inhomogeneous data on SNe and their host galaxies
from the original sources.
For the galaxy data, these
catalogs made large use of the Third Reference Catalogue of Bright
Galaxies (RC3) by \citet{1991rc3..book.....D} and the
HyperLeda\footnote{\texttt{{\tiny http://leda.univ-lyon1.fr/}}}
database \citep{2003A&A...412...45P}
as well as the NASA/IPAC Extragalactic
Database\footnote{\texttt{{\tiny http://ned.ipac.caltech.edu/}}} (NED).
Hence, the data are given with various degrees of accuracy depending
on the accuracy of the original catalog.
Many selection effects and errors
that often arise when data is selected from
different sources and catalogs can significantly bias results
and lead to wrong conclusions.
Quantitative studies of SN progenitors therefore require
a large and well-defined sample of SNe and their host galaxies, and our goal
is to provide the database for such a sample.

The Eighth Data Release\footnote{\texttt{{\tiny http://www.sdss3.org/}}}
(DR8) of the SDSS \citep{2011ApJS..193...29A}
covering over 14000 square degrees with high quality imaging and
spectroscopy makes it finally possible to construct
better samples for studies of
the properties of the host galaxies and environments of SNe.
This large amount of the SDSS data enables statistically meaningful
studies that are only little affected by selection effects.
In this paper, we report the creation of a large
database of several thousand
SNe that exploded in galaxies identified in
the SDSS DR8. We provide identifications of SN host galaxies,
their accurate coordinates, heliocentric redshifts, morphological types,
and activity classes, as well as apparent magnitudes,
diameters, axial ratios, and position angles.
However, our goal is not just to increase the size of the
sample in comparison with previous studies, but also to carry out
morphological classification, as well as individual measurements
of the global parameters of the host galaxies
in a homogenous way.
In addition, we summarize the overall statistical properties
of the sample, analyzing and discussing residual selection
effects and biases that can still affect subsequent studies and results.

An additional motivation for this study is the comparison between
the ASC, SSC, and CBAT databases, to reveal possible inconsistencies
in the listed SNe types and offsets.
Our database includes corrected
data with their uncertainties on the SNe.

This is the first paper of a series and
is organized as follows: Sect.~\ref{data} introduces
the data and describes in detail the reduction techniques.
In Sect.~\ref{Resdiscus}, we give the results and discuss
all the statistical properties and selection effects of the sample.
A summary and perspectives for future uses are finally addressed in
Sect.~\ref{sumandper}.
Throughout this paper, we adopt a cosmological model with
$\Omega_{\rm m}=0.27$, $\Omega_{\rm \Lambda}=0.73$, and the Hubble constant of
$H_0=73 \,\rm km \,s^{-1} \,Mpc^{-1}$
\citep{2007ApJS..170..377S}, consistent with direct determination
based on Cepheid variables and SNe~Ia by \citet{2009ApJ...699..539R}.

Future papers of this series will use this database to
determine how the different types of SN progenitors are correlated
with the global parameters (morphology, size, luminosities etc.)
of the host galaxies as well as on their nuclear properties (activity class,
metallicity, stellar population etc.).

\section{Data and reduction}
\label{data}

\subsection{Supernova catalogs}

The ASC and SSC are
compilations of information about SN discoveries
obtained mainly from reports in the International Astronomical Union (IAU)
Circulars and CBAT, as well as basic information about the host galaxies
generally from the RC3, HyperLeda, and NED.
We used the ASC, updated on
March 2011, and SSC, updated on February 2011,
to obtain and investigate general properties
of SNe (designation, coordinates, offset, type etc.)
and their host galaxies
(name, coordinates, morphology, heliocentric redshift etc.).
Table~\ref{tabl1} displays the distribution of different types of SNe
in updated versions of the ASC and SSC.
Note that SNe~II
include subtypes II~P (Plateau), II~L (Linear),
IIn (narrow-line), and IIb (transitional), but these subtypes are often
absent in the SN catalogs.

\begin{table}[t]
\begin{center}
\caption{Distribution of main SN types in the SN catalogs.
\label{tabl1}}
\begin{tabular}{lcrrrrr}
\hline
\hline
Catalog & Unclassified & \multicolumn{1}{c}{I} & \multicolumn{1}{c}{Ia} & \multicolumn{1}{c}{Ibc} & \multicolumn{1}{c}{II} & \multicolumn{1}{c}{All} \\
& \%& \multicolumn{1}{c}{\%} & \multicolumn{1}{c}{\%} & \multicolumn{1}{c}{\%} & \multicolumn{1}{c}{\%} & \multicolumn{1}{c}{\emph{N}} \\
\hline
ASC & 19.8 & 1.5 & 47.0 & 6.9 & 24.8 & 5609 \\
SSC & 21.1 & 2.2 & 46.1 & 6.8 & 23.8 & 5539 \\
\hline
\end{tabular}
\end{center}
\end{table}

We cross-matched the SN catalogs with the SDSS DR8 galaxies, mainly
using, for the former, the coordinates of the SNe, or of their host galaxies
when the SNe positions were unavailable.

\subsection{The Sloan Digital Sky Survey}

The SDSS is a large photometric  and
spectroscopic survey of the Northern sky (mainly), using
a dedicated 2.5~m telescope with a wide field of view
($3^\circ$) at the Apache Point, New Mexico
\citep[e.g.,][]{2006AJ....131.2332G}.
The first, second, and third phases of the SDSS
(SDSS$-$I, SDSS$-$II, and SDSS$-$III) have produced eight
data releases \citep[see][]{2011ApJS..193...29A}.
The DR8, made publicly available in January 2011,
contains all of the imaging data taken by the SDSS
imaging camera (14555 square degrees of sky),
including two large contiguous zones, one around the Northern Galactic Cap
and one close to the Southern Galactic Cap
(centered on RA $\sim0^{\rm h}$).
The DR8 also contains spectra from
the SDSS spectrograph in an area covering 9274 square degrees.
The photometric survey uses a specially
designed multi-band CCD camera that covers five bands over a wide
wavelength range denoted by $u$, $g$, $r$, $i$, and
$z$ with effective wavelengths
of $3551$~{\AA}, $4686$~{\AA}, $6165$~{\AA}, $7481$~{\AA}, and
$8931$~{\AA}, respectively.
All of the imaging data have been reduced using improved
data processing pipelines
\citep[for more details see][]{2011ApJS..193...29A}.
The SDSS spectra are observed using a pair of fiber-fed
double spectrographs. Each of the two spectrographs collect
spectra on 2 SITe/Tektronix 2048 by 2048 CCDs,
one covering a wavelength range of 3800~{\AA} to 6100~{\AA},
and the other from 5900~{\AA} to 9100~{\AA}.
The resolving power $({\rm \lambda/\Delta\lambda})$ of the SDSS spectra
ranges from 1800 to 2200 \citep{1999AAS...195.8701U}.

Currently, the total spectroscopic sample of the SDSS DR8 consists of
1629129 unique spectra and also 214071 repeated measurements.
The spectra are classified and divided into different categories.
The total galaxy sample of DR8 collects 860836 spectra and
the SDSS Main Galaxy spectroscopic sample consists of 711726
objects with \citeauthor{1976ApJ...209L...1P} magnitudes (corrected for Galactic extinction)
$r \leq 17.77$ \citep{2002AJ....124.1810S}.
Below, the comparisons of physical parameters of the SNe host galaxy sample
are performed with the SDSS Main Galaxy sample.

\subsection{Cross-matching the SNe with the SDSS galaxies}

The number of SNe from the aforementioned SN catalogs
that matched with the SDSS DR8 is 3841 (in 3645 host galaxies).
In several cases, the galaxies do not fully lie in the frame
(field) of the DR8, but are fully covered in the DR7 fields.
Also there are cases when galaxies are in the DR8 field but
with empty (black) images, while they are available in DR7.
We therefore also cross-matched the SN catalogs with the SDSS DR7
and found an additional 35 SNe in 34 host galaxies.
In total, the sample of SNe contains 3876 events in 3679 host galaxies.
The \emph{total sample} includes 163 host galaxies with multiple SNe:
2, 3, 4, 5 and 6 SNe are found in 135, 19, 5, 1 and 1 galaxy, respectively.

\begin{figure}[t]
\begin{center}
\includegraphics[width=0.7\hsize,angle=-90]{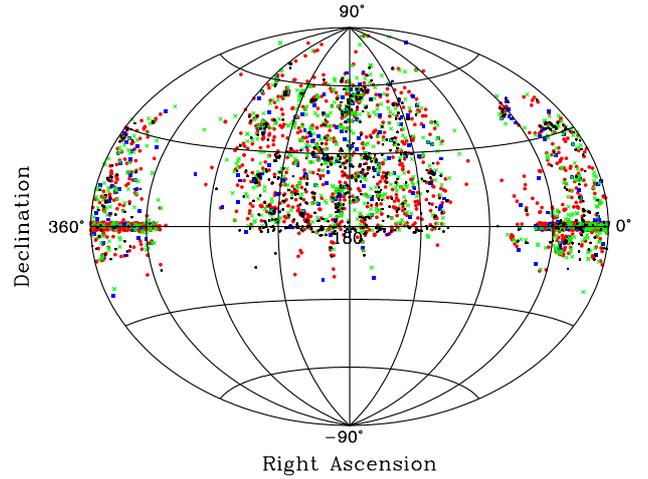}
\end{center}
\caption{The Aitoff projection of equatorial coordinates
of SNe from the \emph{total sample}.
The symbols correspond to different types of SNe:
Ia (\emph{red filled circles}), Ibc (\emph{blue filled squares}),
II (\emph{green crosses}), I and unclassified SNe (\emph{black dots}).}
\label{aitoff}
\end{figure}

The distribution of equatorial coordinates of different types of SNe from
the \emph{total sample} is shown in Fig.~\ref{aitoff}.
The distribution reflects the sky coverage of the final SDSS imaging survey,
and different observational biases for different regions of the sky.
Relatively empty regions in the northern sky are due to obscuration
by the Galactic Plane. The SNe density peak around the celestial equator is
caused by various SN surveys such as the SDSS SN Survey \citep{2008AJ....135..338F}
and ESSENCE \citep{2007ApJ...666..674M}.
These surveys discovered many targets, but because their main focus are on
high-redshift SN Ia, most of their spectroscopically classified SNe
are Ia (red filled circles), although roughly 100 ($\approx$10\%) are
CC~SNe (blue filled squares and green crosses).

We visually inspected all
images of our 3679 cross-matched SN host galaxies from the SDSS Imaging
Server\footnote{\texttt{{\tiny http://skyserver.sdss3.org/dr8/en/tools/chart/list.asp}}},
which builds RGB color images from the $g$, $r$, and $i$ data channels.
This has been done to directly identify the hosts and
to exclude the galaxies that were wrongly selected as
hosts in the SN catalogs.
All the identified hosts are the closest galaxies to the SN
in terms of angular separation and if available also redshift.
We excluded SN-host/SDSS matches when a SN was closer in angular
separation and further in redshift (when available).
We have also performed an extensive
literature search mainly through the NED and HyperLeda databases, as well as
the CBAT website, so as to find any additional data on their
identification as SN host galaxies.
For example, many original sources such as IAUC, CBAT, and unique papers
include data on host galaxy identification, names, morphology,
and description of many details
(spiral arms, nucleus, tidal bridges, peculiarities etc.)
as well as SNe magnitudes in different bands at different epochs,
redshifts and many other information.

We identified $\sim$91\% (3340 hosts with a total of 3536 SNe)
of SN host galaxies that matched with the SDSS.
Nine host galaxies with a total of 10 SNe were
not identified because of a few unavailable RGB images.
The remaining 330 hosts mostly were unidentified
due to insufficient resolution of SDSS images at
the large distance of these objects.
Among the identified hosts, roughly 2000 in the ASC
($\sim$1100 in the SSC) are marked as anonymous.
For these galaxies,
we provide the SDSS names based upon the SDSS fiber targeted at the galaxy nucleus.
In addition to the SDSS names,
we used the multi-band images of the hosts and their appropriate
photometric and spectral data to
provide their accurate (J2000) equatorial coordinates,
apparent $g$-band magnitudes, major axes ($D_{25}$),
axial ratios ($b/a$), position angles (PA), morphological types,
heliocentric redshifts, and nuclear activity classes.

\subsection{Spectroscopic classes of supernovae}

The spectral type of an SN provides a crucial diagnostic of the initial mass,
age and metallicity of
the progenitor, as well as of the explosion mechanism
\citep[e.g.,][]{2003LNP...598...21T,2007AIPC..937..187T}.
We have thus attempted to homogenize the spectral types for SNe that have
been assign different types in different SN catalogs.
In particular, all the SN types taken from the ASC
and SSC were compared.
When the types of given SNe did not match in these two catalogs,
a comprehensive search of the literature
was performed. This search was important because primary
SN classifications can be changed with more accurate subsequent observations.

As an example, SN~1972R is listed as type~Ib: (uncertain type~Ib)
and IPec (peculiar type~I),
in the ASC and SSC, respectively. Following claims regarding possible
errors in the photometry of SNe carried out at Asiago observatory
during the 70's, \cite{1997A&A...317..423P} presented the result
of new photometry of SNe~Ia. They reported SN~1972R as type~Ia
observed at Asiago, not included in their sample but
re-calibrated by Tsvetkov.
As another example, in the ASC, the spectroscopic type of SN~2005az is
listed as Ib. The Nearby Supernova Factory reported that a spectrum
of SN~2005az, obtained, 5 days after its discovery,
with the Integral Field Spectrograph on the
University of Hawaii 2.2~m telescope, shows the SN to be a type~Ib
\citep{2005ATel..451....1A}.
At the same time, in the SSC,
according to \cite{2005IAUC.8504....3Q}, this SN is
listed as type~Ic.
The SN was discovered approximately 17 days before maximum and
spectroscopically classified 3 days after discovery as
a SN~Ic \citep{2005IAUC.8504....3Q}.
Finally, \cite{2011arXiv1110.1377K} updated
the classification using
a comprehensive set of spectra and confirmed that it was
a type~Ic explosion.
We have adopted this latter type, for this SN, in our database.

In addition, to observe the spectroscopic transition that is typical for SN~IIb,
spectra taken over several epochs are required, but
such data is not always included in the SN catalogs.
For instance, SN~2006dj is listed in the ASC as type~Ib, while the SSC and CBAT website
list it as a transitional type~IIb.
Another example is SN~2002au, which is typed as Ia: (uncertain type~Ia)
in the SN catalogs. The spectroscopic type mentioned in
\cite{2002IAUC.7825....1F} is a probable type~Ia.
Recently, \cite{2011MNRAS.412.1419L} analyzed the spectrum observed by
\citeauthor*{2002IAUC.7825....1F}, and suggested type~IIb SN,
which is what we adopt in our database.

SNe~1993R \citep{1993IAUC.5842....2F},
2007jc \citep{2007IAUC.8875....1P},
2007kr \citep{2007CBET.1098....1B},
and 2007me \citep{2007CBET.1102....1B} have shown spectral
properties of type~Ia as well as type~Ic.
For these SNe, we have adopted a peculiar type~I (I~pec) instead of
types~Ia/c that are listed in the ASC.
In addition, for SN~1997ew \citep{1998IAUC.6804....1N},
we adopted a peculiar type~II (II~pec) instead of type~II/Ic that is given in the ASC.

Our sample of SNe includes also a small number of type~Ib SNe
(2000ds, 2001co, 2003dg, 2003dr, 2005E, and 2007ke)
that are calcium-rich
and may have different progenitors from typical SN~Ib
\citep[e.g.,][]{2010Natur.465..322P,2010Natur.465..326K}.
However, the nature of this subtype of SNe is still under debate
\citep[e.g.,][]{2011ApJ...728L..36P,2011ApJ...730..110S}
and therefore, in our
study, we haven't separated these SNe into a separate class.
In addition, the extreme objects described in \cite{2012MSAIS..19...24P}
have not been considered throughout this paper because of their rarity.

We did not include in our sample SN~1984Z, because it is an unconfirmed SN
and is missing from the SSC and CBAT.
SN~1989Z is probably a foreground
variable star \citep{1998IAUC.6911....1B,1998IAUC.6920....2K}, and
SN~2010U was a luminous fast nova \citep{2010ApJ...718L..43H}.
These two objects were also
excluded from our database.
SNe~1954J \citep{2010AJ....139.1451S},
1997bs \citep{2000PASP..112.1532V},
1999bw \citep{1999IAUC.7152....2F},
2000ch \citep{2004PASP..116..326W},
2001ac \citep{2001IAUC.7597....3M},
2002bu \citep{2011MNRAS.415..773S},
2002kg \citep{2006MNRAS.369..390M},
2003gm \citep{2003IAUC.8167....3P},
and 2006bv \citep{2011MNRAS.415..773S} have been shown to actually be
outbursts of Luminous Blue Variables (LBVs) similar to $\eta$-Carinae or
$P$-Cygni, so we have omitted them from our database.

In addition, our SNe sample includes 62~SNe discovered with
the QUasar Equatorial Survey Team (QUEST).
By analyzing the available SDSS spectra of these SNe,
we found that SNe~2001ap
and 2001at \citep{2001IAUC.7608....1S} are indeed
foreground A0 type stars, while SN~2001aw
\citep{2001IAUC.7608....1S} is
a foreground cataclysmic variable.
We removed these 3 objects from our database too, since they
are not true SN explosions.

During the mutual comparison between the SN catalogs,
as well as in our literature search,
we have updated spectroscopic types for 67 SNe.
All the updated SN classifications are labeled by the letter U.
We collected all the available data on
3166 SNe types when they were available in one of the SN
catalogs or in the CBAT.
Our \emph{total sample} consists of
72 SNe~I, 1990 SNe~Ia, 234 SNe~Ibc,
870 SNe~II (including II~P, II~L,
IIn, and IIb),
and 710 unclassified SNe.
All the uncertain (``:'' or ``?'') and
peculiar (``pec'') classifications are flagged.
Types I, Ia, and II include also a few SNe classified
from the light curve only: these SN types are
labeled by ``*'' symbols.

\subsection{Supernova offsets}

The locations of SNe within host galaxies, regardless of their morphological types,
provide a powerful clue toward distinguishing SN progenitor scenarios
\citep[e.g.,][]{1994PASP..106.1276B,1996AJ....111.2017V,
2001AstL...27..411T,2008MNRAS.390.1527A,2009A&A...508.1259H,
2011ApJ...731L...4M,2011A&A...530A..95L}.
The SN location is usually provided by the SN catalogs via its offset
from the host galaxy nucleus, in the East/West (E/W) and North/South (N/S)
directions, in arcsec \citep[e.g.,][]{1999A&AS..139..531B,2004AstL...30..729T}.
Ideally, the equatorial coordinates of an SN can be easily derived
from the coordinates of its host galaxy and its offset.
Alternatively, SN offsets $(\Delta\alpha, \Delta\delta)$ can be simply calculated by
${\rm \Delta\alpha \approx (\alpha_{SN} - \alpha_{g})\cos\delta_{g}}$
and ${\rm \Delta\delta \approx (\delta_{SN} - \delta_{g})}$,
where ${\rm \alpha_{SN}}$ and ${\rm \delta_{SN}}$
are the SN coordinates, ${\rm \alpha_{g}}$ and ${\rm \delta_{g}}$
are the host galaxy coordinates in equatorial system.
However, the astrometric data given in the SN catalogs have various degrees of accuracy
depending on many factors.
Below we present and discuss several examples,
when there are contradictions in SN offsets among different sources.

In the ASC, the coordinates of SN~1987F are
${\rm \alpha_{SN} = 12^{h}41^{m}38\fs99}$,
${\rm \delta_{SN} = +26^\circ04'22\farcs4}$
(J2000.0)\footnote{We always adopt J2000.0 as the system of equatorial coordinates.},
and the SN is located at
$20''$ East and $5''$ South from the nucleus of
PGC~042584 (NGC~4615).
Originally, this SN was discovered on March~22 at about $20''$ East of the nucleus
of host galaxy \citep{1987IAUC.4381....1C}.
On April~23, the object was independently rediscovered at
$24''$ East and $6''$ South of the host nucleus \citep{1987IAUC.4374....1W}.
The latter offset is given in the SSC.
Using the SDSS images, we identified the nucleus of PGC~042584 at
${\rm \alpha_{g} = 12^{h}41^{m}37\fs31}$,
${\rm \delta_{g} = +26^\circ04'22\farcs1}$.
Given the absolute positions of the SN and the nucleus of its host galaxy,
we deduce the offset of SN~1987F to be
$22\farcs6$ East and $0\farcs3$ North.
This leads to a $5''$ difference in the $\Delta\delta$
component of the SN offset from that listed in the SN catalogs.

The ASC lists the position of SN~2001en at
${\rm \alpha_{SN} = 01^{h}25^{m}22\fs90}$,
${\rm \delta_{SN} = +34^\circ01'30\farcs5}$, while, according to
\cite{2001IAUC.7724....1H}, it is located
$6\farcs4$ East and $2\farcs8$ South from
the eastern nucleus of PGC~005268 (NGC~523).
The latter offset is given also in the SSC.
The SDSS image of this galaxy
(see Fig.~\ref{disturb}) suggests that it has a peculiar structure,
and the physical center of PGC~005268,
from which the spiral arms begin, is located at
${\rm \alpha_{g} = 01^{h}25^{m}20\fs75}$,
${\rm \delta_{g} = +34^\circ01'29\farcs8}$.
Taking into account both the coordinates of the SN and its host galaxy,
as well as definition of SN offset as galactocentric distance
in the corresponding directions \citep{1999A&AS..139..531B,2004AstL...30..729T},
the offset for SN~2001en should be corrected to
$26\farcs7$ East and $0\farcs7$ North.
This means that by not taking the center of the host galaxy as
a starting point for the offset calculation, a difference in
the $\Delta\alpha$ component of the SN real offset is about $20''$,
comparing to that reported in the SN catalogs.

In the ASC, the coordinates of SN~2010br are
${\rm \alpha_{SN} = 12^{h}03^{m}10\fs95}$,
${\rm \delta_{SN} = +44^\circ31'43\farcs1}$, while its
offset is about $19\farcs5$ East and $10''$ South
\citep{2010CBET.2245....1M} from the nucleus of
PGC~038068 (NGC~4051).
The SSC lists the coordinates of the SN as
${\rm \alpha_{SN} = 12^{h}03^{m}10\fs96}$,
${\rm \delta_{SN} = +44^\circ31'42\farcs9}$
and offset as $14\farcs3$ East and $9\farcs7$ South
\citep{2010CBET.2245....2M}.
We measure the accurate coordinates of the nucleus of PGC~038068 to be
${\rm \alpha_{g} = 12^{h}03^{m}09\fs61}$,
${\rm \delta_{g} = +44^\circ31'52\farcs6}$,
yielding an offset of
$14\farcs3$ East and $9\farcs5$ South,
which is consistent with that reported in the SSC
\citep{2010CBET.2245....2M},
but different in the $\Delta\alpha$ component of about $5''$
from that reported in the ASC.

The ASC collects 2352 offsets, and 3179 coordinates
for SNe in the sample of identified host galaxies.
At the same time, for these SNe, the SSC includes 2285 offsets,
and 3496 coordinates. In total, 150 SNe have
discrepancies in the offset components $\Delta\alpha$ or $\Delta\delta$
greater than $4''$ from those that we derived from our accurate measurements of
the coordinates of the SNe and their host galaxy nuclei.
The \emph{total sample} also includes 340 SNe with
unidentified hosts in the SDSS.
For these SNe, the two catalogs collect 91 offsets
and 339 coordinates.
Among all SNe of the \emph{total sample}, 40 show offset discrepancies
of more than $4''$, in at least one component,
between the ASC and SSC values.

These examples of inconsistency of SNe offsets clearly show
that the information in the SN catalogs is not always correct,
which can affect any statistical study based on the use of offsets.
For this reason, we have conducted a wide search of SNe offset data
in the literature (mostly CBAT and IAU Circulars) for
all cases where discrepancies greater than
$4''$ were found in $\Delta\alpha$ or $\Delta\delta$.
Our aims were to find correct offsets and to
flag the cases where well-defined data could not be determined.
In general, the ASC offsets are more reliable than the SSC ones, so
offsets from the ASC were mostly adopted if
they agreed with the SSC within $4''$, otherwise we used our own measurements.

In the end, our database contains in total 3599 SNe
with offset data; 2419 SNe with available offsets,
and 1180 SNe with offsets determined by us.
During the investigation, we corrected the offsets of 43 SNe.
We also flagged offsets as uncertain (``:'') for 90 SNe that show
a large dispersion in the offsets
(in the ASC, SSC, CBAT as well as in our determination).
In addition, we calculated unavailable coordinates for 332 SNe,
via correct offsets and accurate coordinates of identified nuclei of
their host galaxies.
We flagged these coordinates as uncertain (``:''), because the precise
determination of the position of the host galaxy nucleus is difficult
and depends on many factors
(e.g., color of image, plate saturation, galaxy peculiarity etc.).
Moreover, the SN catalogs report different offsets with different levels of accuracy.
This information should be considered when analyzing individual accurate
locations of these SNe (e.g., possible associations with \ion{H}{II} regions,
spiral arms etc.).

\subsection{Host galaxy morphologies}
\label{morphdata}

The galaxy classification methodology used in this paper
is based on classification scheme used by
\citet{2007ApJS..170...33P} and is
applied on the SDSS DR7 and DR8 RGB images.
We classified SN host galaxies
using the modified Hubble sequence
(E-E/S0-S0-S0/a-Sa-Sab-Sb-Sbc-Sc-Scd-Sd-Sdm-Sm-Im),
at the same time using morphological information in the SN
catalogs as well as HyperLeda, along
with images for galaxies obtained with NED.
For mostly high-redshift hosts, we classified galaxies as ``S''
when it was not possible to distinguish between the various Hubble
subclasses of lenticulars or spirals.
Some classifications are noted by symbols of uncertainties
adopted from the RC3 scheme. Symbol ``:'' indicates that the
classification is doubtful, ``pec'' indicates that the galaxy
is peculiar (presence of shells, tidal tails etc.), and
finally ``?'' indicates that the classification is highly
uncertain. In addition, the host galaxies in interacting
or merging systems have been flagged as ``\emph{inter}''
or ``\emph{merg}'', respectively.

We classified 2104 host galaxies, corresponding to  63\% of our sample
of galaxies identified in the SDSS.
The remaining 1236 objects were not classified because of their small
angular sizes $(\leq$$5'')$.
In Table~\ref{Gmorph}, we
present the distributions of morphological types and barred structure
of the classified SN host galaxies.
The sample also includes 73 hosts in interacting
(``\emph{inter}''), and 56 hosts in merging (``\emph{merg}'') systems.

In order to test our morphological classification and detection
of bar structure, a set of 100 host galaxies is randomly selected from
the classified sample and re-classified.
We estimated our mean confidence, comparing re-classification
with the earlier classification of the same galaxies.
This step allows estimating relative biases in our classification.
The chances of our failure to detect bars and
misclassification greater than two morphological units are both roughly 2\%.

\subsection{Isophotal measurements of host galaxies}

As one of our goals is to conduct detailed studies of
the SNe distribution in different types of hosts,
we require a galaxy to have well-defined
apparent magnitude, major axis ($D_{25}$),
axial ratio ($b/a$), and position angle (PA).
To measure these parameters on $g$-band images of SDSS DR8,
${\rm 25\,\ mag\,\ arcsec^{-2}}$ isophotes were constructed using
the Graphical Astronomy and Image
Analysis\footnote{{\tiny GAIA is available for download as part of
JAC Starlink Release at \texttt{http://starlink.jach.hawaii.edu}}}
(GAIA) tool.

\begin{table}[t]
\begin{center}
\caption{Distributions of morphological types and
barred structure of classified host galaxies.
\label{Gmorph}}
\begin{tabular}{lrrr}
\hline
\hline
\multicolumn{1}{l}{Morphological type} & \multicolumn{1}{c}{With bars} & \multicolumn{1}{c}{Without bars} & \multicolumn{1}{r}{All} \\
\hline
E & 0 \,\ \,\ \,\ \,\ & 69 \,\ \,\ \,\ \,\ & 69 \\
E/S0 & 0 \,\ \,\ \,\ \,\ & 45 \,\ \,\ \,\ \,\ & 45 \\
S0 & 6 \,\ \,\ \,\ \,\ & 68 \,\ \,\ \,\ \,\ & 74 \\
S0/a & 19 \,\ \,\ \,\ \,\ & 76 \,\ \,\ \,\ \,\ & 95 \\
Sa & 18 \,\ \,\ \,\ \,\ & 48 \,\ \,\ \,\ \,\ & 66 \\
Sab & 30 \,\ \,\ \,\ \,\ & 58 \,\ \,\ \,\ \,\ & 88 \\
Sb & 126 \,\ \,\ \,\ \,\ & 186 \,\ \,\ \,\ \,\ & 312 \\
Sbc & 105 \,\ \,\ \,\ \,\ & 236 \,\ \,\ \,\ \,\ & 341 \\
Sc & 88 \,\ \,\ \,\ \,\ & 297 \,\ \,\ \,\ \,\ & 385 \\
Scd & 23 \,\ \,\ \,\ \,\ & 74 \,\ \,\ \,\ \,\ & 97 \\
Sd & 60 \,\ \,\ \,\ \,\ & 33 \,\ \,\ \,\ \,\ & 93 \\
Sdm & 28 \,\ \,\ \,\ \,\ & 20 \,\ \,\ \,\ \,\ & 48 \\
Sm & 7 \,\ \,\ \,\ \,\ & 20 \,\ \,\ \,\ \,\ & 27 \\
Im & 0 \,\ \,\ \,\ \,\ & 27 \,\ \,\ \,\ \,\ & 27 \\
\hline
All & 510 \,\ \,\ \,\ \,\ & 1257 \,\ \,\ \,\ \,\ & 1767 \\
\hline \\
\end{tabular}
\parbox{\hsize}{\textbf{Notes.} Our sample includes also 337 ``\emph{disk-like}''
galaxies with ``S'' classification, among which
12 show a barred structure.}
\end{center}
\end{table}

GAIA is a highly interactive image display tool with the additional
capability of being extensible to integrate other programs.
Its  photometry package,
PHOTOM\footnote{\texttt{{\tiny
http://www.starlink.ac.uk/docs/sun45.htx/sun45.html}}},
provides a possibility to measure aperture photometry
with a highly interactive
environment for controlling the positions,
sizes and orientations of circular and elliptical apertures.

Starting from the measured ${\rm 25\,\ mag\,\ arcsec^{-2}}$
isophotes, elliptical apertures were centered on
the $g$-band galaxy centroid position (obtained manually).
Major axes ($D_{25}$), axial ratios ($b/a$), and position angles (PA)
of the galaxies were then obtained via PHOTOM in GAIA.
All the elliptical apertures were also checked on host galaxy
image by eye to ensure that they were in good fit with them.
Apparent magnitudes were calculated via the total flux within the
elliptical aperture.
During the photometric measurements we masked
out bright projected/saturated stars.
In addition, apparent magnitudes and major axes
were corrected for Galactic and host galaxy internal extinction
(\citealp{1998ApJ...500..525S,1995A&A...296...64B}, respectively).

The SDSS images of 6 objects were not used, because of the presence of bright
stars and background galaxies projected inside the elliptical apertures.
It was difficult to remove these objects from the
images without leaving residuals.
Also, 31 galaxies have such peculiar geometries that it was not possible to
measure isophotal quantities using elliptical aperture techniques.
Moreover, 37 host galaxies, often with large angular sizes,
were fragmented into separate SDSS fields, and it was not possible to apply
the deprojection technique on the full extent of these galaxies.

In total, we have measured $g$-band magnitudes, major axes ($D_{25}$),
axial ratios ($b/a$), and position angles (PA) for 2030 galaxies among the
2104 morphologically classified ones (96\%).

\begin{figure}[t]
\begin{center}
\includegraphics[width=0.9\hsize]{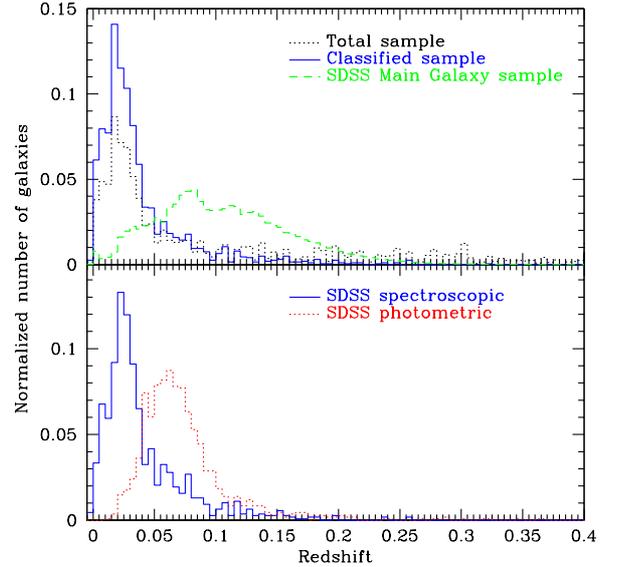}
\end{center}
\caption{\emph{Top}: redshift distributions
of galaxies in the \emph{total} (\emph{black dotted}),
morphologically classified (\emph{blue solid}), and
SDSS Main Galaxy (\emph{green dashed}) samples.
\emph{Bottom}: distribution of SDSS spectroscopic (\emph{blue solid}) and
photometric (\emph{red dotted}) redshifts for
classified host galaxies.
\label{redshift}}
\end{figure}

We now consider possible K$-$corrections to the photometry, but for this we
need to assess the distribution of redshifts.
The redshift completeness of the \emph{total sample} of host
galaxies is high: 3317 of 3679 hosts (90\%) have a redshift;
3232 redshifts were available in
the SN catalogs, but for the 1214 galaxies that also have SDSS redshift
measurements, we adopted the SDSS measurement. Moreover, SDSS provided
redshifts for 85 galaxies that did not have redshifts in the SN catalogs.
The redshift completeness of the morphologically classified sample
is even higher: 2023 of 2104 hosts (96\%) have a redshift.
The majority (93\%) of these host galaxies have $z \leq 0.1$.
In the top panel of Fig.~\ref{redshift},
we separately display the redshift distribution of host galaxies in the
\emph{total} and in morphologically classified samples.
For comparison, the redshift distribution of
$\sim$680000 objects from the SDSS Main Galaxy sample is also shown.

Since the redshifts of classified galaxies are mostly low, their
K$-$corrections are negligible and were ignored.
However, we checked in the PhotoZ table of the SDSS database that the
K$-$corrections of these objects
do not exceed ${\rm 0.2\,\ mag}$ in the $g$-band.
We also compared the redshift distribution
of galaxies, which simultaneously have photometric and spectroscopic redshifts
in the SDSS.
The photometric redshifts are less accurate than spectroscopic redshifts,
and may be sufficient for understanding accuracy
of estimation of the SDSS K$-$correction.
The bottom panel of Fig.~\ref{redshift} clearly shows that photometric
redshifts are overestimated on average by $\sim$0.03,
which suggests that aforementioned value $({\rm 0.2\,\ mag})$
of the SDSS K$-$correction also is overestimated by $\sim$30\%.

To calculate the luminosity distances
and absolute magnitudes of galaxies,
we used the recession velocities both corrected
to the centroid of the Local Group \citep{1977ApJ...217..903Y},
and for infall of the Local Group towards Virgo cluster
\citep{1998A&A...340...21T,2002A&A...393...57T}\footnote{We thus neglect the
peculiar velocities of galaxies relative to the Virgo cluster. There could thus
be non-negligible distance errors for galaxies closer than $z\approx 0.01$,
which accounts for less than 4\% of our total sample (see dashed histogram of
top panel of Fig.~\ref{redshift}).}.
For 48 nearby hosts out of 52 with $z \leq 0.003$,
we were able to find distance estimates via
redshift-independent distance indicators
(e.g., derived from Cepheid variables or the
Tully-Fisher relation) using NED and HyperLeda.

If  spiral galaxies have disks that are not perfectly thin, then the measured
axis ratio will not simply be the $b/a=\cos^{-1} i$ for galaxies viewed
with inclination $i$ relative to their polar axis.
We calculate the inclinations of galaxies following
the classical \cite{1926ApJ....64..321H} formula
\begin{eqnarray}
\sin^2 i= \frac {1-{\rm dex}\left[{-2 \log (a/b)}\right]}{1-{\rm
    dex}\left[{-2 \log r_0}\right]} \,\ ,
\label{galincl}
\end{eqnarray}
where $i$ is the inclination angle in degrees
between the polar axis and the line of sight and $r_0$ is the intrinsic
axis-ratio $a/b$ of galaxies viewed edge-on.
According to \cite{1997A&AS..124..109P},
$\log (r_0)  =0.43+0.053 t$ for $-5\leq t\leq 7$
and $\log (r_0) =0.38$ for $t>7$, where $t$ is
the de~Vaucouleurs morphological type.
We adopt $t = 5$ for galaxies classified as ``S''
classification, because this is the most frequent type in other catalogs.
This simple formula works well for most of disk
galaxies, but probably not for Irregulars and Ellipticals
\citep{1988PASP..100..344V}.

\begin{figure}[t]
\begin{center}$
\begin{array}{cc}
\includegraphics[width=0.48\hsize]{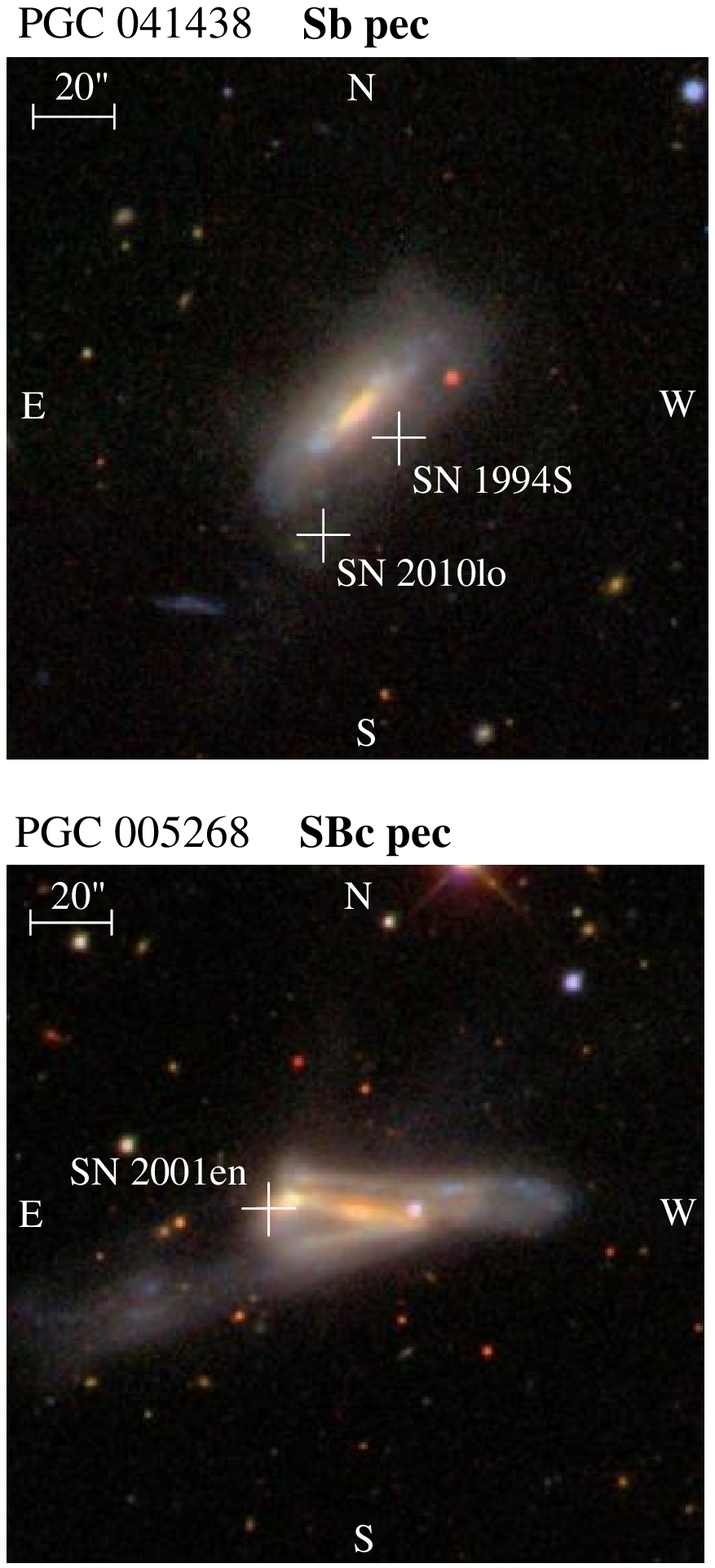} &
\includegraphics[width=0.48\hsize]{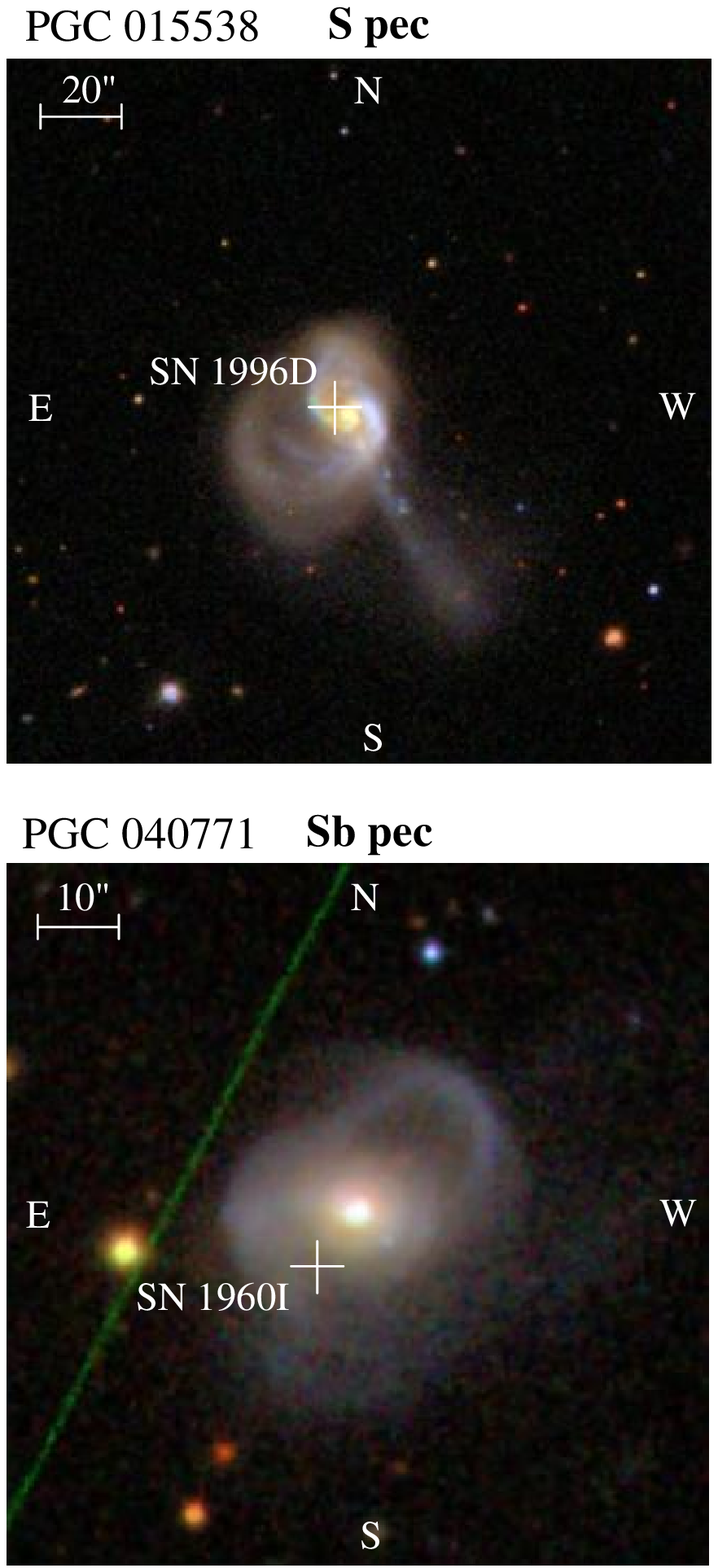}
\end{array}$
\end{center}
\caption{Examples of SDSS DR8 images representing
disturbed (diffused) or lopsided spiral
disk of SN host galaxies. The PGC objects identifiers and
our morphological classification are listed
at the top.
The SN names and positions (marked by a \emph{cross sign})
are also shown.
In all images, north is up and east to the left.}
\label{disturb}
\end{figure}

Many CC SN studies are based on the assumption that these
SNe have young progenitors that are located
in the disks of spiral galaxies
\citep[e.g.,][]{1997AJ....113..197V,2009MNRAS.399..559A,
2009A&A...508.1259H,2010MNRAS.405.2529W}.
The SN radial distances
\citep[e.g.,][]{2009A&A...508.1259H}
as well as metallicity gradients
\citep[e.g.,][]{2009A&A...503..137B}
in disks are usually estimated from the de-projected separations
from host galaxy nuclei, using the inclination correction.
However, in many cases
(clearly seen in 96 hosts), the galaxy disks are
disturbed (diffused) or lopsided and far from the ideal disk structure, which
makes the deprojection less secure.
We therefore flagged these galaxies for
specific investigations in future studies.
In Fig.~\ref{disturb} we show examples of such flagged host galaxies.

\subsection{Nuclear activity levels of host galaxies}

The cross-matching of the SN catalogs with the SDSS DR7 and DR8
provides us with the spectra of the nuclei\footnote{The $3''$ fiber sees out
to a projected radius of only 865 pc for
$z=0.03$ (the median of our sample) but as far as 2.7 kpc for $z=0.1$. So,
the spectra are not limited to the nuclei. Nevertheless, if there is nuclear
activity, this should dominate the non-nuclear emission in the spectrum.}
of SN host galaxies.
We visually inspected the images from the SDSS Imaging Server
to exclude the galaxies whose spectra were offset from their nuclei.
Among the 3340 identified host galaxies, 1206
hosts (with 1287 SNe) have SDSS nuclear spectra\footnote{Only 1214 galaxies
in our sample have SDSS spectra, and 8 of these are not in the nucleus of
the galaxy.}, which were analyzed to diagnose the central power source of the galaxies.

For each narrow emission-line galaxy, we have used the STARLIGHT spectral
synthesis code \citep{2004ApJ...605..105C,2005MNRAS.358..363C} to model the
stellar spectral energy distribution (SED) for each SDSS spectra.
The best-fitting stellar SED was then subtracted from the observed
spectrum in order to isolate the pure emission line spectrum.
This way, in many cases, even weak emission lines that seem to be
absent in the observed spectra could be detected and measured
accurately enough. STARLIGHT uses different techniques, combining
empirical population synthesis and ingredients of evolutionary
synthesis to compute the best-fitting stellar SED. The
best-fitting linear combination of $N_\star$ Single Stellar Populations
(SSPs), is obtained by using a non-uniform sampling of the parameter
space based on the Markov Chain Monte Carlo method, plus an
approach called simulated annealing, and a convergence criteria
similar to that proposed by \cite{gr1992},
to approximately determine the global $\chi^2$ minimum.

\begin{figure}[t]
\begin{center}
\includegraphics[width=0.95\hsize]{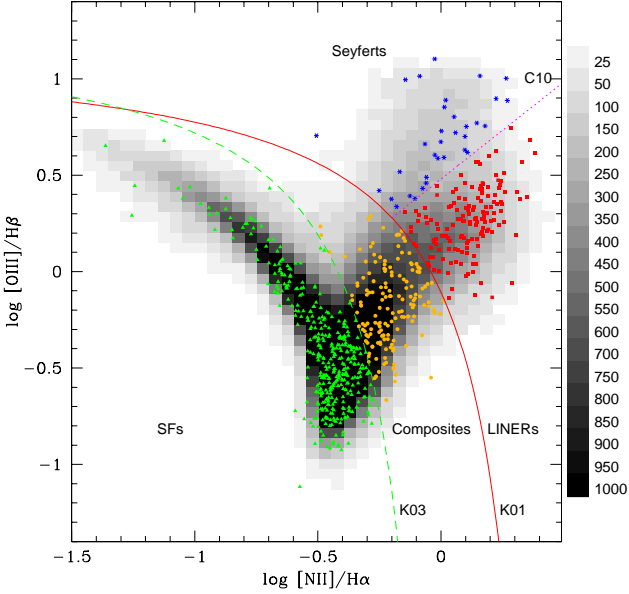}
\end{center}
\caption{BPT spectral diagnostic diagram for 727 host galaxies
with narrow emission-lines. Star forming (SF) and composite galaxies are
displayed with \emph{green triangles} and \emph{orange circles}, respectively.
The \emph{blue asterisks} and \emph{red squares}
respectively are Seyferts and LINERs.
The \emph{red solid} line shows the theoretical demarcation line
separating AGN from SF galaxies proposed by
\citet[K01]{2001ApJ...556..121K}, while the \emph{green dashed} line
is the empirical line proposed by \citet[K03]{2003MNRAS.346.1055K}.
The \emph{magenta dotted} line is the Seyfert-LINER demarcation line
proposed by \citet[C10]{2010MNRAS.403.1036C}.
The underlying density distribution of the SDSS
emission-line galaxies is shown in grayscale.
The grayscale bar in the right represents the number
of galaxies in each density bin.}
\label{BPT_diagram}
\end{figure}

For our analysis, we have chosen  SSPs from
\cite{2003MNRAS.344.1000B}, which are based on the ``Padova
1994'' evolutionary tracks
\citep{1993A&AS...97..851A,1993A&AS..100..647B,1994A&AS..104..365F,
1994A&AS..105...29F,1996A&AS..117..113G} and the \cite{2003PASP..115..763C}
Initial Mass Function (IMF) between 0.1 and $100 \,\ M_\odot$.
The SSP library used here comprises six metallicities
(0.005, 0.02, 0.2, 0.4, 1, and $2.5 \,\ Z_\odot$)
for 25 ages between 1~Myr and 18~Gyr. The intrinsic extinction
has been modeled as a uniform dust screen, adopting the extinction
law by \cite{1989ApJ...345..245C}. Line broadening effects, due to
line-of-sight stellar motions, are accounted
for, in STARLIGHT, by Gaussian convolution.

We classified the spectra, according to enhanced star-forming (SF) or AGN
(Seyfert or LINER), using
the ${\rm [\ion{O}{iii}]\lambda5007/H\beta}$ versus
${\rm [\ion{N}{ii}]\lambda6583/H\alpha}$ standard diagnostic diagram
\citep[][hereafter BPT]{1981PASP...93....5B}.
We adopted the demarcation line proposed by
\citet[hereafter K03]{2003MNRAS.346.1055K} to select SF hosts, and
used the theoretical upper limit of SF galaxies proposed
by \citet[hereafter K01]{2001ApJ...556..121K} to separate AGNs.
The region between the two curves is occupied by
the so-called \emph{composite objects}
\citep[e.g.,][]{2006MNRAS.372..961K,2008ApJ...679...86W},
whose spectra are believed to contain significant
contributions from both SF and AGN.
AGNs were separated by Seyfert-LINERs with the demarcation line
proposed by \citet[hereafter C10]{2010MNRAS.403.1036C}.

We only consider host galaxies with high-quality spectra, i.e. with
signal-to-noise ratio (S/N) $> 5$ in the 4730--4780~{\AA}
continuum, and S/N $\geq 3$ in all emission-lines used in the BPT diagram.
In addition, the SDSS spectra with bad flux calibrations or bad redshift
determinations were also excluded.
Finally, the hosts with broad emission lines
(\emph{FWHM} of ${\rm H\alpha \geq 1000~km~s^{-1}}$), which were assigned
as broad-line AGNs and galaxies without any emission features
were excluded from the BPT diagnostics but considered in studies
of SN host galaxies.
The distribution of 727 narrow emission-line galaxies in the BPT
diagram is illustrated in Fig.~\ref{BPT_diagram}.
For comparison, the density distribution of the SDSS Main Galaxy sample
($\sim$260000 emission-line objects),
which passed our adopted S/N criteria, is also shown.

\begin{figure}[t]
\begin{center}
\includegraphics[width=0.95\hsize]{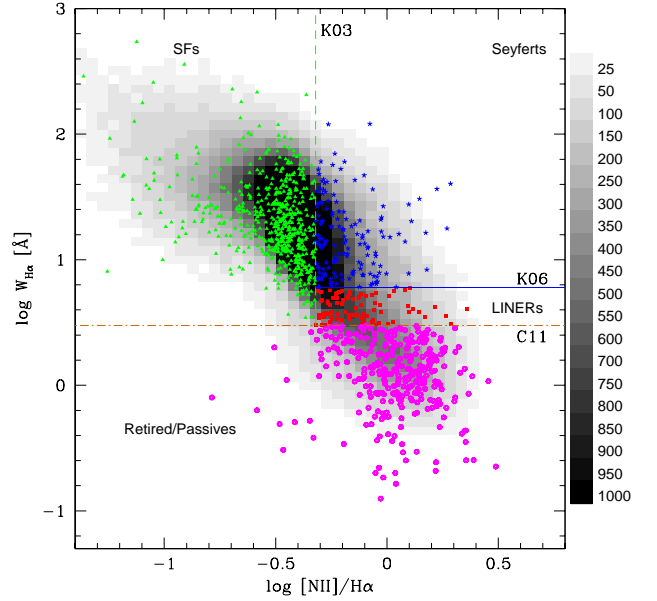}
\end{center}
\caption{WHAN spectral diagnostic diagram for 1106 host galaxies.
SF galaxies are displayed with \emph{green triangles} and
separated by the K03 line.
The \emph{blue asterisks} and \emph{red squares}
respectively are Seyferts and LINERs, which are separated by
\emph{blue solid} line proposed by \citet[K06]{2006MNRAS.372..961K}.
Retired/passive (RP) galaxies are marked by \emph{magenta circles}
and separated by the \emph{orange dashed-dotted} line proposed by
\citet[C11]{2011MNRAS.413.1687C}.
The underlying density distribution of the SDSS
emission-line galaxies is shown in grayscale.
The grayscale bar in the right represents the number
of galaxies in each density bin.}
\label{WHAN}
\end{figure}

However, our strong criteria on S/N prevents us from
including the weak emission-line galaxies.
We therefore also used the WHAN diagram
\citep{2010MNRAS.403.1036C,2011MNRAS.413.1687C}:
the equivalent width of H$\alpha$ (${\rm W_{H\alpha}}$) versus
${\rm [\ion{N}{ii}]\lambda6583/H\alpha}$ to
diagnose the central power source of the hosts including those with weak
emission lines.
A simple transposition strategy was used to plot the K03 demarcation
line in the WHAN diagram \citep{2010MNRAS.403.1036C}.
Here, AGNs were separated by the Seyfert-LINER demarcation line
proposed by \citet[K06]{2006MNRAS.372..961K}.
Also, the line, proposed by \citet[C11]{2011MNRAS.413.1687C},
was used to separate retired/passive (RP) galaxies.
This diagram allows us to plot nearly all (92\%) host
galaxies with available spectra in the SDSS.
The distribution of 1106 galaxies in the WHAN
diagram is presented in Fig.~\ref{WHAN}.

In the BPT diagram, the hosts with SDSS nuclear spectra include
185 narrow-line AGN (33 Seyfert and 152 LINER),
382 SF, and 160 composite galaxies.
The WHAN diagram includes in total
234 narrow-line AGN (151 Seyfert
and 83 LINER), 568 SF, and 304 RP galaxies.
As previously mentioned, we also have 6 broad-line AGNs
and 77 passive (P) galaxies without any emission features.
The activity of nucleus of 17 galaxies
(with 21 SNe) could not be analyzed
due to the poor-quality SDSS spectra.

\subsection{The total sample}

The first 27 (chronologically oldest SNe) and last 3 (most recent SNe) entries
of our \emph{total sample} of SNe and their host galaxies
are shown in Table~\ref{bigbigtable} and the full table is
available electronically.
The full table contains 19 columns for
3876 SNe (3679 host galaxies) and provides the following information:
(Col.~1) designation of SN;
(Col.~2) right ascension of SN (${\rm \alpha_{SN}}$) in degrees;
(Col.~3) declination of SN (${\rm \delta_{SN}}$) in degrees;
(Col.~4) offset (in arcsec) of SN from the host galaxy nucleus;
(Col.~5) spectroscopic type of SN;
(Col.~6) SDSS identification of the host galaxy;
(Col.~7) right ascension of host galaxy (${\rm \alpha_{G}}$) in degrees;
(Col.~8) declination of host galaxy (${\rm \delta_{G}}$) in degrees;
(Col.~9) heliocentric redshift of the host;
(Col.~10) morphological type of the host;
(Col.~11) presence of bar in the host;
(Col.~12) host in interacting (``\emph{inter}'') or in merging (``\emph{merg}'') systems;
(Col.~13) host with disturbed disk structure;
(Col.~14) measured major axis in arcsec (isophotal level of ${\rm
  25\,\ mag\,\ arcsec^{-2}}$ in the $g$-band) of the host;
(Col.~15) axial ratio of the host;
(Col.~16) position angle (in degrees) of the host;
(Col.~17) measured apparent $g$-band magnitude;
(Cols.~18 and 19) nuclear activity class of the host galaxy.

\section{Results and discussion}
\label{Resdiscus}

In this section, we compare some of our measured parameters with previously
available measurements, provide distributions and statistics on several
of them, study their evolution with distance, and finally discuss various
selection effects.
This analysis is important to understand whether our sample of
SN and their host galaxies from the SDSS DR8
is representative of the general population.

Our comparisons with measurements from HyperLeda and SDSS are done by
performing robust linear regressions, with iterative rejection of $3\,\sigma$
outliers, and we also measure a robust estimate of the dispersion, using
the median absolute deviation (MAD) of the
residuals, converting to $\sigma_{\rm MAD} = 1.483\,\rm MAD$, where the
numerical factor is the one appropriate for Gaussian distributions of residuals.

\subsection{Comparison of host galaxy morphological classifications}

It is well known that the central parts of the images of many very bright
galaxies and high surface brightness objects may be over-exposed, which
affects their morphological classification.  Meanwhile, galaxies that are
faint (low surface brightness) can also be misclassified, because of the lack
of precise morphological details in the image.

Using a homogenous sample of 604 SNe,
\cite{2002PASP..114..820V,2003PASP..115.1280V,2005PASP..117..773V}
classified the SN host galaxies from the Lick Observatory Supernova
Search (LOSS) in the David Dunlap Observatory (DDO) morphological
type system. They suggested that to understand
the dependence of SN type on the host galaxy population,
it is more important to obtain accurate morphological classifications
than it is to increase the size of the sample.
For example, among $\sim$800 morphologically classified hosts of CC~SNe,
\cite{2008A&A...488..523H} found 22 cases where the host had been classified
as E or S0. Following a detailed morphological analysis, they found that
among these 22 early-type objects, 17 are in fact misclassified spiral
galaxies, one is a misclassified irregular, and one is
a misclassified ring galaxy, leaving only 3 early-type galaxies\footnote{
One of these 3 SNe, SN~2005md,
reported by \cite{2005CBET..332....1L} and initially classified by
\cite{2005IAUC.8650....2M} as a probable young type~IIb SN,
was shown to be in fact a new Galactic cataclysmic
variable \citep{2010ATel.2750....1L}.}.
In this respect, the host morphology is a crucial parameter in the study of
SN progenitors.

To present a detailed numerical comparison of our morphological
classification of SN host galaxies
with those given in the SN catalogs (mainly from RC3),
we introduce our $t$-type values that we will use in this study.
In comparison to the standard RC3 classification, we have grouped
Ellipticals and Lenticulars into broader classes:
cE and E galaxies are typed together, as well as
S0$^{-}$, S0$^{0}$ and S0$^{+}$ galaxies.
Indeed, using the SDSS images, it was not possible to visually
distinguish the differences between these subclasses.
Table~\ref{ttype} shows the relation between our $t$-types and those of the RC3.

\begin{figure}[h]
\begin{center}
\includegraphics[width=0.9\hsize]{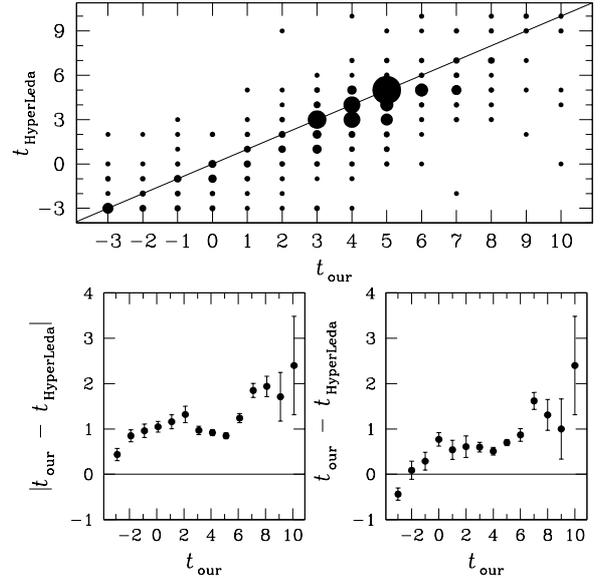}
\end{center}
\caption{\emph{Top}: comparison of
HyperLeda (reordered) versus our
$t$ morphological types for 1313 host galaxies with available
classifications. Point sizes are keyed to the number of objects.
\emph{Bottom}: the distributions of differences of our and
HyperLeda $t$-types according to our classification.
The error bars for the mean values in each bin are presented.
The \emph{solid lines} in each figure are added to visually
better illustrate the deviation in classifications.}
\label{morphRC3}
\end{figure}

In the top panel of Fig.~\ref{morphRC3}, we show the comparison of our
morphological classifications for the 1313 hosts from our
classified sample that are also present in the HyperLeda database (when both
classifications are more accurate than just ``S'': 1767 for our sample among
the total of 2104 classified galaxies).
The reference system of the morphological classification in HyperLeda
is generally the RC3.
The comparisons with RC3 types were performed after converting the RC3
numerical ($t$) classifications to our scheme of $t$ versus type
(see Table~\ref{ttype}).
Point sizes in the figure correspond to the number of
hosts in each morphological bin.
In the bottom panel of Fig.~\ref{morphRC3}, we present the
distributions of the differences between the  HyperLeda $t$-types and ours.

\begin{table}[t]
\begin{center}
\caption{Relation between the RC3 morphological types and $t$ values with ours.
\label{ttype}}
\begin{tabular}{lrclr}
\hline
\hline
\multicolumn{2}{c}{Ours} & & \multicolumn{2}{c}{RC3} \\
\cline{1-2}
\cline{4-5}
\multicolumn{1}{c}{Type} & \multicolumn{1}{c}{$t$} & &
\multicolumn{1}{c}{Type} & \multicolumn{1}{c}{$t$} \\
\hline
E & --3 & & cE & --6 \\
E & --3 & & E & --5 \\
E/S0 & --2 & & E$^{+}$ & --4 \\
S0 & --1 & & S0$^{-}$ & --3 \\
S0 & --1 & & S0$^{0}$ & --2 \\
S0 & --1 & & S0$^{+}$ & --1 \\
S0/a & 0 & & S0/a & 0 \\
Sa & 1 & & Sa & 1 \\
Sab & 2 & & Sab & 2 \\
Sb & 3 & & Sb & 3 \\
Sbc & 4 & & Sbc & 4 \\
Sc & 5 & & Sc & 5 \\
Scd & 6 & & Scd & 6 \\
Sd & 7 & & Sd & 7 \\
Sdm & 8 & & Sdm & 8 \\
Sm & 9 & & Sm & 9 \\
Im & 10 & & Im & 10 \\
\hline
\end{tabular}
\end{center}
\end{table}
\begin{figure*}[ht]
\begin{center}$
\begin{array}{rlrl}
\includegraphics[width=0.24\hsize]{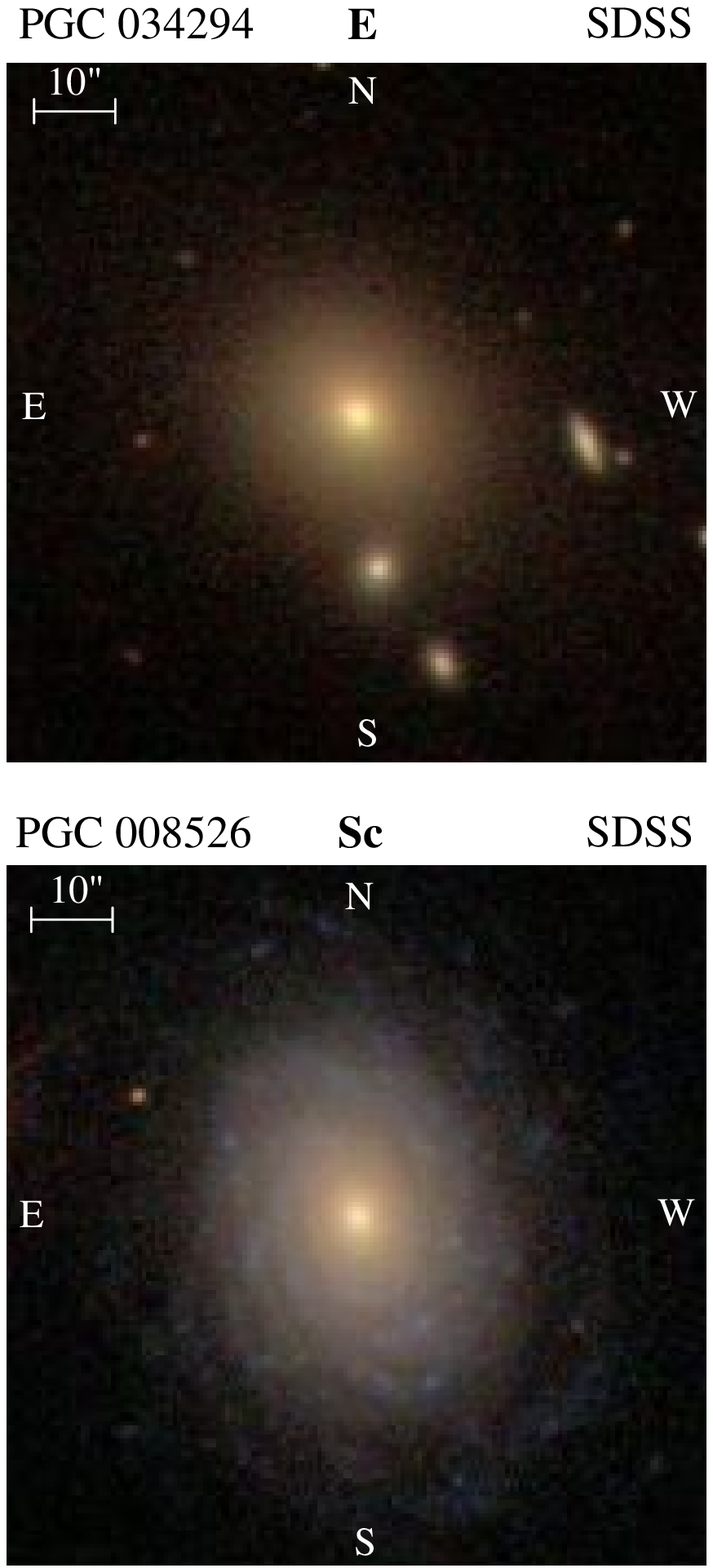} &
\includegraphics[width=0.24\hsize]{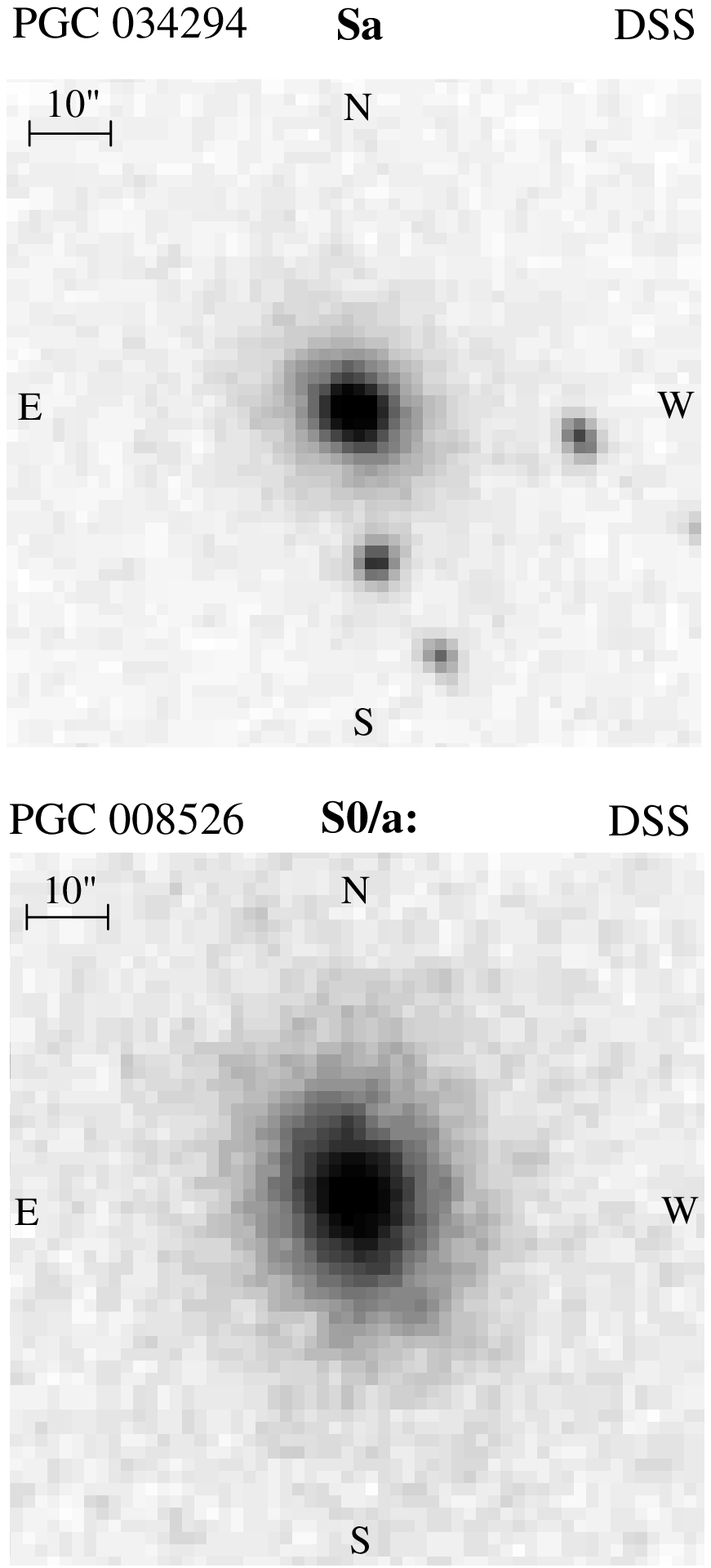} &
\includegraphics[width=0.24\hsize]{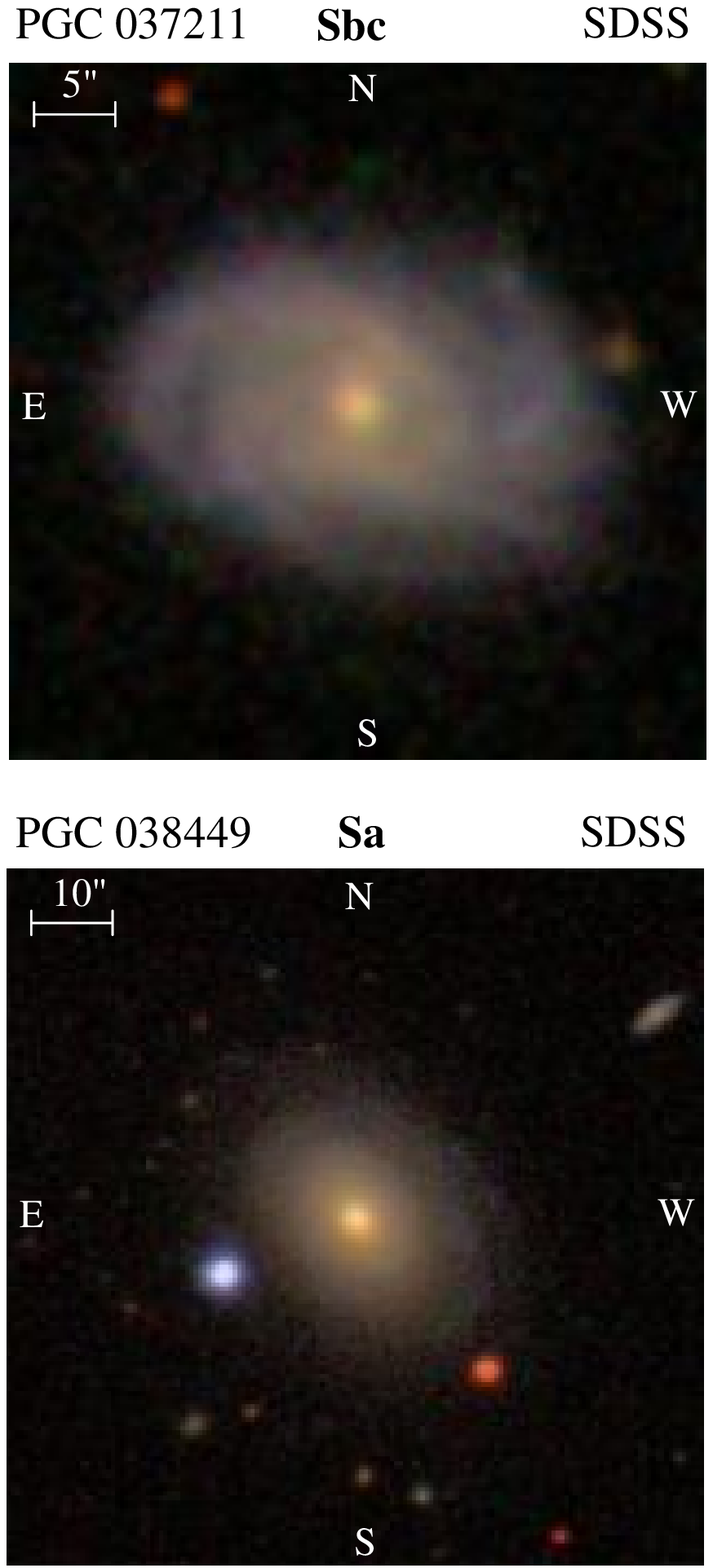} &
\includegraphics[width=0.24\hsize]{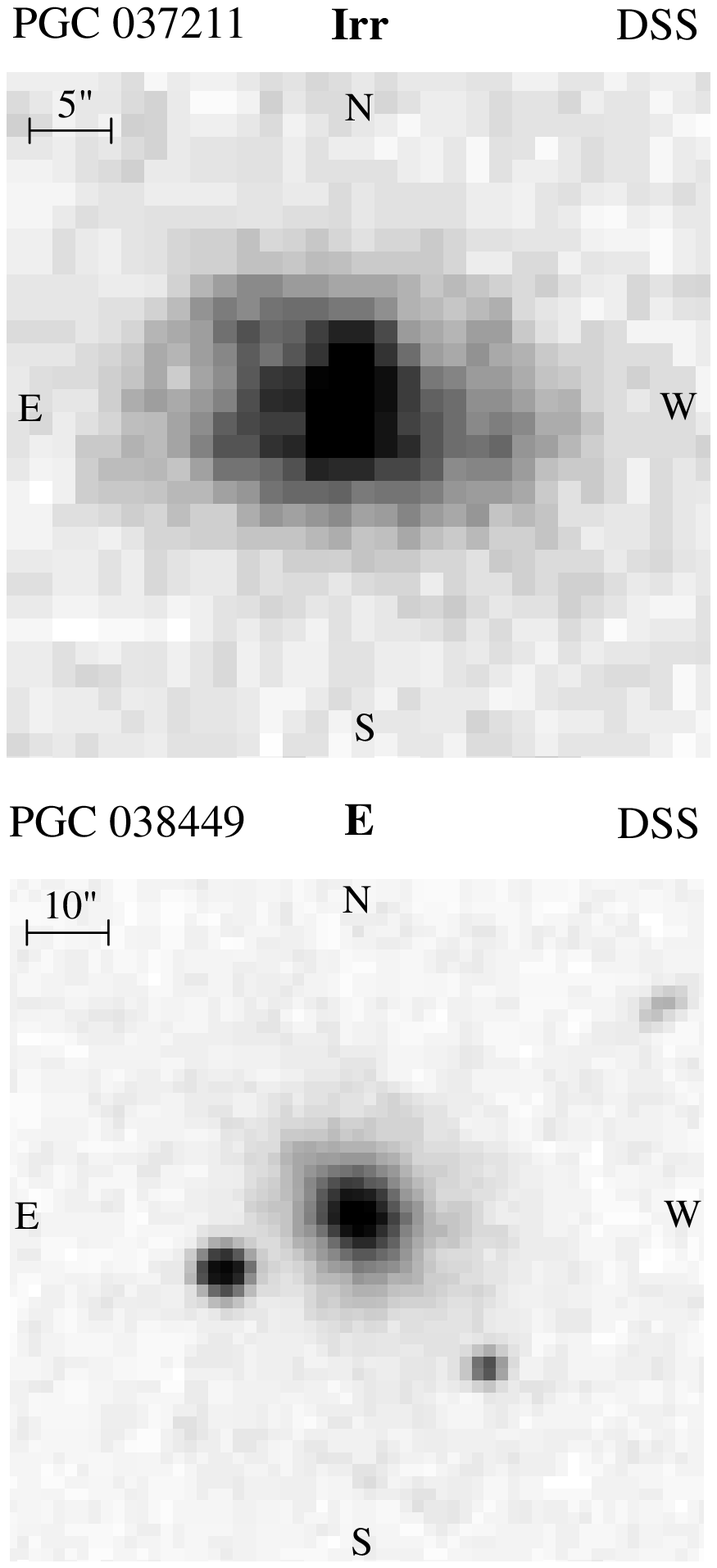}
\end{array}$
\end{center}
\caption{Examples of the SDSS DR8 and DSS~I images of SN host galaxies
representing the cases when $| t_{\rm our}-t_{\rm HyperLeda} | \geq 3$.
The PGC object identifier is listed at the top
with our (\emph{left}) and the HyperLeda (mainly selected from the RC3, \emph{right})
classifications. In all images, north is up and east to the left.}
\label{fmorph}
\end{figure*}

Inspection of Fig.~\ref{morphRC3} shows a trend for our
classifications to be later overall (except E) in comparison with those of
HyperLeda.
The mean deviation in classifications is $0.65 \pm 0.04$ $t$-types.
Meanwhile, the mean absolute deviation is $1.05 \pm 0.03$ $t$-types.
Table~\ref{tab:compare} shows the results of a linear regression between the
HyperLeda types and ours. The relation between the two measures of $t$ has a
slope significantly different than unity, with a best fit value of 0.93, and
the residuals from this trend have a dispersion close to 1.5 types.

A similar trend was already found by \cite{2010ApJS..186..427N},
who recently released a morphological catalog of 14034 visually
classified galaxies $(0.01 < z < 0.1)$.

\begin{table}[t]
\begin{center}
\caption{Comparisons of HyperLeda and SDSS~DR7 measurements with ours for
  SN host galaxies. \label{tab:compare}}
\tabcolsep 4pt
\begin{tabular}{llrrr@{\hspace{5.5mm}}}
\hline
\hline
Quantity & Ref. cat. & \multicolumn{1}{c}{$a$} & \multicolumn{1}{c}{$b$} & \multicolumn{1}{c}{Dispersion} \\
\hline
$t$ & HyperLeda & $0.93\pm0.01$ & $-0.29\pm0.05$ & 1.48 \\
$\log a$ & HyperLeda & $0.93\pm0.00$ & $0.06\pm0.01$ & 0.06 \\
$\log a$ & SDSS & $0.87\pm0.01$ & $0.25\pm0.01$ & 0.07 \\
$b/a$ & HyperLeda & $0.94\pm0.01$ & $0.04\pm0.01$ & 0.08 \\
$b/a$ & SDSS & $0.97\pm0.01$ & $0.01\pm0.01$ & 0.07 \\
PA & HyperLeda & $1.00\pm0.00$ & $-0.94\pm0.49$ & 5.72 \\
PA & SDSS & $1.00\pm0.00$ & $-0.56\pm0.47$ & 4.57 \\
mag & HyperLeda & $0.98\pm0.00$ & $0.66\pm0.06$ & 0.22 \\
mag & SDSS & $0.87\pm0.00$ & $2.01\pm0.06$ & 0.25 \\
\hline \\
\end{tabular}
\parbox{\hsize}{\textbf{Notes.}
Columns~3 ($a$) and 4 ($b$) represent the robust linear fits
(with iterative rejection of outliers) of $x_{\rm theirs} = a\,x_{\rm ours}+b$.
The last column (Dispersion) is computed as 1.483 times the median
absolute deviation (which corresponds to $\sigma$ for Gaussian
distributions) of the residuals from our best-fit trend.
Cases with no measurements (given arbitrary values such as 99)
have been discarded. The SDSS magnitude is cModelMag in the $g$-band.
The PA dispersion is in degrees.}
\end{center}
\end{table}
\begin{figure}[t]
\begin{center}
\includegraphics[width=0.9\hsize]{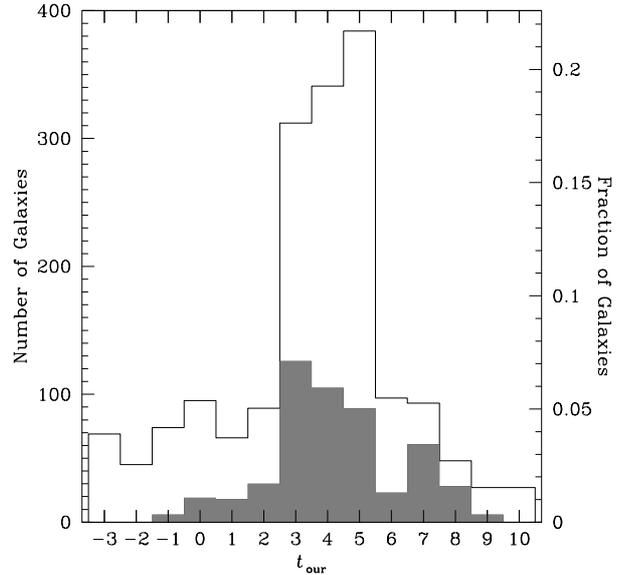}
\end{center}
\caption{Distributions of morphological types and
presence of bars (\emph{shaded region}) in
1767 classified hosts.}
\label{morphbar}
\end{figure}

We found that approximately 10\% (125 hosts)
of the 1313 galaxies in common with HyperLeda have
$t$-types that are dramatically different
$(|t_{\rm our}-t_{\rm HyperLeda}| \geq 3)$.
In Fig.~\ref{fmorph}, we present several extreme cases where the difference between
morphological type codes is $\geq 3$. Color images are taken from the SDSS
DR8 on which our classification was performed, while the grayscale images
are from the photographic plates given in the
Digitized Sky Survey~I (DSS~I), as the RC3 classification is mostly based on
these or similar plates.
In many cases, photographic plates suffer from the narrow dynamical range
that causes saturation as well as underexposure, and also from their
non-linear response functions \citep[e.g.,][]{1995MNRAS.274.1107N}.
A detailed comparative study of the SDSS and DSS images of hosts, when
$|t_{\rm our}-t_{\rm HyperLeda}| \geq 3$, allows us to emphasize that
in nearly all cases, the overexposure as well as low resolution of
the photographic plates cause late-type galaxies of high surface brightness
to be misclassified as early-type in the RC3.
We have found a handful of cases with the opposite trend:
E/S0 misclassified as spirals in RC3,
mainly due to the heterogeneous nature of
morphological data sets in the HyperLeda.

\cite{2011A&A...532A..74B} have recently released the EFIGI
(Extraction de Formes Id\'{e}alis\'{e}es de Galaxies en Imagerie)
catalogue; a multi-wavelength database specifically designed to
densely sample all Hubble types. Their imaging data were obtained
from the SDSS DR4 for a sample of 4458 PGC galaxies.
This catalog includes 453 galaxies from our sample.
We found very good agreement between
the EFIGI  morphological classifications and ours.

\begin{figure}[t]
\begin{center}
\includegraphics[width=0.95\hsize]{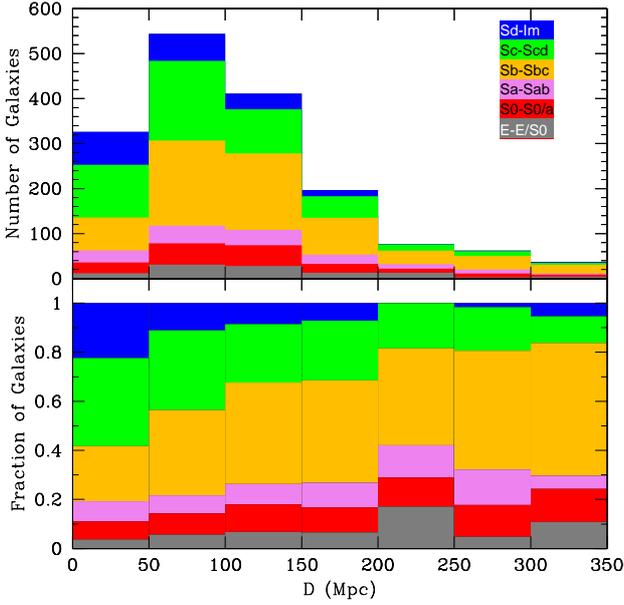}
\end{center}
\caption{\emph{Top}: distribution of  morphological types as
a function of distance.
The types of SN host galaxies have been grouped into the
following broad classes: E and E/S0 (\emph{gray}), S0 and S0/a (\emph{red}),
Sa and Sab (\emph{magenta}), Sb and Sbc (\emph{orange}), Sc and Scd (\emph{green}),
and hosts Sd to Im (\emph{blue}).
\emph{Bottom}: fractional
distribution of morphological types (with the same morphological groups)
as a function of distance.}
\label{dmorph}
\end{figure}

Fig.~\ref{morphbar} shows the distribution (see Table~\ref{Gmorph})
of SN host galaxies with respect to our $t$-type.
This histogram shows that the intermediate types Sb, Sbc,
and Sc are the most frequent.
Only Sm and Im types, which are intrinsically faint,
have fewer than 30 galaxies per type.

In the top panel of Fig.~\ref{dmorph} we present the histogram of morphological types as
a function of distance.
The types of host galaxies have been grouped into the
following broad classes: E and E/S0 (gray), S0 and S0/a (red),
Sa and Sab (magenta), Sb and Sbc (orange), Sc and Scd (green),
and hosts Sd to Im (blue). The bottom panel of Fig.~\ref{dmorph} shows the fractional
distribution of types of the same morphological groups as a function of distance.
Host galaxies with $D < 200~{\rm Mpc}$ include nearly the whole range in morphological
types. At the same time, the late-type hosts are preferentially distributed in the
lower distance bins, while early-types are more populated at the higher distances.
At distances $D < 200~{\rm Mpc}$,  Sd to Im type hosts represent
12\% of the classified galaxies.

At large distances, spiral galaxies can be under-represented, because spiral
arms are difficult to resolve
\citep[see, e.g.,][]{2010ApJS..186..427N,2011A&A...532A..74B}.
Hence, the most massive/luminous early-type galaxies
prefer the higher distances, while the least massive/luminous late-type galaxies
are more populated at the lower distances.
This is also the selection effect on the SN type: bright SNe Ia, exploding
also in E galaxies, can be found more easily than fainter CC~SNe,
exploding only in late-type galaxies
\citep[e.g.,][]{2011MNRAS.412.1419L}.
Therefore, the sample of CC~SNe hosts is closer on
average than the sample of SNe Ia host galaxies.

\subsection{Comparison of presence of bars in host galaxies}

A proper detection of barred structures of hosts is very important when
constraining the nature of the SNe progenitors by comparing their
distribution within host galaxies with the distributions of stellar
populations and ionized gas in the disks
\citep[e.g.,][]{2005AJ....129.1369P,2009A&A...508.1259H}.
Below, we have carried out a comparative study to find differences
in the detection of barred structures of host galaxies between
our classified sample and HyperLeda.

\begin{figure*}
\begin{center}$
\begin{array}{rlrl}
\includegraphics[width=0.24\hsize]{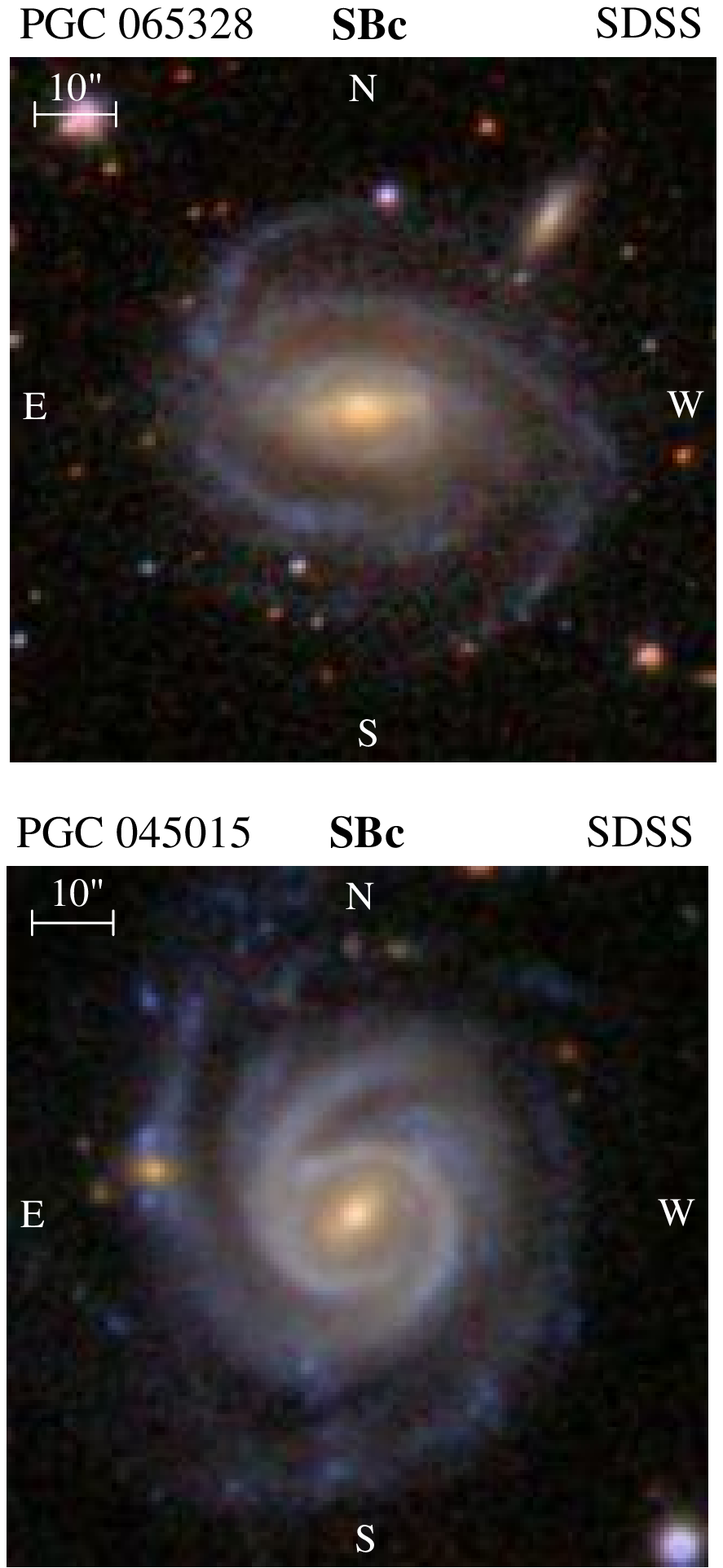} &
\includegraphics[width=0.24\hsize]{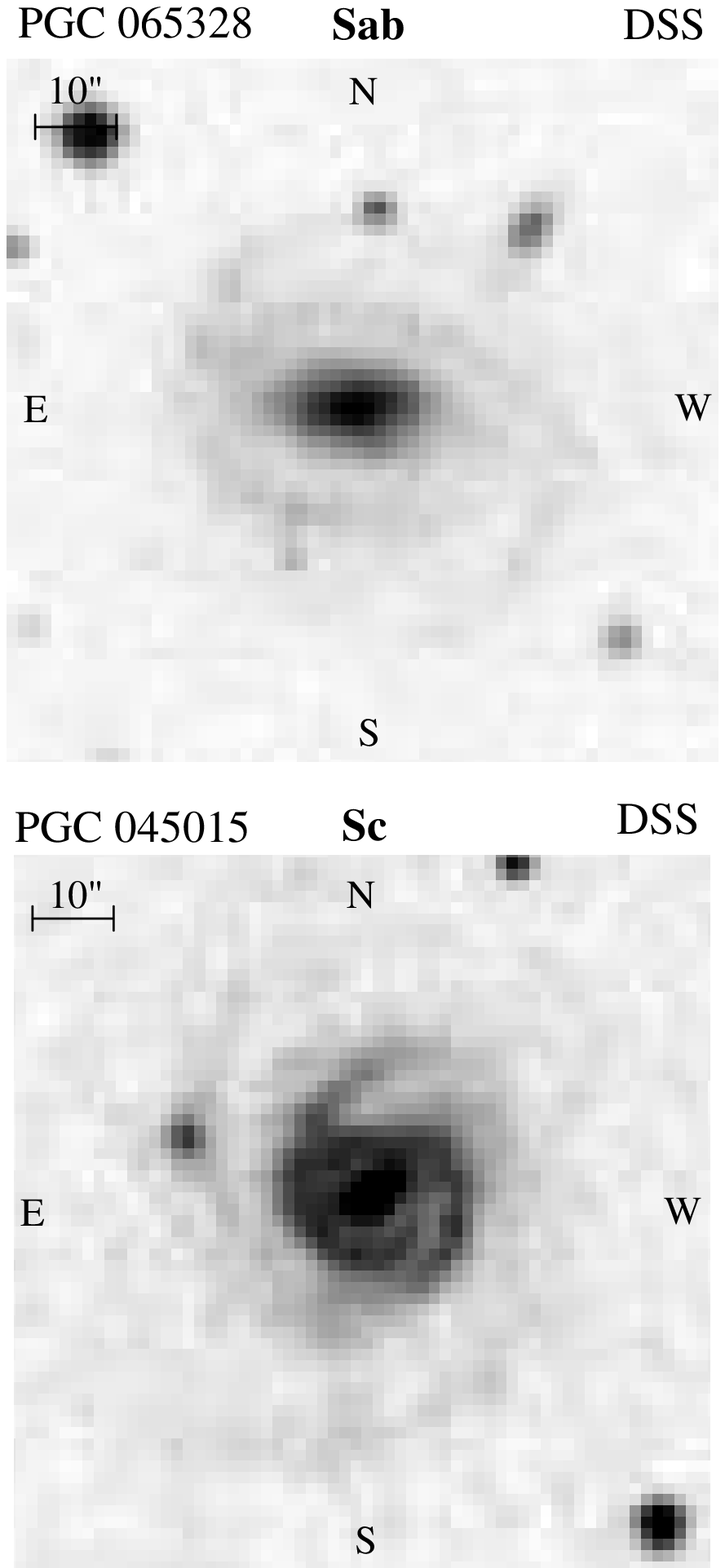} &
\includegraphics[width=0.24\hsize]{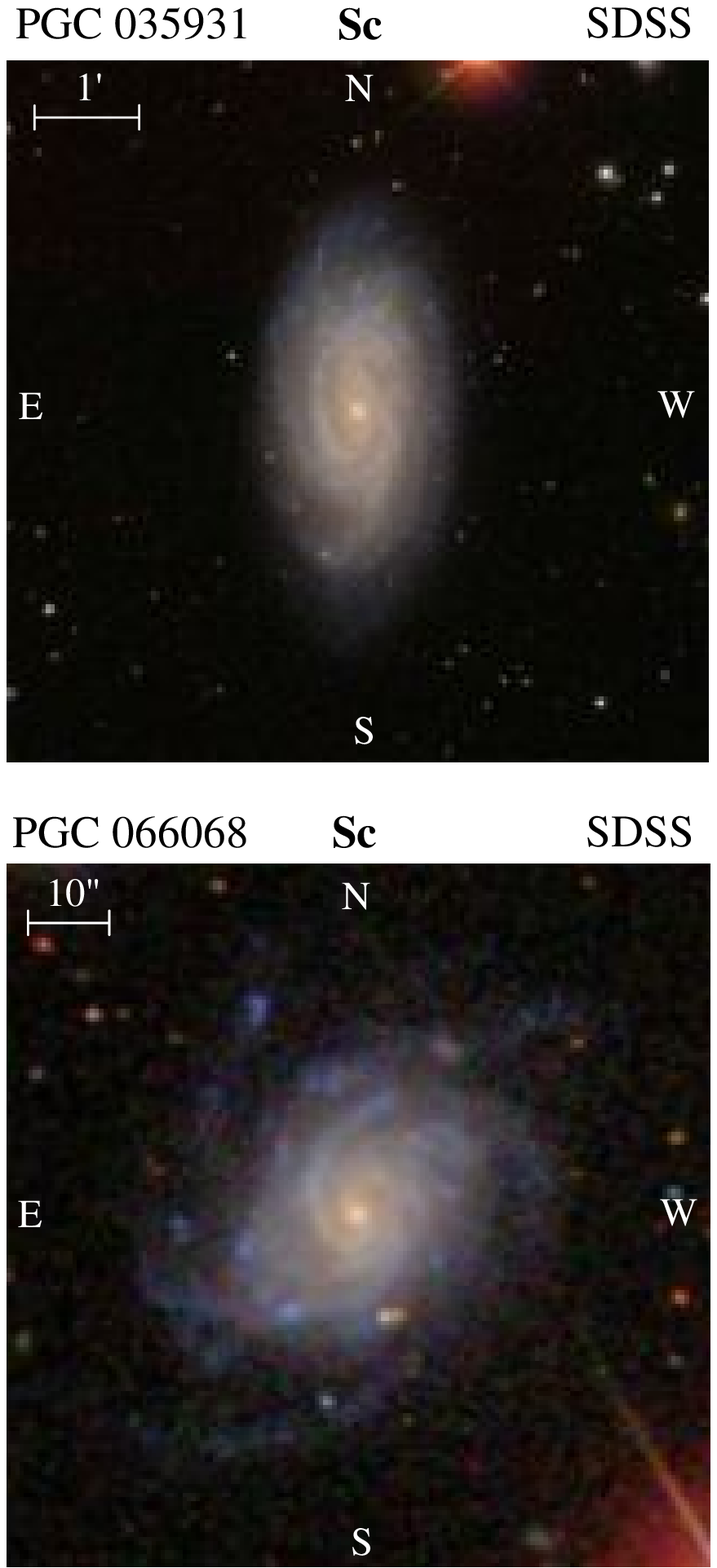} &
\includegraphics[width=0.24\hsize]{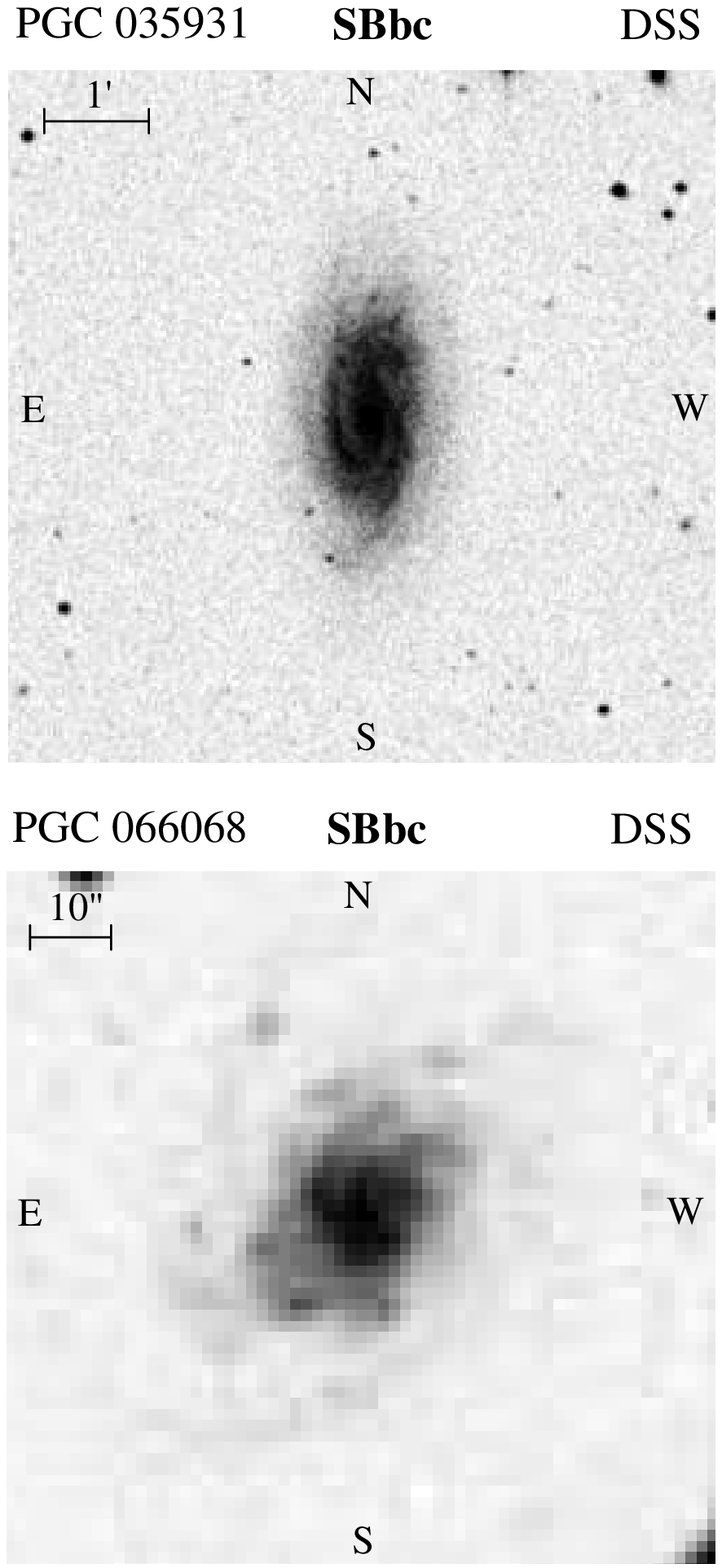}
\end{array}$
\end{center}
\caption{Examples of SDSS DR8 and DSS~I images of SN host galaxies
representing the cases of discrepancies in bar detection between HyperLeda
and us. The PGC object identifier is listed at the top
with our (\emph{left}) and the HyperLeda (mainly selected from the RC3, \emph{right})
classifications. In the two galaxies on the \emph{left}, we detect bars, while
HyperLeda does not, while in the two galaxies on the \emph{right}, HyperLeda
claims that there is a bar while we dot not see any on our higher-resolution
3-band images. In all images, north is up and east to the left.}
\label{bar}
\end{figure*}

Fig.~\ref{morphbar} presents the distribution
(see Table~\ref{Gmorph}) of hosts with or without
bars as a function of $t$-type.
Roughly 29\% of our 1767 classified galaxies have bars.
The barred fraction is highest in types Sb, Sbc, and Sc.
A detailed comparison with HyperLeda reveals that in 378 galaxies
among our 1767 in common (21\%), HyperLeda fails to detect the bar that we
visually detect on the SDSS images or conversely detects a bar when we don't.
In Fig.~\ref{bar}, we present examples of hosts galaxies with discrepancies
in bar detection between HyperLeda and us.
Given their superior angular resolution and 3-colour representations, the SDSS
images offer a much more reliable source for bar detection than do the
plate-based images on which most of the HyperLeda classifications were performed.

Inspecting these cases of discrepancies in bar detection, we conclude
that HyperLeda fails to show bars in both high central surface brightness
early-type galaxies and
late-type galaxies with low surface brightness bars.
We found that bars tend to be incorrectly detected in HyperLeda
galaxies of high inclination ($i>70^\circ$).
The remaining cases of detection discrepancies are again due to the
heterogeneous nature of the HyperLeda data sets.
Note that we may also have missed
weak bars because of inclination effects, or that in
some cases the SDSS images of hosts may be too shallow to detect bars.
For instance, among our S0-Sm galaxies with inclinations $i<70^\circ$, the
average bar fraction is $(37\pm1)\%$ whereas for hosts with inclinations
$\geq70^\circ$ the average bar fraction is only $(11\pm2)\%$.

\subsection{Comparison of isophotal measurements of host galaxies}

We checked the differences of our measurements of $g$-band major axes
with the HyperLeda $B$-band diameters $(D_{25})$
as well as with the SDSS $g$-band isophotal major axes (isoA).
Table~\ref{tab:compare} shows the results of a linear regression between
the HyperLeda, SDSS isophotal measurements and ours.

\begin{figure*}[t]
\begin{center}
\includegraphics[width=0.85\hsize]{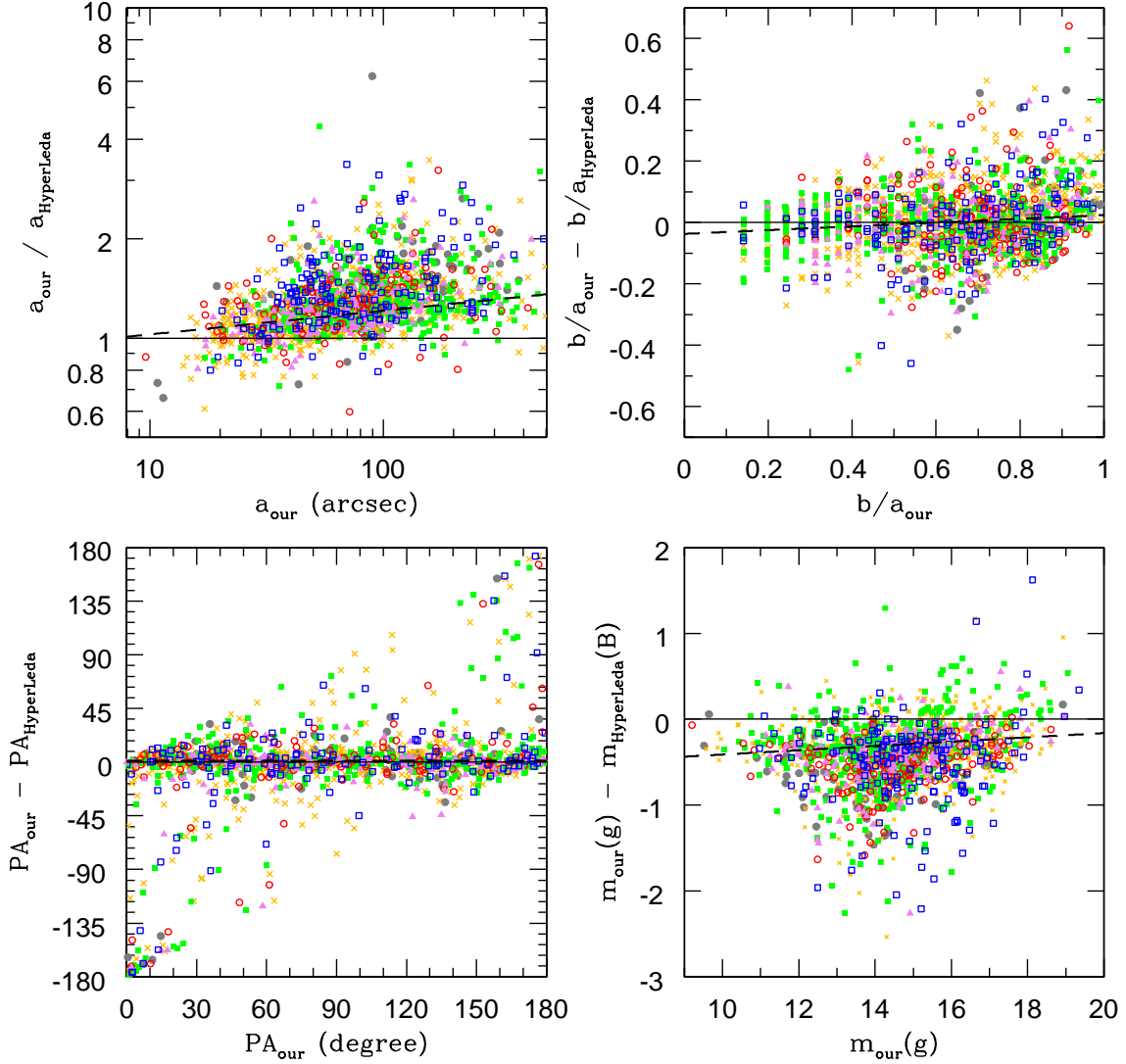}
\end{center}
\caption{\emph{Top left}: comparison between our measurements of major axes of E-Im galaxies and
those of the HyperLeda in the $B$-band.
\emph{Top right}: comparison between measured axial ratios and those of the
HyperLeda.
\emph{Bottom left}: comparison between our measurements of position angles
and those of the HyperLeda.
\emph{Bottom right}: comparison between our measurements of apparent $g$-band magnitudes
and that in the $B$-band of the HyperLeda.
The color coding corresponds to Fig.~\ref{dmorph}:
E-E/S0 (\emph{gray filled circles}), S0-S0/a (\emph{red open circles}),
Sa-Sab (\emph{magenta triangles}), Sb-Sbc (\emph{orange crosses}),
Sc-Scd (\emph{green filled squares}),
and Sd-Im (\emph{blue open squares}).
The \emph{solid lines} in each figure are added to visually
better illustrate the deviations.
The \emph{dashed lines} are best fit linear trends from Table~\ref{tab:compare}.
\label{galourHL}}
\end{figure*}

\begin{figure*}[t]
\begin{center}
\includegraphics[width=0.85\hsize]{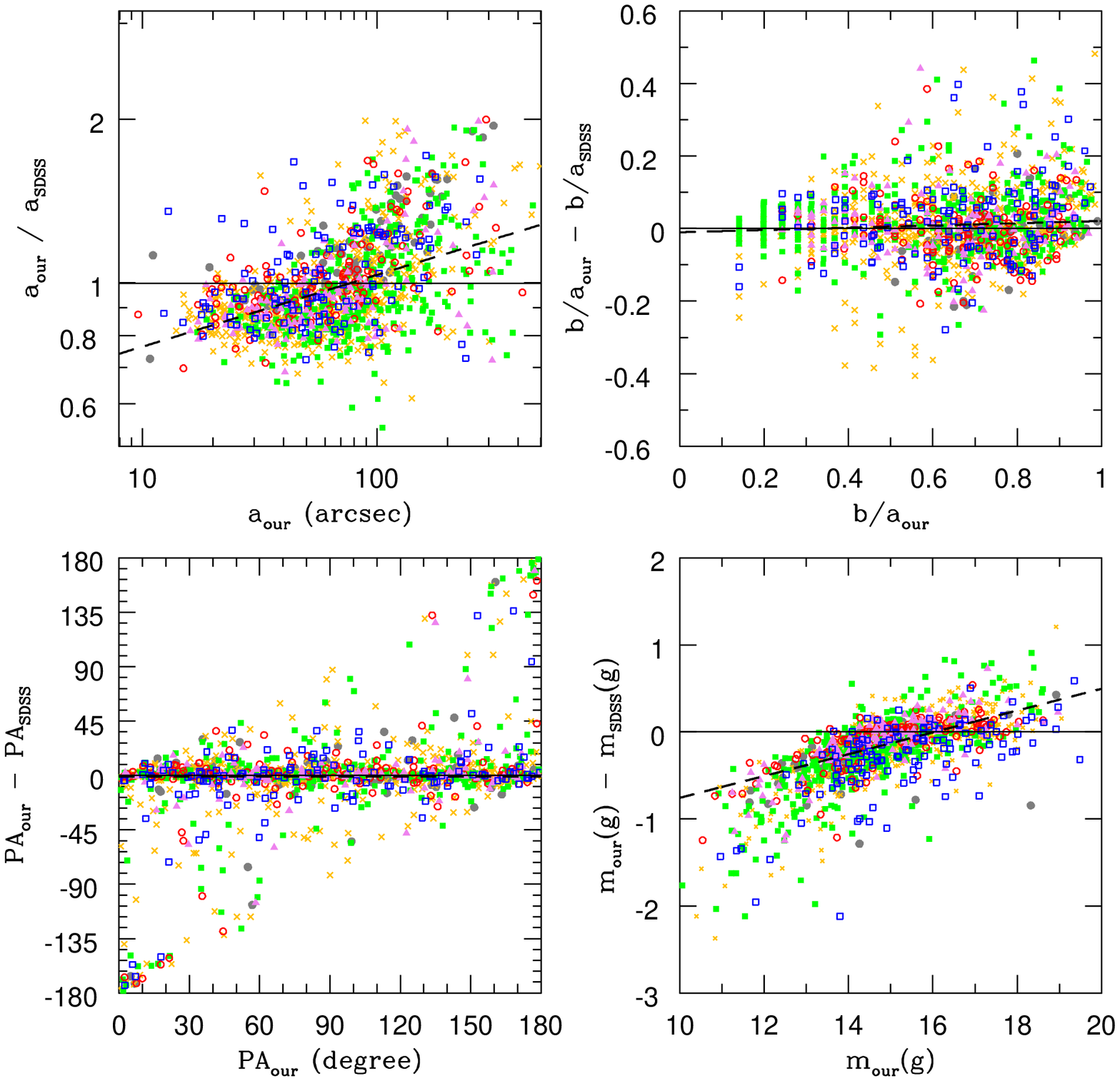}
\end{center}
\caption{\emph{Top left}: comparison between our measurements of major axes of E-Im galaxies and
isophotal major axes in the $g$-band of the SDSS DR7.
\emph{Top right}: comparison between measured axial ratios and those of the
SDSS DR7.
\emph{Bottom left}: comparison between our measurements of position angles
and those of the SDSS DR7.
\emph{Bottom right}: comparison between our measurements of apparent $g$-band magnitudes
and Composite Model Magnitudes of the same band in the SDSS DR7.
The color and symbol coding corresponds to Fig.~\ref{galourHL}.
The \emph{solid lines} in each figure are added to visually
better illustrate the deviations.
The \emph{dashed lines} are best fit linear trends from Table~\ref{tab:compare}.
\label{galourSDSS}}
\end{figure*}

In the top left panels of Figs.~\ref{galourHL} and \ref{galourSDSS}
we show the discrepancy of
the major diameters of the host galaxies between the samples.
Our measured $g$-band diameters are systematically larger than
the HyperLeda $D_{25}$ diameters in $B$-band for all
the morphological types of hosts.
Our diameters are greater than those in the HyperLeda
on average by a factor of $1.32\pm0.01$.
This level of discrepancy is not unexpected, given that our sizes are
measured at the $\mu_g=25\,\rm mag\,arcsec^{-2}$ isophotal level.
With the transformation equation, $B=g+0.39\,(g-r)+0.21$
(\citealp{2005AJ....130..873J}, for
all stars with $R-I<1.15)$, and given our mean $g-r\simeq 0.7$ color,
our $g$-band measurements are performed at the equivalent of the
$\left\langle\mu_B\right\rangle\simeq 25.48$ isophote, hence our greater host galaxy sizes.

For the data comparisons with SDSS, we used DR7 instead of DR8,
because DR7 includes isophotal photometric quantities.
Before comparing the samples of classified E-Im galaxies
we excluded 96 objects with unreliable SDSS measurements that
differ from our measurements by more than a factor of 2.
We considered the data for these galaxies as incorrectly measured in SDSS.
Our diameters are in good agreement with isoA,
and greater only on average by a factor of $1.01\pm0.01$.
In general, the SDSS measurements are unreliable for objects larger
than $\sim$100~arcsec \citep[e.g.,][]{2011A&A...532A..74B}.
Still, for objects smaller than $100''$, we find scatters corresponding to
factors of 33\% (with HyperLeda) and 16\% (with SDSS~DR7).

In the top right panels of Figs.~\ref{galourHL} and \ref{galourSDSS}
we present the differences
of our measured axial ratios and that of the HyperLeda
as well as isoB/isoA of the SDSS DR7.
The majority ($\sim$94\%) of E-Im galaxies, for which the axial ratios are available
in our and other measurements, these axial ratios are consistent within 0.2.
The mean deviation of our measurements from that of the HyperLeda is $0.005\pm0.003$,
and from isoB/isoA is $0.015\pm0.003$. The MADs are 0.055 with HyperLeda and
0.052 with SDSS~DR7.
In fact, after correction for trends, the residuals show a robust dispersion of $\sim$0.1.
There is no dependence of residuals on
the morphological types of host galaxies.

The position angles (PAs) of the major axes were determined at
the same (${\mu_g = \rm 25\,\ mag\,\ arcsec^{-2}}$) isophotal level as the
measurements of angular diameters.
Comparisons of our PA measurements with the HyperLeda and
SDSS DR7 (isoPhi) determinations are shown in
the bottom left panels of Figs.~\ref{galourHL} and~\ref{galourSDSS}.

There are non-negligible fractions of
cases where $\rm PA_{\rm our} + PA_{\rm theirs} = 180^\circ$, especially for
$\rm PA_{\rm our} \simeq 0^\circ$ or $180^\circ$, for which small errors can (for
the correct sign) flip the PA to $180^\circ$ minus its true value.
To avoid these unfair extreme outliers in our comparisons,
we redistributed the values of differences of PAs such that we considered
the $\Delta$PA$-$180$^\circ$ or $\Delta$PA$+$180$^\circ$
when the differences were $> 90^\circ$ or $< -90^\circ$,
respectively.


With these corrections, the mean difference between  HyperLeda's PAs and ours
is $0\fdg9\pm0\fdg5$, while the MAD is $4\fdg9$.
The comparison with the PAs from SDSS DR7 (isoPhi) yields a mean deviation of
$0\fdg5\pm0\fdg6$ and a MAD of $4\fdg3$.
In both cases, 85\% of the host galaxies have PAs consistent within
$20^\circ$ with those of HyperLeda or SDSS.
The scatter in the bottom left panel of Fig.~\ref{galourHL}
may be due to the fact that
the HyperLeda values correspond to measurements made at
$\mu_{\rm B}={\rm 25\,\ mag\,\ arcsec^{-2}}$,
whereas ours are made at typically lower surface brightness thresholds:
$\langle\mu_{\rm B}\rangle=$$\,\ \sim$${\rm 25.48\,\ mag\,\ arcsec^{-2}}$.
Inspections of the remaining cases with large discrepancies show that
they are mostly contributed by peculiar and low surface brightness galaxies or objects in
interacting/merging systems.
In addition, the PA is hard to determine when the galaxy is face-on,
because for any given elliptical aperture it is ill-defined;
97\% of galaxies with $b/a\leq0.5$
are consistent within $20^\circ$ of PA, whereas only 80\% of hosts
with $b/a>0.5$ have the same consistency.
Again, there is no dependence of discrepancies on
the morphological types of galaxies.
For more detailed inspection and explanation of this effect
see \cite{2007ApJS..170...33P}.

We also compared our $g$-band magnitude measurements to the HyperLeda $B$-band
and to the SDSS DR7 Composite Model Magnitude (cModelMag) determinations in the
$g$-band, which are measured from the linear combination
of the exponential and de~Vaucouleurs profiles that fit best
the $g$-band SDSS images.
Also, there is excellent agreement between cModelMag and
\cite{1976ApJ...209L...1P} magnitudes of galaxies.
Although, the cModelMag and Petrosian magnitudes are not identical,
there is an offset of 0.05-0.1 mag
but this is within errors of our elliptical aperture measurements.
The results of magnitude comparisons are presented in the bottom right panels
of Figs.~\ref{galourHL} and \ref{galourSDSS}.
The relation between the HyperLeda magnitudes and ours is quite linear with a
slope of 0.98 (our magnitudes are slightly brighter relative to theirs at the
bright end).
The residuals from our robust linear fit between the two magnitudes is 0.22 mag.

The mean difference between our magnitudes and those of HyperLeda is $-0.42\pm0.01$,
while the mean absolute difference is $0.47\pm0.01$.
It is clear that most of the measurements agree well with each other once the
magnitudes are brought into the same system.
For $\sim$65\% of galaxies our and HyperLeda magnitude differences are less
than 0.5~mag, which corresponds to 0.02~mag after converting the $g$-band into $B$-band.
Only for $\sim$7\% of the hosts galaxies the magnitude difference is equal or larger
than 1~mag, hence 0.52~mag after the conversion.
These galaxies are mostly of types Sc to Im.

The mean difference between our magnitudes and the SDSS cModelMag photometry is $-0.21\pm0.01$.
The MAD is 0.19 mag.
There is a trend where we measure brighter magnitudes than SDSS for the
brightest galaxies. Despite some curvature in the relation between the two
magnitude estimates, we fit a lines, and find
$m_{\rm SDSS} = 0.87\,m_{\rm ours}$,
with a dispersion of 0.25 mag about this relation.

For $\sim$81\% of galaxies our and the SDSS magnitude differences are less than 0.5~mag.
For $\sim$5\% of the objects the magnitude differences are equal to or larger than 1~mag.
For faint galaxies, the SDSS measurements algorithm overestimates the cModelMag fluxes,
while for bright galaxies it underestimates the fluxes.
This trend is stronger for late-type galaxies.
Again, we explain the presence of the large scatters,
especially for the bright galaxies,
by the unreliable SDSS photometric measurements for objects larger than
$\sim$100~arcsec \citep[e.g.,][]{2011A&A...532A..74B}.
We also performed the same analysis using the cModelMag of SDSS DR8
instead of DR7, and found the same behavior for the photometric bias.

\begin{table*}[t]
\begin{center}
  \caption{Distribution of SN types according to
the morphological classification of the host galaxies.}
\label{bigtable}
  \begin{tabular}{lrrrrrrrrrrrrrrrrr} \hline
    \hline
  &\multicolumn{1}{c}{E}&\multicolumn{1}{c}{E/S0}&\multicolumn{1}{c}{S0}&\multicolumn{1}{c}{S0/a}&\multicolumn{1}{c}{Sa}
  &\multicolumn{1}{c}{Sab}&\multicolumn{1}{c}{Sb}&\multicolumn{1}{c}{Sbc}&\multicolumn{1}{c}{Sc}&\multicolumn{1}{c}{Scd}
  &\multicolumn{1}{c}{Sd}&\multicolumn{1}{c}{Sdm}&\multicolumn{1}{c}{Sm}&\multicolumn{1}{c}{Im}&\multicolumn{1}{c}{S}
  &\multicolumn{1}{c}{Unclassified}&\multicolumn{1}{r}{All}\\
  \hline
    I&4&2&3&5&4&4&8&6&8&1&1&1&3&0&4&18&72\\
    Ia&61&35&51&73&31&50&122&125&114&28&22&9&2&7&199&1061&1990\\
    Ib&1&0&1&0&1&2&6&9&14&5&4&3&2&1&5&9&63\\
    Ib/c&0&0&0&1&3&2&5&12&10&1&2&3&0&0&4&8&51\\
    Ic&0&0&0&0&1&4&15&30&18&9&10&1&1&0&8&23&120\\
    II&0&0&2&4&9&9&98&121&182&39&39&15&9&12&55&116&710\\
    IIb&0&0&0&0&2&0&4&9&10&5&3&5&2&1&6&3&50\\
    IIn&0&0&0&1&2&0&11&12&28&9&3&3&3&2&9&27&110\\
    Unclassified&9&9&18&16&17&21&68&69&65&14&17&10&6&4&55&312&710\\
  \hline
  All&75&46&75&100&70&92&337&393&449&111&101&50&28&27&345&1577&3876\\
    \hline \\
  \end{tabular}
\parbox{\hsize}{\textbf{Notes.} All SNe types include uncertain (``:'' or ``?'') and peculiar (``pec'')
classifications. Type~II SNe include subtypes II~P and II~L.
Types~I, Ia, and II include also few SNe classified
from the light curve only,
these SNe are labeled by ``*'' symbols in the \emph{total sample}.}
\end{center}
\end{table*}
\begin{figure}[t]
\begin{center}
\includegraphics[width=\hsize]{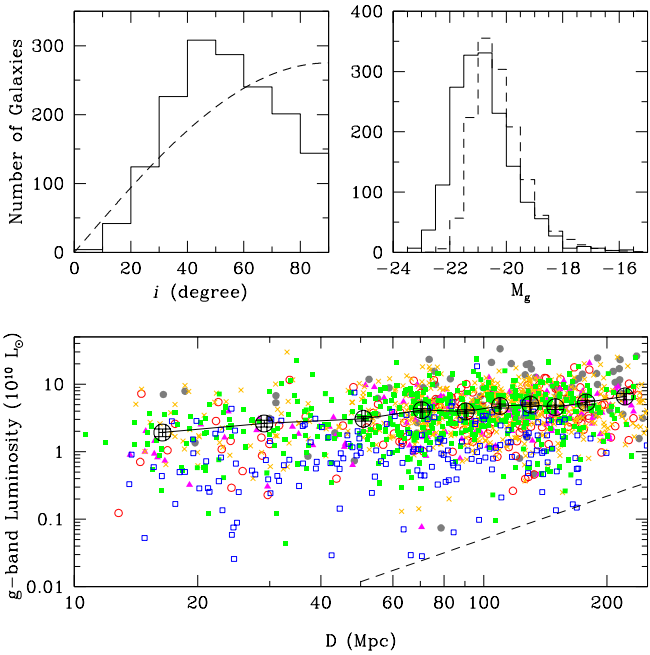}
\end{center}
\caption{\emph{Top left}: distribution of the inclination angles
for disk (S0-Sm) galaxies.
The \emph{dashed} curve represents the expected random distribution.
\emph{Top right}: distribution
of corrected $g$-band absolute magnitudes for the classified
(E-Im) galaxies.
The \emph{dashed} histogram shows the distribution of $g$-band
absolute magnitudes of the SDSS Main Galaxy sample
(the values are divided by 500 for the sake of clarity).
\emph{Bottom}: the $g$-band luminosity
of the same galaxies as a function of their distance.
The color and symbol coding corresponds to Fig.~\ref{galourHL}.
The average luminosity in different distance bins is overplotted
as big open circles with error bars of the mean values.
The dashed line represents the selection
limit of the SDSS Main Galaxy spectroscopic
sample ($r\leq 17.77$) for the extinction corrected Petrosian magnitude,
assuming $g-r=0.64$.}
\label{galaxyfdata}
\end{figure}

In addition, we checked the influence of nuclear activity on the
discrepancies of the photometric measurements.
Indeed, since the SDSS total magnitudes for extended objects (ModelMag and cModelMag)
are based on single-component fits,
we could expect that such fitting will perform poorly for AGNs with
relatively bright nuclei.
The result is negative: there is no dependence of the discrepancies
on the nuclear activity of the host galaxies.

\begin{figure}[t]
\begin{center}
\includegraphics[width=\hsize]{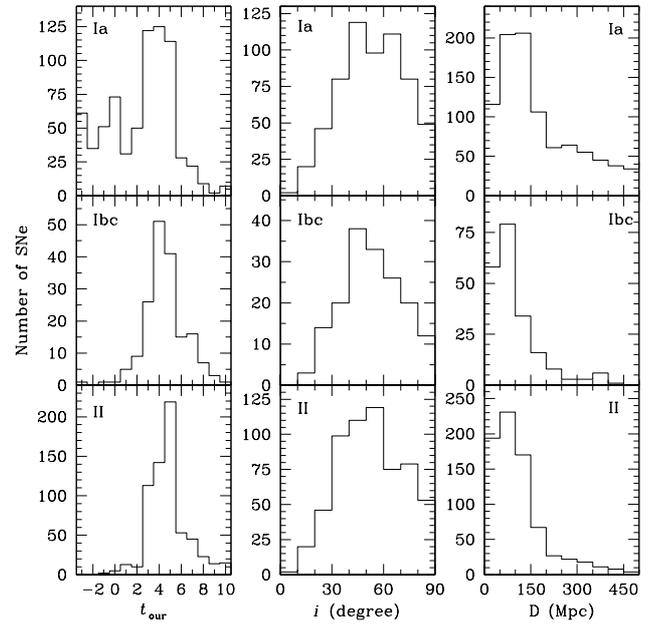}
\end{center}
\caption{\emph{Left}: distribution
of different types of SNe as a function of the host galaxy $t$-type.
\emph{Middle}: distribution of SNe as a function of
the inclination of the spiral hosts.
\emph{Right}: distribution of different types of
SNe as a function of the distance of their host galaxy.}
\label{dSNtype}
\end{figure}

The top left panel of Fig.~\ref{galaxyfdata} shows the distribution of
inclination angles for morphologically classified disk
(S0-Sm) galaxies. There is a  clear deficit of SN host galaxies having
small and large inclinations.
We thus share the view with \cite{2011MNRAS.412.1419L},
who found a deficit of LOSS galaxies with small
inclinations and explained this deficit by limits of
the precision on the major and minor axes.
Indeed, it is very difficult to measure inclinations
smaller than $20^\circ$ from elliptical aperture
measurements applied to nearly face-on galaxies.
The lack of galaxies with large inclinations can be explained
by a bias in the discovery of SNe
(see also the middle panel of Fig.~\ref{dSNtype})
in highly inclined spirals
\citep[e.g.,][]{1988A&A...190...10C,1991ARA&A..29..363V,1997A&A...322..431C}.

The top right panel of Fig.~\ref{galaxyfdata} presents the distribution
of corrected $g$-band absolute magnitudes for the classified galaxies.
In comparison with the distribution of SDSS galaxies (dashed histogram), the
SN host galaxies are more luminous.
A Kolmogorov-Smirnov (KS) test indicates that the more luminous distribution of
SN host galaxy magnitudes, relative to the SDSS galaxies in general, cannot
be obtained by chance with more than 0.1\% probability.
This distribution again suffers from a selection effect on SN productivity,
since the rate of SNe depends on the luminosity or stellar content of
the host galaxies \citep[e.g.,][]{1991ARA&A..29..363V,1997A&A...322..431C,
2011MNRAS.412.1473L,2011Ap.....54..301H}.
Therefore, our sample of classified host galaxies
is biased toward bright galaxies.
For comparison, the distribution of $g$-band absolute magnitudes of
the SDSS Main Galaxy sample is also shown.

The bottom panel of Fig.~\ref{galaxyfdata} shows the $g$-band luminosity
of the same galaxies as a function of their distance.
Galaxy luminosities were derived from absolute magnitudes,
assuming that $g$-band absolute magnitude of the Sun is 5.45
\citep{2003ApJ...592..819B}.
The luminosities of late-type (Sd-Im) hosts are on average 5 times lower
than those of early-type (E-E/S0) galaxies.
This trend is clearly seen in the bottom panel of Fig.~\ref{galaxyfdata}.
The average $g$-band luminosities in different distance bins
are also plotted.
At greater distances, the low-luminosity host galaxies are lost due to flux
limitations.
Thus, Malmquist bias causes the average $g$-band luminosity to increase with
increasing distance.
The databases become progressively incomplete for low-luminosity
galaxies at greater distances.
This was already mentioned
by \cite{2011MNRAS.412.1419L} for the LOSS galaxy sample.

\begin{figure}[t]
\begin{center}$
\begin{array}{cc}
\includegraphics[width=0.48\hsize]{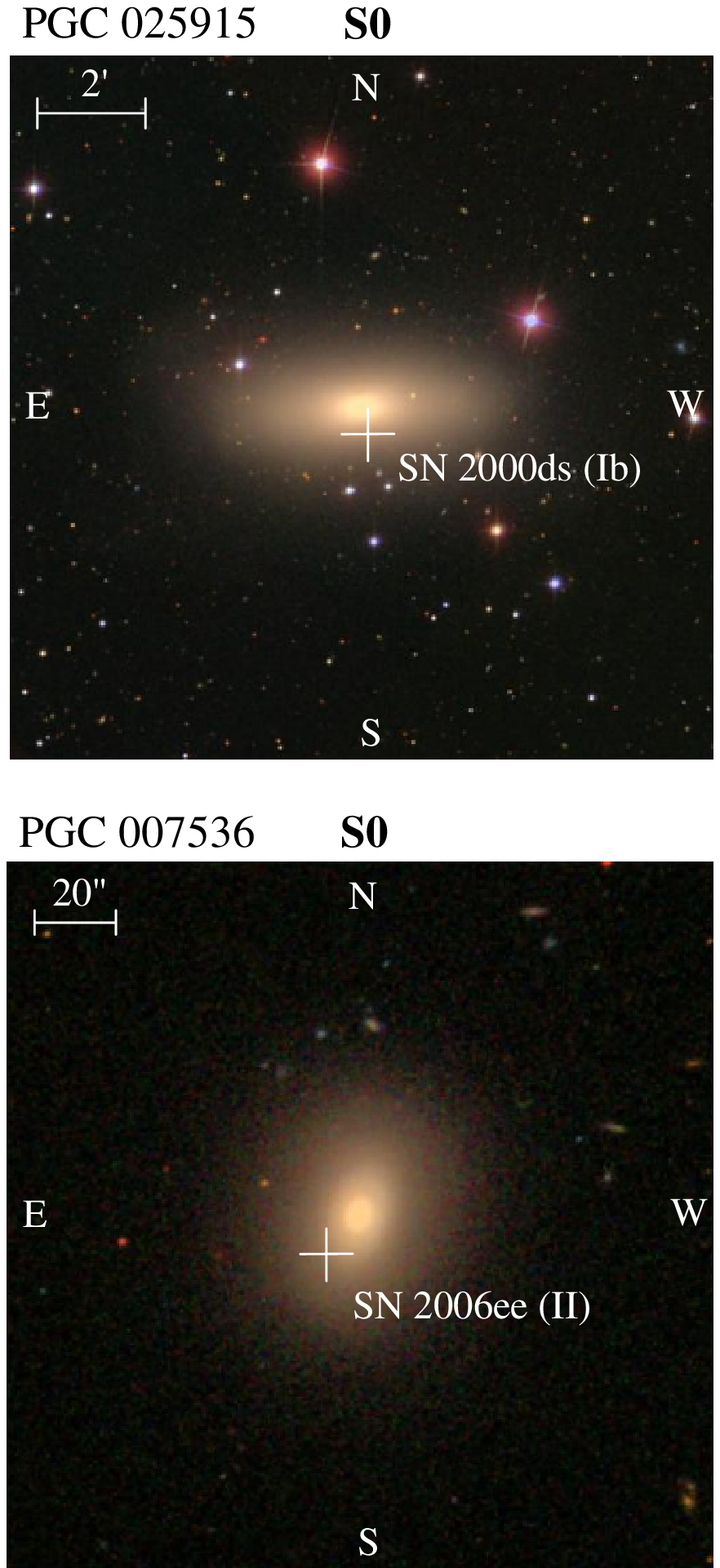} &
\includegraphics[width=0.48\hsize]{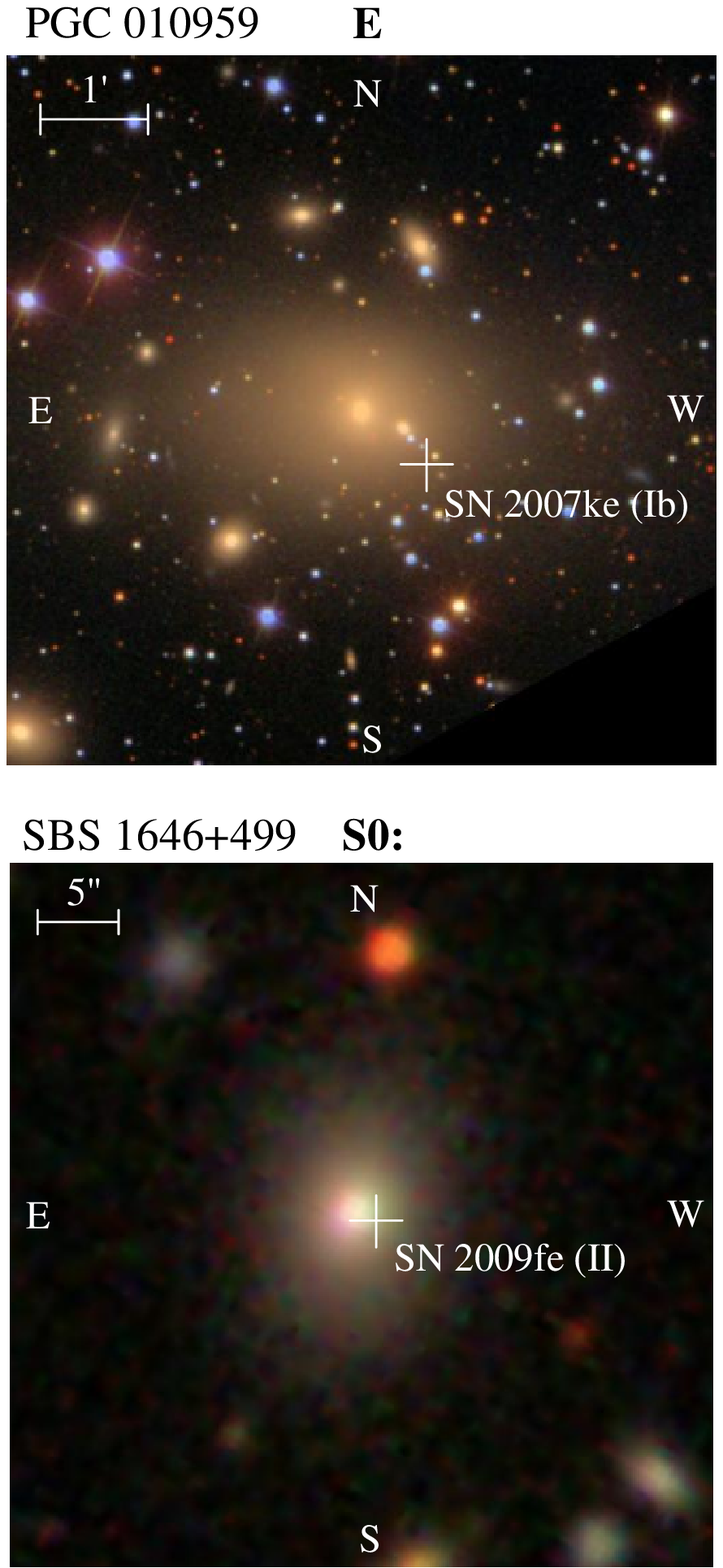}
\end{array}$
\end{center}
\caption{SDSS images of
early-type (E or S0) host galaxies with CC~SNe.
The objects identifiers are listed
at the top with our morphological classification.
The SN names, types, and positions (marked by cross sign)
are also shown.
In all images, north is up and east to the left.}
\label{CCinES0}
\end{figure}

\subsection{Distribution of SN types}

In Table~\ref{bigtable}, we present the distribution of SN types according
to the morphology of their host galaxies
in an analogous fashion as in Table~5 of \cite{1999A&AS..139..531B}.
It is clear that most SNe are found in spiral hosts.
The percentage of unclassified SNe is $\sim$18\%
of the \emph{total sample}.
Approximately half of the unclassified SNe
were discovered before 2000.

The left panel of Fig.~\ref{dSNtype} shows the number distribution of
SN types as a function of host morphology.
There is a significant difference between the distribution of the SN~Ia
hosts and that of CC SN hosts, while types~Ibc and II SN hosts have
similar distributions. These trends were previously reported
by \cite{2002PASP..114..820V,2003PASP..115.1280V,
2005PASP..117..773V} and \cite{2011MNRAS.412.1419L}.

\begin{table}[t]
\begin{center}
\caption{Distribution of SN types
according to the level of nuclear activity of the host galaxy.
\label{actSNe}}
\begin{tabular}{llrrrrr}
\hline
\hline
Diagram & Activity & Ia & Ibc & II & Unclassified & All \\
\hline
& Sy & 20 & 3 & 14 & 4 & 41 \\
& LINER & 77 & 8 & 41 & 40 & 166 \\
BPT& C & 78 & 12 & 64 & 27 & 181 \\
& SF & 110 & 49 & 157 & 96 & 412 \\ \cline{2-7}
& All & 285 & 72 & 276 & 167 & 800 \\
\hline
& Sy & 83 & 7 & 43 & 35 & 168 \\
& LINER & 38 & 7 & 24 & 19 & 88 \\
WHAN& SF & 199 & 66 & 218 & 121 & 604 \\
& RP & 235 & 13 & 72 & 86 & 406 \\ \cline{2-7}
& All & 555 & 93 & 357 & 261 & 1266 \\
\hline \\
\end{tabular}
\parbox{\hsize}{\textbf{Notes.}
Each of the Seyfert (Sy) rows additionally include 6 BL~AGN hosts.
The RP row includes 77 P galaxies.
The column of unclassified SNe includes also type I SNe.}
\end{center}
\end{table}

It is generally believed that hosts of CC~SNe are objects with
young stellar populations (generally spiral or irregular galaxies),
while the old stellar population of early-type galaxies can
produce only SNe Ia \citep[e.g.,][]{1991ARA&A..29..363V,1999A&A...351..459C}.
Nevertheless, among the morphologically classified
host galaxies of CC~SNe in our sample, we have found 4 cases (2000ds [Ib] in
PGC~025915, 2006ee [II] in PGC~007536, 2007ke [Ib] in PGC~010959,
and 2009fe [II] in SBS~1646+499) in which the host has been classified
as E or S0, in apparent contradiction to this conventional view.
Fig.~\ref{CCinES0} presents the cases of CC~SNe in early-type hosts.

\cite{2008A&A...488..523H} already reported and investigated in detail two cases
of such CC~SNe in early-type galaxies (2000ds and 2006ee).
The host galaxy of SN~2000ds \citep{2000IAUC.7511....2F} has been confirmed
to be an S0,
with a central region showing dust and a disky central gas distribution
\citep[e.g.,][]{2008A&A...488..523H}.
According to the outer isophotal structure and radial surface brightness
profile of the host of SN~2006ee \citep{2006IAUC.8741....1P},
this must be an S0 galaxy.
It has been shown that the surface brightness distribution
has some small degree of asymmetry in the region to
the south-southwest of the nucleus \citep{2008A&A...488..523H}.
Here, we suspected the presence of an embryonic spiral arm.
We classified the host galaxy of SN~2007ke \citep{2007CBET.1101....1F} as type E,
this classification is also given in both NED and HyperLeda.
It is in interacting system and is a member of the cluster of galaxies.
The host galaxy of SN~2009fe \citep{2009CBET.1819....1K}
is classified as an uncertain type S0.
The same morphological type is reported in \cite{2011Ap.....54...15G}.
In NED, it is classified as a blazar (Seyfert~1).
This object also shows 1.4 GHz radio continuum emission \citep{1998AJ....115.1693C}.
However, more detailed inspection on high resolution images is still required.
In principle, the galaxy could have some diffuse spiral arms and be classified
as S0/a but due to insufficient resolution of the SDSS image at the distance
of this object, it has been classified as an uncertain type S0.

The presence of CC~SNe in early-type galaxies can be
interpreted as an additional indication that residual star formation
episodes also take place in E or S0 galaxies, due to merging/accretion or
interaction with close neighbors.
Recently, \cite{2009MNRAS.394.1713K} have found that the recent star formation
is likely to be driven by minor mergers, which seems to fit
with our interpretation as well.
Meanwhile, using rest-frame UV photometry of
early-type galaxies in the nearby Universe, \cite{2007ApJS..173..619K}
suggested that low-level recent star formation is widespread in
nearby early-type galaxies.
The situation is also very similar at intermediate redshifts
\citep{2008MNRAS.388...67K}.
Hence, the detection of SNe II in early-type galaxies is expected, but at
lower frequency than type~Ia.

The middle panel of Fig.~\ref{dSNtype} shows the number distribution of
different types of SNe as a function of the inclination of S0-Sm hosts
(see also the top left panel of Fig.~\ref{galaxyfdata} and its explanation).
The KS test suggests that there is no significant
difference between the distributions of types~Ia and CC~SNe.
The same behavior occurs when comparing
type~Ia SNe separately with types~Ib and II SNe.
A similar trend was mentioned by \cite{1997A&A...322..431C}.

The right panel of Fig.~\ref{dSNtype} shows the number distribution of
SN types as a function of distance.
All the major types of SNe are peaked at 50-100 Mpc.
It is clear that
the sample of SNe is largely incomplete beyond $\sim$100 Mpc.
The distributions of types~Ibc and II~SNe are similar and display
a sharp decline beyond 100 Mpc.
Type Ia SNe, because of their comparatively high luminosity and
the presence of dedicated surveys (e.g., SDSS SN Survey, ESSENCE etc.),
are discovered at much greater distances than CC~SNe.
A similar behavior was also found by \cite{2011MNRAS.412.1419L}.

Table~\ref{actSNe} displays the numbers of different types of SNe in hosts
with different levels of nuclear activity.
It is important to note that
nuclear activity is affected by aperture bias
\citep[e.g.,][]{2005PASP..117..227K}.
The SDSS spectra are taken with a fixed fiber size $(3'')$.
For a nearby galaxy, the SDSS fiber covers the central nuclear region or its part,
while for more distant case it covers a larger fraction of the galaxy
(e.g., 120 pc at $z=0.004$ but 2.7 kpc at $z=0.1$).
Hence, the activity data can be contaminated by the emission of circumnuclear regions or
from the whole galaxy. The effect also depends on galaxy size,
as for dwarf galaxies the fiber will cover a larger fraction of the total emission.

\section{Summary and perspectives}
\label{sumandper}

In a series of papers, we will analyze
how the different types of SNe events and their spatial distribution are
correlated with the properties of the nuclei and global physical
parameters of the host galaxies, as well as with the
nearby and distant environments of these galaxies.
In this first paper, we report the creation of
large and well-defined database of 3876 SNe and their 3679 host
galaxies that are located on the SDSS DR8 coverage.
This database has been created to increase the size
of the sample relative to previous works, and to carry out morphological
classification, as well as individual measurements of the global
parameters of SN host galaxies to a more homogenous and detailed level.
Throughout this work, we analyzed and discussed many selection effects
that can bias future analyses.

We identified 91\%
of the host galaxy sample (3340 hosts with 3536 SNe)
among which the SDSS identification of
$\sim$1100 hosts have been done for the first time.
Using the SDSS multi-band images,
photometric and spectral data, we
provided accurate coordinates,
heliocentric redshifts,
morphological types, nuclear activity classes
(RP, SF, C, LINER, and Sy),
apparent $g$-band magnitudes, major axes ($D_{25}$),
axial ratios ($b/a$), and position angles (PA)
of the host galaxies.

During the mutual comparison of spectroscopic classification of SNe
taken from the ASC and SSC and the literature search, we updated
spectroscopic types for 67 SNe.
We collected all the available data on 3166 SNe types when it was
present in one of the SN catalogs or in the CBAT.
Our \emph{total sample} consists of 72 SNe~I, 1990 SNe~Ia, 234 SNe~Ibc,
870 SNe~II, and 710 unclassified SNe.
In addition, the sample includes 3599 SNe with offset data.
We corrected offsets of 43 SNe and calculated (relatively uncertain) coordinates
for 332 SNe with unavailable astrometry via offsets and accurate
coordinates of the identified nuclei of their host galaxies.

The morphological classification is available for 2104 host galaxies,
which is 63\% of our sample of identified galaxies and
includes 73 hosts in interacting (``\emph{inter}''),
and 56 hosts in merging (``\emph{merg}'') systems.
The \emph{total sample} of host galaxies collects heliocentric redshifts
for 3317 (90\%) galaxies.
The $g$-band magnitudes, $D_{25}$, $b/a$, and PA are available for
2030 hosts of the morphologically classified sample of galaxies.

We also provided information on the nuclear activity of 1106 host
galaxies by using the BPT and WHAN diagrams.
The database also includes 6 broad-line AGNs and 77 passive (P)
galaxies without any emission features.

The creation of this large database will minimize
possible selection effects and errors that often arise when data is collected
from different sources and catalogs.
In future papers of this series, we will largely use all
photometric and spectral data to
constrain the nature of SN progenitors, using their distribution
within host galaxies of different morphological types and levels of nuclear activity.

\begin{acknowledgements}

A.A.H. and A.R.P. acknowledge the hospitality of the
Institut d'Astrophysique de Paris (France) during their
stay as visiting scientists supported by the Collaborative
Bilateral Research Project of the State Committee of Science (SCS)
of the Republic of Armenia and the French
Centre National de la Recherch\'{e} Scientifique (CNRS).
V.Zh.A. and J.M.G. are supported by grants SFRH/BPD/70574/2010 and
SFRH/BPD/66958/2009 from FCT (Portugal), respectively.
V.Zh.A. would further like to thank for the support by the ERC under
the FP7/EC through a Starting Grant agreement number 239953.
This work was made possible in part by a research grant from the
Armenian National Science and Education Fund (ANSEF)
based in New York, USA.
Special thanks to Richard~Trilling,
who kindly agreed to edit the manuscript.
This research made use of the Asiago Supernova Catalogue (ASC), which is
available at \texttt{http://web.oapd.inaf.it/supern/cat/},
the Sternberg Astronomical Institute (SAI) Supernova Catalogue, available at
\texttt{http://www.sai.msu.su/sn/sncat/}, website of the Central
Bureau for Astronomical Telegrams (CBAT), available at
\texttt{http://www.cbat.eps.harvard.edu/lists/Supernovae.html},
the HyperLeda database (\texttt{http://leda.univ-lyon1.fr/}), and
the NASA/IPAC Extragalactic
Database (NED), which is
available at \texttt{http://ned.ipac.caltech.edu/}, and
operated by the Jet Propulsion
Laboratory, California Institute of Technology,
under contract with the National
Aeronautics and Space Administration.
GAIA was created by the now closed Starlink UK project,
funded by the Particle Physics and Astronomy Research Council (PPARC)
and has been more recently supported by the Joint Astronomy Center Hawaii
funded again by PPARC and more recently by its successor organization
the Science and Technology Facilities Council (STFC).
The GAIA home page is \texttt{http://astro.dur.ac.uk/\~{}pdraper/gaia/gaia.html}.
Funding for SDSS-III has been provided by the Alfred P. Sloan Foundation,
the Participating Institutions, the National Science Foundation,
and the US Department of Energy Office of Science.
The SDSS-III web site is \texttt{http://www.sdss3.org/}.
SDSS-III is managed by the Astrophysical Research Consortium for the
Participating Institutions of the SDSS-III Collaboration including the
University of Arizona, the Brazilian Participation Group,
Brookhaven National Laboratory, University of Cambridge,
University of Florida, the French Participation Group,
the German Participation Group, the Instituto de Astrofisica de Canarias,
the Michigan State/Notre Dame/JINA Participation Group,
Johns Hopkins University, Lawrence Berkeley National Laboratory,
Max Planck Institute for Astrophysics, New Mexico State University,
New York University, Ohio State University, Pennsylvania State University,
University of Portsmouth, Princeton University, the Spanish Participation Group,
University of Tokyo, University of Utah, Vanderbilt University,
University of Virginia, University of Washington, and Yale University.

\end{acknowledgements}

\begin{center}
\begin{landscape}
\tabcolsep 5pt
\begin{table}
\caption{Properties of the \emph{total sample}$^{a}$ of SNe and their host galaxies.}
\label{bigbigtable}
\scriptsize
\begin{tabular}{lrrcllrrrllllrrrrll}
\hline
\hline
\multicolumn{1}{c}{SN$^{b}$}&\multicolumn{1}{c}{${\rm \alpha_{SN}}$$^{c}$}&\multicolumn{1}{c}{${\rm \delta_{SN}}$$^{c}$}&\multicolumn{1}{c}{Offset$^{d}$}&\multicolumn{1}{c}{SN type$^{e}$}&
\multicolumn{1}{c}{Galaxy$^{f}$}&\multicolumn{1}{c}{${\rm \alpha_{G}}$$^{g}$}&\multicolumn{1}{c}{${\rm \delta_{G}}$$^{g}$}&\multicolumn{1}{c}{Redshift$^{h}$}&\multicolumn{1}{c}{Morph.$^{i}$}&
\multicolumn{1}{c}{Bar}&\multicolumn{1}{c}{``\emph{inter}/\emph{merg}''}&\multicolumn{1}{c}{\emph{Disturbed}$^{j}$}&\multicolumn{1}{c}{$D_{25}$}&\multicolumn{1}{c}{$b/a$}&
\multicolumn{1}{c}{PA}&\multicolumn{1}{c}{$m_{\rm g}$}&\multicolumn{1}{c}{BPT$^{k}$}&\multicolumn{1}{c}{WHAN$^{k}$}
\\
\multicolumn{1}{c}{(1)} & \multicolumn{1}{c}{(2)} & \multicolumn{1}{c}{(3)} & \multicolumn{1}{c}{(4)}
&\multicolumn{1}{c}{(5)} &\multicolumn{1}{c}{(6)} &\multicolumn{1}{c}{(7)} &\multicolumn{1}{c}{(8)}
&\multicolumn{1}{c}{(9)} &\multicolumn{1}{c}{(10)} &\multicolumn{1}{c}{(11)} &\multicolumn{1}{c}{(12)}
&\multicolumn{1}{c}{(13)} &\multicolumn{1}{c}{(14)} &\multicolumn{1}{c}{(15)} &\multicolumn{1}{c}{(16)}
&\multicolumn{1}{c}{(17)} &\multicolumn{1}{c}{(18)} &\multicolumn{1}{c}{(19)}  \\
\hline
1885A & 10.67917\,\ & 41.26778\,\ & 15W 4S & I & PGC 2557 & 10.68330 & 41.26890 & --0.00103 & Sb &  &  &  &  &  &  &  &  & \\
1895A & 186.82042\,\ & 9.41806\,\ & 75E 11S &  & J122711.61+092514.4 & 186.79838 & 9.42067 & 0.00148 & Sa & B &  &  & 271.8 & 0.558 & 96.4 & 11.97 & SF & SF \\
1901A & 122.81292\,\ & 25.21139\,\ & 19E 7N &  & J081113.48+251224.4 & 122.80617 & 25.20678 & 0.01361 & Sc &  & inter & D & 205.4 & 0.560 & 167.6 & 12.49 & SF & SF \\
1901B & 185.69833\,\ & 15.82361\,\ & 110W 4N: & I & J122254.91+154920.2 & 185.72879 & 15.82228 & 0.00526 & Sc &  &  &  &  &  &  &  &  & \\
1907A & 191.51161: & --8.65264: & 10W 11N & I & J124603.46--083920.5 & 191.51442 & --8.65569 & 0.00468 & Sab &  &  &  & 124.8 & 0.456 & 121.6 & 13.63 &  & \\
1909A & 210.51292\,\ & 54.46611\,\ & 620W 408N: & II P: & J140312.52+542056.2 & 210.80217 & 54.34894 & 0.00082 & Sc &  &  &  &  &  &  &  & SF & SF \\
1912A & 140.48902: & 50.98214: & 50W 20N &  & J092202.66+505835.7 & 140.51108 & 50.97658 & 0.00212 & Sb &  &  &  & 637.9 & 0.527 & 135.2 & 9.90 &  & \\
1914A & 185.73750\,\ & 15.79167\,\ & 24E 111S: &  & J122254.91+154920.2 & 185.72879 & 15.82228 & 0.00526 & Sc &  &  &  &  &  &  &  &  & \\
1915A & 188.54740: & 2.65158: & 44E 8S &  & J123408.44+023913.7 & 188.53517 & 2.65381 & 0.00550 & Sbc &  &  &  & 407.4 & 0.456 & 64.2 & 11.08 &  & \\
1919A & 187.70292\,\ & 12.41778\,\ & 15W 100N: & I & J123049.41+122328.1 & 187.70588 & 12.39114 & 0.00420 & E &  &  &  & 661.3 & 0.822 & 145.7 & 9.65 &  & \\
1920A & 128.81620: & 28.47478: & 19W 5N & II: \,\ U & J083517.33+282824.2 & 128.82221 & 28.47339 & 0.00718 & Sbc pec &  &  &  & 162.1 & 0.780 & 46.1 & 12.63 &  & \\
1921B & 154.58227: & 41.37986: & 32E 160S & II & J101816.90+412527.5 & 154.57042 & 41.42431 & 0.00198 & Sc &  &  &  &  &  &  &  &  & \\
1921C & 154.59968: & 41.35875: & 79E 236S & I & J101816.90+412527.5 & 154.57042 & 41.42431 & 0.00198 & Sc &  &  &  &  &  &  &  &  & \\
1926A & 185.47577: & 4.49294: & 11W 69N & II L & J122154.92+042825.6 & 185.47883 & 4.47378 & 0.00523 & Sbc & B &  &  &  &  &  &  &  & \\
1926B & 248.08729: & 19.83972: & 0E 48N &  & J163220.95+194935.0 & 248.08729 & 19.82639 & 0.00791 & Sc & B &  &  & 171.9 & 0.475 & 176.0 & 12.10 &  & \\
1936A & 184.98363: & 5.35136: & 0E 29N & II P & J121956.07+052035.9 & 184.98363 & 5.34331 & 0.00783 & Sc & B &  &  & 157.8 & 0.640 & 6.1 & 12.11 &  & \\
1936B &  &  &  &  & J012109.32+154140.9 & 20.28883 & 15.69469 & 0.01711 & Sc & B &  &  & 49.3 & 0.661 & 168.8 & 15.40 & SF & SF \\
1937A & 182.78646: & 50.49614: & 42E 42N & II P: & J121104.35+502904.1 & 182.76813 & 50.48447 & 0.00256 & Sb &  &  &  & 447.2 & 0.199 & 65.0 & 12.11 &  & \\
1937C & 196.47167\,\ & 37.60194\,\ & 30E 40N & Ia & J130548.70+373613.0 & 196.45292 & 37.60361 & 0.00105 & Sm &  &  &  &  &  &  &  &  & \\
1937F & 154.57227: & 41.38292: & 5E 149S & II P: & J101816.90+412527.5 & 154.57042 & 41.42431 & 0.00198 & Sc &  &  &  &  &  &  &  &  & \\
1938B & 132.35024: & 19.06969: & 31E 19S &  & J084921.87+190429.9 & 132.34113 & 19.07497 & 0.01398 & E &  &  &  & 231.2 & 0.720 & 111.8 & 12.01 &  & \\
1938C & 198.98408: & 25.42906: & 31E 20S &  & J131553.89+252604.6 & 198.97454 & 25.43461 & 0.00312 & Sm &  &  &  & 112.3 & 0.312 & 32.1 & 15.49 &  & \\
1939A & 190.70035: & 2.69331: & 26W 20N & Ia & J124249.82+024115.9 & 190.70758 & 2.68775 & 0.00303 & E &  &  &  &  &  &  &  & LINER & RP \\
1939B & 190.50938: & 11.63217: & 0E 53S & I & J124202.25+113848.8 & 190.50938 & 11.64689 & 0.00147 & E/S0 &  &  &  &  &  &  &  &  & \\
1939D & 14.39366: & --4.99947: & 9W 11N &  & J005735.08--050009.1 & 14.39617 & --5.00253 & 0.01888 & Sbc & B &  &  & 64.9 & 0.704 & 47.1 & 14.32 &  & \\
1940A & 229.04268: & 56.24267: & 137E 310S & II L & J151553.77+561943.6 & 228.97404 & 56.32878 & 0.00222 & Sc &  &  &  &  &  &  &  &  & \\
1940B & 192.63995: & 25.53353: & 95E 118N & II P & J125026.57+253002.7 & 192.61071 & 25.50075 & 0.00402 & Sb & B &  &  & 755.7 & 0.760 & 35.6 & 9.58 &  & \\
\multicolumn{1}{c}{...} & \multicolumn{1}{c}{...} & \multicolumn{1}{c}{...} & \multicolumn{1}{c}{...} & \multicolumn{1}{l}{...} & \multicolumn{1}{c}{...} & \multicolumn{1}{c}{...} & \multicolumn{1}{c}{...} & \multicolumn{1}{c}{...} & \multicolumn{1}{l}{...} & \multicolumn{1}{l}{...} & \multicolumn{1}{l}{...} & \multicolumn{1}{l}{...} & \multicolumn{1}{c}{...} & \multicolumn{1}{c}{...} & \multicolumn{1}{c}{...} & \multicolumn{1}{c}{...} & \multicolumn{1}{c}{...} & \multicolumn{1}{c}{...} \\
\multicolumn{1}{c}{...} & \multicolumn{1}{c}{...} & \multicolumn{1}{c}{...} & \multicolumn{1}{c}{...} & \multicolumn{1}{l}{...} & \multicolumn{1}{c}{...} & \multicolumn{1}{c}{...} & \multicolumn{1}{c}{...} & \multicolumn{1}{c}{...} & \multicolumn{1}{l}{...} & \multicolumn{1}{l}{...} & \multicolumn{1}{l}{...} & \multicolumn{1}{l}{...} & \multicolumn{1}{c}{...} & \multicolumn{1}{c}{...} & \multicolumn{1}{c}{...} & \multicolumn{1}{c}{...} & \multicolumn{1}{c}{...} & \multicolumn{1}{c}{...} \\
2011T & 257.51508\,\ & 32.50033\,\ & 22.9W 0.9N & Ia & J171005.44+322959.7 & 257.52267 & 32.49992 & 0.02917 & E/S0 &  &  &  & 80.7 & 0.510 & 111.1 & 14.42 & LINER & RP \\
2011V & 141.91150\,\ & 28.79089\,\ & 35E 27S & IIb: & J092736.09+284755.5 & 141.90038 & 28.79875 & 0.01382 & Sdm &  &  &  & 107.7 & 0.921 & 37.6 & 14.50 &  & LINER \\
2011W & 8.21350\,\ & 11.73542\,\ & 12E 4.8S & II: & J003250.47+114412.2 & 8.21029 & 11.73672 & 0.01636 & Sb &  &  &  & 59.9 & 0.390 & 107.8 & 15.23 &  & \\
\hline
\end{tabular}

\medskip
\parbox{0.97\hsize}{\textbf{Notes.}
$^{(a)}$~Only the 30 entries are shown. The full table is available at the CDS.
See text for details concerning the explanations of the columns.
The list is arranged chronologically according to the date of SN discovery.
$^{(b)}$~SN designation, the symbols ``?'' denote an unconfirmed SN.
$^{(c)}$~Equatorial coordinates of SN at the 2000.0 epoch, in degrees.
The symbols ``:'' denote uncertain coordinates.
$^{(d)}$~SN offset from host galaxy nucleus in arcsec, in the E/W and N/S direction respectively.
The symbols ``:'' denote uncertain offset.
$^{(e)}$~SN type, mostly from spectroscopy. All the updated SN classifications are labeled by the letter U.
In a few cases, marked by ``*'', types have been inferred from the light curve.
Uncertainties in SN type are marked by ``:'' and ``?'' (highly uncertain).
$^{(f)}$~Host galaxy SDSS identification. An alternative name is mentioned when SDSS
identification was not possible to obtain.
In a few cases, where the association with a definite host galaxy was not
possible (multiple galaxy systems, etc.), we have added ``:'' symbol.
Unidentified (anonymous) galaxies are listed with the letter A.
$^{(g)}$~Equatorial coordinates of host galaxy at the 2000.0 epoch, in degrees.
$^{(h)}$~Heliocentric redshift of host galaxy from different sources
(mostly from the SN catalogues and SDSS spectra).
$^{(i)}$~Morphological type of host galaxy. Symbol ``:'' indicates that the classification is doubtful, ``pec''
indicates that the galaxy is peculiar, and finally ``?'' indicates that
the classification is highly uncertain.
$^{(j)}$~SNe host galaxies with disturbed disk structures.
$^{(k)}$~The activity of host in the BPT and WHAN diagrams,
which includes narrow-line AGN (Seyfert (Sy) or LINER),
SF, composite (C), and retired/passive (RP) galaxies.
The WHAN column includes also broad-line AGN (BL~AGN), and passive (P) galaxies.}
\end{table}
\end{landscape}
\end{center}


\begin{thebibliography}{132}
\expandafter\ifx\csname natexlab\endcsname\relax\def\natexlab#1{#1}\fi

\bibitem[{{Aihara} {et~al.}(2011){Aihara}, {Allende Prieto}, {An}, {Anderson},
  {Aubourg}, {Balbinot}, {Beers}, {Berlind}, {Bickerton}, {Bizyaev}, {Blanton},
  {Bochanski}, {Bolton}, {Bovy}, {Brandt}, {Brinkmann}, {Brown}, {Brownstein},
  {Busca}, {Campbell}, {Carr}, {Chen}, {Chiappini}, {Comparat}, {Connolly},
  {Cortes}, {Croft}, {Cuesta}, {da Costa}, {Davenport}, {Dawson}, {Dhital},
  {Ealet}, {Ebelke}, {Edmondson}, {Eisenstein}, {Escoffier}, {Esposito},
  {Evans}, {Fan}, {Femen{\'{\i}}a Castell{\'a}}, {Font-Ribera}, {Frinchaboy},
  {Ge}, {Gillespie}, {Gilmore}, {Gonz{\'a}lez Hern{\'a}ndez}, {Gott}, {Gould},
  {Grebel}, {Gunn}, {Hamilton}, {Harding}, {Harris}, {Hawley}, {Hearty}, {Ho},
  {Hogg}, {Holtzman}, {Honscheid}, {Inada}, {Ivans}, {Jiang}, {Johnson},
  {Jordan}, {Jordan}, {Kazin}, {Kirkby}, {Klaene}, {Knapp}, {Kneib},
  {Kochanek}, {Koesterke}, {Kollmeier}, {Kron}, {Lampeitl}, {Lang}, {Le Goff},
  {Lee}, {Lin}, {Long}, {Loomis}, {Lucatello}, {Lundgren}, {Lupton}, {Ma},
  {MacDonald}, {Mahadevan}, {Maia}, {Makler}, {Malanushenko}, {Malanushenko},
  {Mandelbaum}, {Maraston}, {Margala}, {Masters}, {McBride}, {McGehee},
  {McGreer}, {M{\'e}nard}, {Miralda-Escud{\'e}}, {Morrison}, {Mullally},
  {Muna}, {Munn}, {Murayama}, {Myers}, {Naugle}, {Fausti Neto}, {Cuong Nguyen},
  {Nichol}, {O'Connell}, {Ogando}, {Olmstead}, {Oravetz}, {Padmanabhan},
  {Palanque-Delabrouille}, {Pan}, {Pandey}, {P{\^a}ris}, {Percival},
  {Petitjean}, {Pfaffenberger}, {Pforr}, {Phleps}, {Pichon}, {Pieri}, {Prada},
  {Price-Whelan}, {Raddick}, {Ramos}, {Reyl{\'e}}, {Rich}, {Richards}, {Rix},
  {Robin}, {Rocha-Pinto}, {Rockosi}, {Roe}, {Rollinde}, {Ross}, {Ross},
  {Rossetto}, {S{\'a}nchez}, {Sayres}, {Schlegel}, {Schlesinger}, {Schmidt},
  {Schneider}, {Sheldon}, {Shu}, {Simmerer}, {Simmons}, {Sivarani}, {Snedden},
  {Sobeck}, {Steinmetz}, {Strauss}, {Szalay}, {Tanaka}, {Thakar}, {Thomas},
  {Tinker}, {Tofflemire}, {Tojeiro}, {Tremonti}, {Vandenberg}, {Vargas
  Maga{\~n}a}, {Verde}, {Vogt}, {Wake}, {Wang}, {Weaver}, {Weinberg}, {White},
  {White}, {Yanny}, {Yasuda}, {Yeche}, \& {Zehavi}}]{2011ApJS..193...29A}
{Aihara}, H., {Allende Prieto}, C., {An}, D., {et~al.} 2011, \apjs, 193, 29

\bibitem[{{Aldering} {et~al.}(2005){Aldering}, {Lee}, {Loken}, {Nugent},
  {Perlmutter}, {Scalzo}, {Thomas}, {Wang}, {Bonnaud}, {Pecontal}, {Blanc},
  {Bongard}, {Copin}, {Gangler}, {Sauge}, {Smadja}, {Kessler}, {Antilogus},
  {Garavini}, {Gilles}, {Pain}, {Baltay}, {Rabinowitz}, \&
  {Bauer}}]{2005ATel..451....1A}
{Aldering}, G., {Lee}, B.~C., {Loken}, S., {et~al.} 2005, The Astronomer's
  Telegram, 451, 1

\bibitem[{{Alongi} {et~al.}(1993){Alongi}, {Bertelli}, {Bressan}, {Chiosi},
  {Fagotto}, {Greggio}, \& {Nasi}}]{1993A&AS...97..851A}
{Alongi}, M., {Bertelli}, G., {Bressan}, A., {et~al.} 1993, \aaps, 97, 851

\bibitem[{{Anderson} \& {James}(2008)}]{2008MNRAS.390.1527A}
{Anderson}, J.~P. \& {James}, P.~A. 2008, \mnras, 390, 1527

\bibitem[{{Anderson} \& {James}(2009)}]{2009MNRAS.399..559A}
{Anderson}, J.~P. \& {James}, P.~A. 2009, \mnras, 399, 559

\bibitem[{{Arcavi} {et~al.}(2010){Arcavi}, {Gal-Yam}, {Kasliwal}, {Quimby},
  {Ofek}, {Kulkarni}, {Nugent}, {Cenko}, {Bloom}, {Sullivan}, {Howell},
  {Poznanski}, {Filippenko}, {Law}, {Hook}, {J{\"o}nsson}, {Blake}, {Cooke},
  {Dekany}, {Rahmer}, {Hale}, {Smith}, {Zolkower}, {Velur}, {Walters},
  {Henning}, {Bui}, {McKenna}, \& {Jacobsen}}]{2010ApJ...721..777A}
{Arcavi}, I., {Gal-Yam}, A., {Kasliwal}, M.~M., {et~al.} 2010, \apj, 721, 777

\bibitem[{{Baillard} {et~al.}(2011){Baillard}, {Bertin}, {de Lapparent},
  {Fouqu{\'e}}, {Arnouts}, {Mellier}, {Pell{\'o}}, {Leborgne}, {Prugniel},
  {Makarov}, {Makarova}, {McCracken}, {Bijaoui}, \&
  {Tasca}}]{2011A&A...532A..74B}
{Baillard}, A., {Bertin}, E., {de Lapparent}, V., {et~al.} 2011, \aap, 532, A74

\bibitem[{{Baldwin} {et~al.}(1981){Baldwin}, {Phillips}, \&
  {Terlevich}}]{1981PASP...93....5B}
{Baldwin}, J.~A., {Phillips}, M.~M., \& {Terlevich}, R. 1981, \pasp, 93, 5

\bibitem[{{Barbon} {et~al.}(1999){Barbon}, {Buond{\'{\i}}}, {Cappellaro}, \&
  {Turatto}}]{1999A&AS..139..531B}
{Barbon}, R., {Buond{\'{\i}}}, V., {Cappellaro}, E., \& {Turatto}, M. 1999,
  \aaps, 139, 531

\bibitem[{{Bartunov} {et~al.}(1992){Bartunov}, {Makarova}, \&
  {Tsvetkov}}]{1992A&A...264..428B}
{Bartunov}, O.~S., {Makarova}, I.~N., \& {Tsvetkov}, D.~I. 1992, \aap, 264, 428

\bibitem[{{Bartunov} {et~al.}(1994){Bartunov}, {Tsvetkov}, \&
  {Filimonova}}]{1994PASP..106.1276B}
{Bartunov}, O.~S., {Tsvetkov}, D.~Y., \& {Filimonova}, I.~V. 1994, \pasp, 106,
  1276

\bibitem[{{Bassett} {et~al.}(2007{\natexlab{a}}){Bassett}, {Becker}, {Bizyaev},
  {Brewington}, {Choi}, {Cinabro}, {D'Andrea}, {Dembicky}, {Depoy}, {Dilday},
  {Doi}, {Eastman}, {Frieman}, {Gall}, {Garnavich}, {Goobar}, {Hogan},
  {Holtzman}, {Im}, {Jha}, {Konishi}, {Krzesinski}, {Lampeitl}, {Kessler},
  {Ketzeback}, {Long}, {Malanushenko}, {Marriner}, {McMillan}, {Miknaitis},
  {Morokuma}, {Mosher}, {Nichol}, {Oravetz}, {Pan}, {Ostman}, {Prieto},
  {Richmond}, {Riess}, {Romani}, {Sako}, {Schneider}, {Simmons}, {Smith},
  {Snedden}, {Sollerman}, {Stritzinger}, {Takanashi}, {Tokita}, {Taylor}, {van
  der Heyden}, {Watters}, {Yasuda}, {Wheeler}, {Zheng}, {Bender}, {Hopp},
  {Kollatschny}, {Assef}, {Peeples}, {Molla}, {Castander}, {Miquel},
  {McGinnis}, {Challis}, {Narayan}, \& {Kirshner}}]{2007CBET.1098....1B}
{Bassett}, B., {Becker}, A., {Bizyaev}, D., {et~al.} 2007{\natexlab{a}},
  Central Bureau Electronic Telegrams, 1098, 1

\bibitem[{{Bassett} {et~al.}(2007{\natexlab{b}}){Bassett}, {Becker}, {Bizyaev},
  {Brewington}, {Choi}, {Cinabro}, {D'Andrea}, {Dembicky}, {Depoy}, {Dilday},
  {Doi}, {Eastman}, {Frieman}, {Garnavich}, {Goobar}, {Hogan}, {Holtzman},
  {Im}, {Jha}, {Kessler}, {Ketzeback}, {Konishi}, {Krzesinski}, {Lampeitl},
  {Leloudas}, {Long}, {Malanushenko}, {Marriner}, {McMillan}, {Miknaitis},
  {Morokuma}, {Mosher}, {Nichol}, {Oravetz}, {Pan}, {Ostman}, {Prieto},
  {Richmond}, {Riess}, {Romani}, {Sako}, {Schneider}, {Simmons}, {Smith},
  {Snedden}, {Sollerman}, {Stritzinger}, {Takanashi}, {Tokita}, {Taylor}, {van
  der Heyden}, {Watters}, {Yasuda}, {Wheeler}, {Zheng}, {Watson}, {Filippenko},
  {Silverman}, {Foley}, {Modjaz}, {McGinnis}, {Aragon-Salamanca}, {Bremer},
  {Turatto}, {Ruiz-Lapuente}, {Castander}, {Romer}, {Collins}, {Lucey}, \&
  {Edge}}]{2007CBET.1102....1B}
{Bassett}, B., {Becker}, A., {Bizyaev}, D., {et~al.} 2007{\natexlab{b}},
  Central Bureau Electronic Telegrams, 1102, 1

\bibitem[{{Bettoni} {et~al.}(1998){Bettoni}, {Hjorth}, \&
  {Benetti}}]{1998IAUC.6911....1B}
{Bettoni}, D., {Hjorth}, J., \& {Benetti}, S. 1998, \iaucirc, 6911, 1

\bibitem[{{Blanton} {et~al.}(2003){Blanton}, {Hogg}, {Bahcall}, {Brinkmann},
  {Britton}, {Connolly}, {Csabai}, {Fukugita}, {Loveday}, {Meiksin}, {Munn},
  {Nichol}, {Okamura}, {Quinn}, {Schneider}, {Shimasaku}, {Strauss}, {Tegmark},
  {Vogeley}, \& {Weinberg}}]{2003ApJ...592..819B}
{Blanton}, M.~R., {Hogg}, D.~W., {Bahcall}, N.~A., {et~al.} 2003, \apj, 592,
  819

\bibitem[{{Boissier} \& {Prantzos}(2009)}]{2009A&A...503..137B}
{Boissier}, S. \& {Prantzos}, N. 2009, \aap, 503, 137

\bibitem[{{Bottinelli} {et~al.}(1995){Bottinelli}, {Gouguenheim}, {Paturel}, \&
  {Teerikorpi}}]{1995A&A...296...64B}
{Bottinelli}, L., {Gouguenheim}, L., {Paturel}, G., \& {Teerikorpi}, P. 1995,
  \aap, 296, 64

\bibitem[{{Bressan} {et~al.}(2002){Bressan}, {Della Valle}, \&
  {Marziani}}]{2002MNRAS.331L..25B}
{Bressan}, A., {Della Valle}, M., \& {Marziani}, P. 2002, \mnras, 331, L25

\bibitem[{{Bressan} {et~al.}(1993){Bressan}, {Fagotto}, {Bertelli}, \&
  {Chiosi}}]{1993A&AS..100..647B}
{Bressan}, A., {Fagotto}, F., {Bertelli}, G., \& {Chiosi}, C. 1993, \aaps, 100,
  647

\bibitem[{{Bruzual} \& {Charlot}(2003)}]{2003MNRAS.344.1000B}
{Bruzual}, G. \& {Charlot}, S. 2003, \mnras, 344, 1000

\bibitem[{{Cappellaro} {et~al.}(1999){Cappellaro}, {Evans}, \&
  {Turatto}}]{1999A&A...351..459C}
{Cappellaro}, E., {Evans}, R., \& {Turatto}, M. 1999, \aap, 351, 459

\bibitem[{{Cappellaro} \& {Turatto}(1988)}]{1988A&A...190...10C}
{Cappellaro}, E. \& {Turatto}, M. 1988, \aap, 190, 10

\bibitem[{{Cappellaro} {et~al.}(1997){Cappellaro}, {Turatto}, {Tsvetkov},
  {Bartunov}, {Pollas}, {Evans}, \& {Hamuy}}]{1997A&A...322..431C}
{Cappellaro}, E., {Turatto}, M., {Tsvetkov}, D.~Y., {et~al.} 1997, \aap, 322,
  431

\bibitem[{{Cardelli} {et~al.}(1989){Cardelli}, {Clayton}, \&
  {Mathis}}]{1989ApJ...345..245C}
{Cardelli}, J.~A., {Clayton}, G.~C., \& {Mathis}, J.~S. 1989, \apj, 345, 245

\bibitem[{{Chabrier}(2003)}]{2003PASP..115..763C}
{Chabrier}, G. 2003, \pasp, 115, 763

\bibitem[{{Cherepashchuk} {et~al.}(1987){Cherepashchuk}, {Metlova}, {Wheeler},
  {Kirshner}, {Crotts}, {McMahan}, {Wegner}, \&
  {Swanson}}]{1987IAUC.4381....1C}
{Cherepashchuk}, A.~M., {Metlova}, N., {Wheeler}, J.~C., {et~al.} 1987,
  \iaucirc, 4381, 1

\bibitem[{{Cid Fernandes} {et~al.}(2004){Cid Fernandes}, {Gonz{\'a}lez
  Delgado}, {Schmitt}, {Storchi-Bergmann}, {Martins}, {P{\'e}rez}, {Heckman},
  {Leitherer}, \& {Schaerer}}]{2004ApJ...605..105C}
{Cid Fernandes}, R., {Gonz{\'a}lez Delgado}, R.~M., {Schmitt}, H., {et~al.}
  2004, \apj, 605, 105

\bibitem[{{Cid Fernandes} {et~al.}(2005){Cid Fernandes}, {Mateus}, {Sodr{\'e}},
  {Stasi{\'n}ska}, \& {Gomes}}]{2005MNRAS.358..363C}
{Cid Fernandes}, R., {Mateus}, A., {Sodr{\'e}}, L., {Stasi{\'n}ska}, G., \&
  {Gomes}, J.~M. 2005, \mnras, 358, 363

\bibitem[{{Cid Fernandes} {et~al.}(2011){Cid Fernandes}, {Stasi{\'n}ska},
  {Mateus}, \& {Vale Asari}}]{2011MNRAS.413.1687C}
{Cid Fernandes}, R., {Stasi{\'n}ska}, G., {Mateus}, A., \& {Vale Asari}, N.
  2011, \mnras, 413, 1687

\bibitem[{{Cid Fernandes} {et~al.}(2010){Cid Fernandes}, {Stasi{\'n}ska},
  {Schlickmann}, {Mateus}, {Vale Asari}, {Schoenell}, \&
  {Sodr{\'e}}}]{2010MNRAS.403.1036C}
{Cid Fernandes}, R., {Stasi{\'n}ska}, G., {Schlickmann}, M.~S., {et~al.} 2010,
  \mnras, 403, 1036

\bibitem[{{Condon} {et~al.}(1998){Condon}, {Cotton}, {Greisen}, {Yin},
  {Perley}, {Taylor}, \& {Broderick}}]{1998AJ....115.1693C}
{Condon}, J.~J., {Cotton}, W.~D., {Greisen}, E.~W., {et~al.} 1998, \aj, 115,
  1693

\bibitem[{{de Vaucouleurs} {et~al.}(1991){de Vaucouleurs}, {de Vaucouleurs},
  {Corwin}, {Buta}, {Paturel}, \& {Fouqu{\'e}}}]{1991rc3..book.....D}
{de Vaucouleurs}, G., {de Vaucouleurs}, A., {Corwin}, Jr., H.~G., {et~al.}
  1991, {Third Reference Catalogue of Bright Galaxies. Volume I: Explanations
  and references. Volume II: Data for galaxies between 0$^{h}$ and 12$^{h}$.
  Volume III: Data for galaxies between 12$^{h}$ and 24$^{h}$.}, ed. {de
  Vaucouleurs, G., de Vaucouleurs, A., Corwin, H.~G., Jr., Buta, R.~J.,
  Paturel, G., \& Fouqu{\'e}, P.}

\bibitem[{{Fagotto} {et~al.}(1994{\natexlab{a}}){Fagotto}, {Bressan},
  {Bertelli}, \& {Chiosi}}]{1994A&AS..104..365F}
{Fagotto}, F., {Bressan}, A., {Bertelli}, G., \& {Chiosi}, C.
  1994{\natexlab{a}}, \aaps, 104, 365

\bibitem[{{Fagotto} {et~al.}(1994{\natexlab{b}}){Fagotto}, {Bressan},
  {Bertelli}, \& {Chiosi}}]{1994A&AS..105...29F}
{Fagotto}, F., {Bressan}, A., {Bertelli}, G., \& {Chiosi}, C.
  1994{\natexlab{b}}, \aaps, 105, 29

\bibitem[{{Filippenko} \& {Chornock}(2000)}]{2000IAUC.7511....2F}
{Filippenko}, A.~V. \& {Chornock}, R. 2000, \iaucirc, 7511, 2

\bibitem[{{Filippenko} \& {Chornock}(2002)}]{2002IAUC.7825....1F}
{Filippenko}, A.~V. \& {Chornock}, R. 2002, \iaucirc, 7825, 1

\bibitem[{{Filippenko} {et~al.}(1999){Filippenko}, {Li}, \&
  {Modjaz}}]{1999IAUC.7152....2F}
{Filippenko}, A.~V., {Li}, W.~D., \& {Modjaz}, M. 1999, \iaucirc, 7152, 2

\bibitem[{{Filippenko} \& {Matheson}(1993)}]{1993IAUC.5842....2F}
{Filippenko}, A.~V. \& {Matheson}, T. 1993, \iaucirc, 5842, 2

\bibitem[{{Filippenko} {et~al.}(2007){Filippenko}, {Silverman}, {Foley},
  {Modjaz}, {Papovich}, {Willmer}, {Blondin}, \& {Brown}}]{2007CBET.1101....1F}
{Filippenko}, A.~V., {Silverman}, J.~M., {Foley}, R.~J., {et~al.} 2007, Central
  Bureau Electronic Telegrams, 1101, 1

\bibitem[{{F{\"o}rster} \& {Schawinski}(2008)}]{2008MNRAS.388L..74F}
{F{\"o}rster}, F. \& {Schawinski}, K. 2008, \mnras, 388, L74

\bibitem[{{Frieman} {et~al.}(2008){Frieman}, {Bassett}, {Becker}, {Choi},
  {Cinabro}, {DeJongh}, {Depoy}, {Dilday}, {Doi}, {Garnavich}, {Hogan},
  {Holtzman}, {Im}, {Jha}, {Kessler}, {Konishi}, {Lampeitl}, {Marriner},
  {Marshall}, {McGinnis}, {Miknaitis}, {Nichol}, {Prieto}, {Riess}, {Richmond},
  {Romani}, {Sako}, {Schneider}, {Smith}, {Takanashi}, {Tokita}, {van der
  Heyden}, {Yasuda}, {Zheng}, {Adelman-McCarthy}, {Annis}, {Assef},
  {Barentine}, {Bender}, {Blandford}, {Boroski}, {Bremer}, {Brewington},
  {Collins}, {Crotts}, {Dembicky}, {Eastman}, {Edge}, {Edmondson}, {Elson},
  {Eyler}, {Filippenko}, {Foley}, {Frank}, {Goobar}, {Gueth}, {Gunn},
  {Harvanek}, {Hopp}, {Ihara}, {Ivezi{\'c}}, {Kahn}, {Kaplan}, {Kent},
  {Ketzeback}, {Kleinman}, {Kollatschny}, {Kron}, {Krzesi{\'n}ski}, {Lamenti},
  {Leloudas}, {Lin}, {Long}, {Lucey}, {Lupton}, {Malanushenko}, {Malanushenko},
  {McMillan}, {Mendez}, {Morgan}, {Morokuma}, {Nitta}, {Ostman}, {Pan},
  {Rockosi}, {Romer}, {Ruiz-Lapuente}, {Saurage}, {Schlesinger}, {Snedden},
  {Sollerman}, {Stoughton}, {Stritzinger}, {Subba Rao}, {Tucker}, {Vaisanen},
  {Watson}, {Watters}, {Wheeler}, {Yanny}, \& {York}}]{2008AJ....135..338F}
{Frieman}, J.~A., {Bassett}, B., {Becker}, A., {et~al.} 2008, \aj, 135, 338

\bibitem[{{Gelman} \& {Rubin}(1992)}]{gr1992}
{Gelman}, A. \& {Rubin}, D.~B. 1992, Statistical Science, 7, 457

\bibitem[{{Girardi} {et~al.}(1996){Girardi}, {Bressan}, {Chiosi}, {Bertelli},
  \& {Nasi}}]{1996A&AS..117..113G}
{Girardi}, L., {Bressan}, A., {Chiosi}, C., {Bertelli}, G., \& {Nasi}, E. 1996,
  \aaps, 117, 113

\bibitem[{{Gunn} {et~al.}(2006){Gunn}, {Siegmund}, {Mannery}, {Owen}, {Hull},
  {Leger}, {Carey}, {Knapp}, {York}, {Boroski}, {Kent}, {Lupton}, {Rockosi},
  {Evans}, {Waddell}, {Anderson}, {Annis}, {Barentine}, {Bartoszek}, {Bastian},
  {Bracker}, {Brewington}, {Briegel}, {Brinkmann}, {Brown}, {Carr},
  {Czarapata}, {Drennan}, {Dombeck}, {Federwitz}, {Gillespie}, {Gonzales},
  {Hansen}, {Harvanek}, {Hayes}, {Jordan}, {Kinney}, {Klaene}, {Kleinman},
  {Kron}, {Kresinski}, {Lee}, {Limmongkol}, {Lindenmeyer}, {Long}, {Loomis},
  {McGehee}, {Mantsch}, {Neilsen}, {Neswold}, {Newman}, {Nitta}, {Peoples},
  {Pier}, {Prieto}, {Prosapio}, {Rivetta}, {Schneider}, {Snedden}, \&
  {Wang}}]{2006AJ....131.2332G}
{Gunn}, J.~E., {Siegmund}, W.~A., {Mannery}, E.~J., {et~al.} 2006, \aj, 131,
  2332

\bibitem[{{Gyulzadyan} {et~al.}(2011){Gyulzadyan}, {McLean}, {Adibekyan},
  {Allen}, {Kunth}, {Petrosian}, \& {Stepanian}}]{2011Ap.....54...15G}
{Gyulzadyan}, M., {McLean}, B., {Adibekyan}, V.~Z., {et~al.} 2011,
  Astrophysics, 54, 15

\bibitem[{{Hakobyan}(2008)}]{2008Ap.....51...69H}
{Hakobyan}, A.~A. 2008, Astrophysics, 51, 69

\bibitem[{{Hakobyan} {et~al.}(2009){Hakobyan}, {Mamon}, {Petrosian}, {Kunth},
  \& {Turatto}}]{2009A&A...508.1259H}
{Hakobyan}, A.~A., {Mamon}, G.~A., {Petrosian}, A.~R., {Kunth}, D., \&
  {Turatto}, M. 2009, \aap, 508, 1259

\bibitem[{{Hakobyan} {et~al.}(2011){Hakobyan}, {Petrosian}, {Mamon}, {McLean},
  {Kunth}, {Turatto}, {Cappellaro}, {Mannucci}, {Allen}, {Panagia}, \& {Della
  Valle}}]{2011Ap.....54..301H}
{Hakobyan}, A.~A., {Petrosian}, A.~R., {Mamon}, G.~A., {et~al.} 2011,
  Astrophysics, 54, 301

\bibitem[{{Hakobyan} {et~al.}(2008){Hakobyan}, {Petrosian}, {McLean}, {Kunth},
  {Allen}, {Turatto}, \& {Barbon}}]{2008A&A...488..523H}
{Hakobyan}, A.~A., {Petrosian}, A.~R., {McLean}, B., {et~al.} 2008, \aap, 488,
  523

\bibitem[{{Han} {et~al.}(2010){Han}, {Park}, {Choi}, \&
  {Park}}]{2010ApJ...724..502H}
{Han}, D.-H., {Park}, C., {Choi}, Y.-Y., \& {Park}, M.-G. 2010, \apj, 724, 502

\bibitem[{{Herrero-Illana} {et~al.}(2012){Herrero-Illana}, {P{\'e}rez-Torres},
  \& {Alberdi}}]{2012A&A...540L...5H}
{Herrero-Illana}, R., {P{\'e}rez-Torres}, M.~{\'A}., \& {Alberdi}, A. 2012,
  \aap, 540, L5

\bibitem[{{Hubble}(1926)}]{1926ApJ....64..321H}
{Hubble}, E.~P. 1926, \apj, 64, 321

\bibitem[{{Humphreys} {et~al.}(2010){Humphreys}, {Prieto}, {Rosenfield},
  {Helton}, {Kochanek}, {Stanek}, {Khan}, {Szczygiel}, {Mogren}, {Fesen},
  {Milisavljevic}, {Williams}, {Murphy}, {Dalcanton}, \&
  {Gilbert}}]{2010ApJ...718L..43H}
{Humphreys}, R.~M., {Prieto}, J.~L., {Rosenfield}, P., {et~al.} 2010, \apjl,
  718, L43

\bibitem[{{Hutchings} \& {Li}(2001)}]{2001IAUC.7724....1H}
{Hutchings}, D. \& {Li}, W.~D. 2001, \iaucirc, 7724, 1

\bibitem[{{Iye} \& {Kodaira}(1975)}]{1975PASJ...27..411I}
{Iye}, M. \& {Kodaira}, K. 1975, \pasj, 27, 411

\bibitem[{{Jester} {et~al.}(2005){Jester}, {Schneider}, {Richards}, {Green},
  {Schmidt}, {Hall}, {Strauss}, {Vanden Berk}, {Stoughton}, {Gunn},
  {Brinkmann}, {Kent}, {Smith}, {Tucker}, \& {Yanny}}]{2005AJ....130..873J}
{Jester}, S., {Schneider}, D.~P., {Richards}, G.~T., {et~al.} 2005, \aj, 130,
  873

\bibitem[{{Kasliwal} {et~al.}(2009){Kasliwal}, {Kulkarni}, {Quimby}, {Nugent},
  {Howell}, {Cooke}, {Cenko}, {Gal-Yam}, {Law}, {Levitan}, {Ofek}, \&
  {Poznanski}}]{2009CBET.1819....1K}
{Kasliwal}, M.~M., {Kulkarni}, S.~R., {Quimby}, R., {et~al.} 2009, Central
  Bureau Electronic Telegrams, 1819, 1

\bibitem[{{Kauffmann} {et~al.}(2003){Kauffmann}, {Heckman}, {Tremonti},
  {Brinchmann}, {Charlot}, {White}, {Ridgway}, {Brinkmann}, {Fukugita}, {Hall},
  {Ivezi{\'c}}, {Richards}, \& {Schneider}}]{2003MNRAS.346.1055K}
{Kauffmann}, G., {Heckman}, T.~M., {Tremonti}, C., {et~al.} 2003, \mnras, 346,
  1055

\bibitem[{{Kaviraj} {et~al.}(2008){Kaviraj}, {Khochfar}, {Schawinski}, {Yi},
  {Gawiser}, {Silk}, {Virani}, {Cardamone}, {van Dokkum}, \&
  {Urry}}]{2008MNRAS.388...67K}
{Kaviraj}, S., {Khochfar}, S., {Schawinski}, K., {et~al.} 2008, \mnras, 388, 67

\bibitem[{{Kaviraj} {et~al.}(2009){Kaviraj}, {Peirani}, {Khochfar}, {Silk}, \&
  {Kay}}]{2009MNRAS.394.1713K}
{Kaviraj}, S., {Peirani}, S., {Khochfar}, S., {Silk}, J., \& {Kay}, S. 2009,
  \mnras, 394, 1713

\bibitem[{{Kaviraj} {et~al.}(2007){Kaviraj}, {Schawinski}, {Devriendt},
  {Ferreras}, {Khochfar}, {Yoon}, {Yi}, {Deharveng}, {Boselli}, {Barlow},
  {Conrow}, {Forster}, {Friedman}, {Martin}, {Morrissey}, {Neff},
  {Schiminovich}, {Seibert}, {Small}, {Wyder}, {Bianchi}, {Donas}, {Heckman},
  {Lee}, {Madore}, {Milliard}, {Rich}, \& {Szalay}}]{2007ApJS..173..619K}
{Kaviraj}, S., {Schawinski}, K., {Devriendt}, J.~E.~G., {et~al.} 2007, \apjs,
  173, 619

\bibitem[{{Kawabata} {et~al.}(2010){Kawabata}, {Maeda}, {Nomoto},
  {Taubenberger}, {Tanaka}, {Deng}, {Pian}, {Hattori}, \&
  {Itagaki}}]{2010Natur.465..326K}
{Kawabata}, K.~S., {Maeda}, K., {Nomoto}, K., {et~al.} 2010, \nat, 465, 326

\bibitem[{{Kazarian}(1997)}]{1997Ap.....40..296K}
{Kazarian}, M.~A. 1997, Astrophysics, 40, 296

\bibitem[{{Kelly} \& {Kirshner}(2011)}]{2011arXiv1110.1377K}
{Kelly}, P.~L. \& {Kirshner}, R.~P. 2011, arXiv:1110.1377

\bibitem[{{Kewley} {et~al.}(2001){Kewley}, {Dopita}, {Sutherland}, {Heisler},
  \& {Trevena}}]{2001ApJ...556..121K}
{Kewley}, L.~J., {Dopita}, M.~A., {Sutherland}, R.~S., {Heisler}, C.~A., \&
  {Trevena}, J. 2001, \apj, 556, 121

\bibitem[{{Kewley} {et~al.}(2006){Kewley}, {Groves}, {Kauffmann}, \&
  {Heckman}}]{2006MNRAS.372..961K}
{Kewley}, L.~J., {Groves}, B., {Kauffmann}, G., \& {Heckman}, T. 2006, \mnras,
  372, 961

\bibitem[{{Kewley} {et~al.}(2005){Kewley}, {Jansen}, \&
  {Geller}}]{2005PASP..117..227K}
{Kewley}, L.~J., {Jansen}, R.~A., \& {Geller}, M.~J. 2005, \pasp, 117, 227

\bibitem[{{Kushida} {et~al.}(1998){Kushida}, {Nakano}, {Kushida}, \&
  {Aoki}}]{1998IAUC.6920....2K}
{Kushida}, R., {Nakano}, S., {Kushida}, Y., \& {Aoki}, M. 1998, \iaucirc, 6920,
  2

\bibitem[{{Leaman} {et~al.}(2011){Leaman}, {Li}, {Chornock}, \&
  {Filippenko}}]{2011MNRAS.412.1419L}
{Leaman}, J., {Li}, W., {Chornock}, R., \& {Filippenko}, A.~V. 2011, \mnras,
  412, 1419

\bibitem[{{Leloudas} {et~al.}(2011){Leloudas}, {Gallazzi}, {Sollerman},
  {Stritzinger}, {Fynbo}, {Hjorth}, {Malesani}, {Micha{\l}owski},
  {Milvang-Jensen}, \& {Smith}}]{2011A&A...530A..95L}
{Leloudas}, G., {Gallazzi}, A., {Sollerman}, J., {et~al.} 2011, \aap, 530, A95

\bibitem[{{Lennarz} {et~al.}(2012){Lennarz}, {Altmann}, \&
  {Wiebusch}}]{2012A&A...538A.120L}
{Lennarz}, D., {Altmann}, D., \& {Wiebusch}, C. 2012, \aap, 538, A120

\bibitem[{{Leonard}(2010)}]{2010ATel.2750....1L}
{Leonard}, D.~C. 2010, The Astronomer's Telegram, 2750, 1

\bibitem[{{Li} {et~al.}(2011){Li}, {Chornock}, {Leaman}, {Filippenko},
  {Poznanski}, {Wang}, {Ganeshalingam}, \& {Mannucci}}]{2011MNRAS.412.1473L}
{Li}, W., {Chornock}, R., {Leaman}, J., {et~al.} 2011, \mnras, 412, 1473

\bibitem[{{Li} {et~al.}(2005){Li}, {Wang}, \& {Bian}}]{2005CBET..332....1L}
{Li}, W., {Wang}, X.-F., \& {Bian}, F.-Y. 2005, Central Bureau Electronic
  Telegrams, 332, 1

\bibitem[{{Maksym} {et~al.}(2010){Maksym}, {Elenin}, \&
  {Schwartz}}]{2010CBET.2245....1M}
{Maksym}, A., {Elenin}, L., \& {Schwartz}, M. 2010, Central Bureau Electronic
  Telegrams, 2245, 1

\bibitem[{{Mannucci} {et~al.}(2005){Mannucci}, {Della Valle}, {Panagia},
  {Cappellaro}, {Cresci}, {Maiolino}, {Petrosian}, \&
  {Turatto}}]{2005A&A...433..807M}
{Mannucci}, F., {Della Valle}, M., {Panagia}, N., {et~al.} 2005, \aap, 433, 807

\bibitem[{{Matheson} {et~al.}(2001){Matheson}, {Jha}, {Challis}, {Kirshner}, \&
  {Calkins}}]{2001IAUC.7597....3M}
{Matheson}, T., {Jha}, S., {Challis}, P., {Kirshner}, R., \& {Calkins}, M.
  2001, \iaucirc, 7597, 3

\bibitem[{{Maund} {et~al.}(2006){Maund}, {Smartt}, {Kudritzki}, {Pastorello},
  {Nelemans}, {Bresolin}, {Patat}, {Gilmore}, \& {Benn}}]{2006MNRAS.369..390M}
{Maund}, J.~R., {Smartt}, S.~J., {Kudritzki}, R.-P., {et~al.} 2006, \mnras,
  369, 390

\bibitem[{{Maxwell} {et~al.}(2010){Maxwell}, {Graham}, {Parker}, {Sadavoy},
  {Pritchet}, {Hsiao}, \& {Balam}}]{2010CBET.2245....2M}
{Maxwell}, A.~J., {Graham}, M.~L., {Parker}, A., {et~al.} 2010, Central Bureau
  Electronic Telegrams, 2245, 2

\bibitem[{{Miknaitis} {et~al.}(2007){Miknaitis}, {Pignata}, {Rest},
  {Wood-Vasey}, {Blondin}, {Challis}, {Smith}, {Stubbs}, {Suntzeff}, {Foley},
  {Matheson}, {Tonry}, {Aguilera}, {Blackman}, {Becker}, {Clocchiatti},
  {Covarrubias}, {Davis}, {Filippenko}, {Garg}, {Garnavich}, {Hicken}, {Jha},
  {Krisciunas}, {Kirshner}, {Leibundgut}, {Li}, {Miceli}, {Narayan}, {Prieto},
  {Riess}, {Salvo}, {Schmidt}, {Sollerman}, {Spyromilio}, \&
  {Zenteno}}]{2007ApJ...666..674M}
{Miknaitis}, G., {Pignata}, G., {Rest}, A., {et~al.} 2007, \apj, 666, 674

\bibitem[{{Modjaz} {et~al.}(2011){Modjaz}, {Kewley}, {Bloom}, {Filippenko},
  {Perley}, \& {Silverman}}]{2011ApJ...731L...4M}
{Modjaz}, M., {Kewley}, L., {Bloom}, J.~S., {et~al.} 2011, \apjl, 731, L4

\bibitem[{{Modjaz} {et~al.}(2005){Modjaz}, {Kirshner}, {Challis}, {Blondin}, \&
  {Berlind}}]{2005IAUC.8650....2M}
{Modjaz}, M., {Kirshner}, R., {Challis}, P., {Blondin}, S., \& {Berlind}, P.
  2005, \iaucirc, 8650, 2

\bibitem[{{Naim} {et~al.}(1995){Naim}, {Lahav}, {Buta}, {Corwin}, {de
  Vaucouleurs}, {Dressler}, {Huchra}, {van den Bergh}, {Raychaudhury}, {Sodre},
  \& {Storrie-Lombardi}}]{1995MNRAS.274.1107N}
{Naim}, A., {Lahav}, O., {Buta}, R.~J., {et~al.} 1995, \mnras, 274, 1107

\bibitem[{{Nair} \& {Abraham}(2010)}]{2010ApJS..186..427N}
{Nair}, P.~B. \& {Abraham}, R.~G. 2010, \apjs, 186, 427

\bibitem[{{Navasardyan} {et~al.}(2001){Navasardyan}, {Petrosian}, {Turatto},
  {Cappellaro}, \& {Boulesteix}}]{2001MNRAS.328.1181N}
{Navasardyan}, H., {Petrosian}, A.~R., {Turatto}, M., {Cappellaro}, E., \&
  {Boulesteix}, J. 2001, \mnras, 328, 1181

\bibitem[{{Nugent} {et~al.}(1998){Nugent}, {Aldering}, {Castro}, {Nunes}, \&
  {Quimby}}]{1998IAUC.6804....1N}
{Nugent}, P., {Aldering}, G., {Castro}, P., {Nunes}, N., \& {Quimby}, R. 1998,
  \iaucirc, 6804, 1

\bibitem[{{Pastorello}(2012)}]{2012MSAIS..19...24P}
{Pastorello}, A. 2012, Memorie della Societa Astronomica Italiana Supplementi,
  19, 24

\bibitem[{{Patat} {et~al.}(1997){Patat}, {Barbon}, {Cappellaro}, \&
  {Turatto}}]{1997A&A...317..423P}
{Patat}, F., {Barbon}, R., {Cappellaro}, E., \& {Turatto}, M. 1997, \aap, 317,
  423

\bibitem[{{Patat} {et~al.}(2003){Patat}, {Pastorello}, \&
  {Aceituno}}]{2003IAUC.8167....3P}
{Patat}, F., {Pastorello}, A., \& {Aceituno}, J. 2003, \iaucirc, 8167, 3

\bibitem[{{Paturel} {et~al.}(1997){Paturel}, {Andernach}, {Bottinelli}, {di
  Nella}, {Durand}, {Garnier}, {Gouguenheim}, {Lanoix}, {Marthinet}, {Petit},
  {Rousseau}, {Theureau}, \& {Vauglin}}]{1997A&AS..124..109P}
{Paturel}, G., {Andernach}, H., {Bottinelli}, L., {et~al.} 1997, \aaps, 124,
  109

\bibitem[{{Paturel} {et~al.}(2003){Paturel}, {Petit}, {Prugniel}, {Theureau},
  {Rousseau}, {Brouty}, {Dubois}, \& {Cambr{\'e}sy}}]{2003A&A...412...45P}
{Paturel}, G., {Petit}, C., {Prugniel}, P., {et~al.} 2003, \aap, 412, 45

\bibitem[{{Perets} {et~al.}(2011){Perets}, {Gal-Yam}, {Crockett}, {Anderson},
  {James}, {Sullivan}, {Neill}, \& {Leonard}}]{2011ApJ...728L..36P}
{Perets}, H.~B., {Gal-Yam}, A., {Crockett}, R.~M., {et~al.} 2011, \apjl, 728,
  L36

\bibitem[{{Perets} {et~al.}(2010){Perets}, {Gal-Yam}, {Mazzali}, {Arnett},
  {Kagan}, {Filippenko}, {Li}, {Arcavi}, {Cenko}, {Fox}, {Leonard}, {Moon},
  {Sand}, {Soderberg}, {Anderson}, {James}, {Foley}, {Ganeshalingam}, {Ofek},
  {Bildsten}, {Nelemans}, {Shen}, {Weinberg}, {Metzger}, {Piro}, {Quataert},
  {Kiewe}, \& {Poznanski}}]{2010Natur.465..322P}
{Perets}, H.~B., {Gal-Yam}, A., {Mazzali}, P.~A., {et~al.} 2010, \nat, 465, 322

\bibitem[{{Petrosian} {et~al.}(2007){Petrosian}, {McLean}, {Allen}, \&
  {MacKenty}}]{2007ApJS..170...33P}
{Petrosian}, A., {McLean}, B., {Allen}, R.~J., \& {MacKenty}, J.~W. 2007,
  \apjs, 170, 33

\bibitem[{{Petrosian} {et~al.}(2005){Petrosian}, {Navasardyan}, {Cappellaro},
  {McLean}, {Allen}, {Panagia}, {Leitherer}, {MacKenty}, \&
  {Turatto}}]{2005AJ....129.1369P}
{Petrosian}, A., {Navasardyan}, H., {Cappellaro}, E., {et~al.} 2005, \aj, 129,
  1369

\bibitem[{{Petrosian} \& {Turatto}(1990)}]{1990A&A...239...63P}
{Petrosian}, A.~R. \& {Turatto}, M. 1990, \aap, 239, 63

\bibitem[{{Petrosian} \& {Turatto}(1995)}]{1995A&A...297...49P}
{Petrosian}, A.~R. \& {Turatto}, M. 1995, \aap, 297, 49

\bibitem[{{Petrosian}(1976)}]{1976ApJ...209L...1P}
{Petrosian}, V. 1976, \apjl, 209, L1

\bibitem[{{Pojmanski}(2007)}]{2007IAUC.8875....1P}
{Pojmanski}, G. 2007, \iaucirc, 8875, 1

\bibitem[{{Prantzos} \& {Boissier}(2003)}]{2003A&A...406..259P}
{Prantzos}, N. \& {Boissier}, S. 2003, \aap, 406, 259

\bibitem[{{Prieto} {et~al.}(2008){Prieto}, {Stanek}, \&
  {Beacom}}]{2008ApJ...673..999P}
{Prieto}, J.~L., {Stanek}, K.~Z., \& {Beacom}, J.~F. 2008, \apj, 673, 999

\bibitem[{{Puckett} {et~al.}(2006){Puckett}, {Peoples}, {Joubert}, {Madison},
  {Mostardi}, {Khandrika}, {Li}, \& {Foley}}]{2006IAUC.8741....1P}
{Puckett}, T., {Peoples}, M., {Joubert}, N., {et~al.} 2006, \iaucirc, 8741, 1

\bibitem[{{Quimby} {et~al.}(2005){Quimby}, {Hoeflich}, {Wheeler}, {Gerardy},
  {Shetrone}, \& {Terrazas}}]{2005IAUC.8504....3Q}
{Quimby}, R., {Hoeflich}, P., {Wheeler}, J.~C., {et~al.} 2005, \iaucirc, 8504,
  3

\bibitem[{{Riess} {et~al.}(2009){Riess}, {Macri}, {Casertano}, {Sosey},
  {Lampeitl}, {Ferguson}, {Filippenko}, {Jha}, {Li}, {Chornock}, \&
  {Sarkar}}]{2009ApJ...699..539R}
{Riess}, A.~G., {Macri}, L., {Casertano}, S., {et~al.} 2009, \apj, 699, 539

\bibitem[{{Schaefer}(2001)}]{2001IAUC.7608....1S}
{Schaefer}, B.~E. 2001, \iaucirc, 7608, 1

\bibitem[{{Schlegel} {et~al.}(1998){Schlegel}, {Finkbeiner}, \&
  {Davis}}]{1998ApJ...500..525S}
{Schlegel}, D.~J., {Finkbeiner}, D.~P., \& {Davis}, M. 1998, \apj, 500, 525

\bibitem[{{Smartt}(2009)}]{2009ARA&A..47...63S}
{Smartt}, S.~J. 2009, \araa, 47, 63

\bibitem[{{Smith} {et~al.}(2011){Smith}, {Li}, {Silverman}, {Ganeshalingam}, \&
  {Filippenko}}]{2011MNRAS.415..773S}
{Smith}, N., {Li}, W., {Silverman}, J.~M., {Ganeshalingam}, M., \&
  {Filippenko}, A.~V. 2011, \mnras, 415, 773

\bibitem[{{Smith} {et~al.}(2010){Smith}, {Miller}, {Li}, {Filippenko},
  {Silverman}, {Howard}, {Nugent}, {Marcy}, {Bloom}, {Ghez}, {Lu}, {Yelda},
  {Bernstein}, \& {Colucci}}]{2010AJ....139.1451S}
{Smith}, N., {Miller}, A., {Li}, W., {et~al.} 2010, \aj, 139, 1451

\bibitem[{{Spergel} {et~al.}(2007){Spergel}, {Bean}, {Dor{\'e}}, {Nolta},
  {Bennett}, {Dunkley}, {Hinshaw}, {Jarosik}, {Komatsu}, {Page}, {Peiris},
  {Verde}, {Halpern}, {Hill}, {Kogut}, {Limon}, {Meyer}, {Odegard}, {Tucker},
  {Weiland}, {Wollack}, \& {Wright}}]{2007ApJS..170..377S}
{Spergel}, D.~N., {Bean}, R., {Dor{\'e}}, O., {et~al.} 2007, \apjs, 170, 377

\bibitem[{{Strauss} {et~al.}(2002){Strauss}, {Weinberg}, {Lupton}, {Narayanan},
  {Annis}, {Bernardi}, {Blanton}, {Burles}, {Connolly}, {Dalcanton}, {Doi},
  {Eisenstein}, {Frieman}, {Fukugita}, {Gunn}, {Ivezi{\'c}}, {Kent}, {Kim},
  {Knapp}, {Kron}, {Munn}, {Newberg}, {Nichol}, {Okamura}, {Quinn}, {Richmond},
  {Schlegel}, {Shimasaku}, {SubbaRao}, {Szalay}, {Vanden Berk}, {Vogeley},
  {Yanny}, {Yasuda}, {York}, \& {Zehavi}}]{2002AJ....124.1810S}
{Strauss}, M.~A., {Weinberg}, D.~H., {Lupton}, R.~H., {et~al.} 2002, \aj, 124,
  1810

\bibitem[{{Suh} {et~al.}(2011){Suh}, {Yoon}, {Jeong}, \&
  {Yi}}]{2011ApJ...730..110S}
{Suh}, H., {Yoon}, S.-c., {Jeong}, H., \& {Yi}, S.~K. 2011, \apj, 730, 110

\bibitem[{{Terry} {et~al.}(2002){Terry}, {Paturel}, \&
  {Ekholm}}]{2002A&A...393...57T}
{Terry}, J.~N., {Paturel}, G., \& {Ekholm}, T. 2002, \aap, 393, 57

\bibitem[{{Theureau} {et~al.}(1998){Theureau}, {Rauzy}, {Bottinelli}, \&
  {Gouguenheim}}]{1998A&A...340...21T}
{Theureau}, G., {Rauzy}, S., {Bottinelli}, L., \& {Gouguenheim}, L. 1998, \aap,
  340, 21

\bibitem[{{Tsvetkov} {et~al.}(2001){Tsvetkov}, {Blinnikov}, \&
  {Pavlyuk}}]{2001AstL...27..411T}
{Tsvetkov}, D.~Y., {Blinnikov}, S.~I., \& {Pavlyuk}, N.~N. 2001, Astronomy
  Letters, 27, 411

\bibitem[{{Tsvetkov} {et~al.}(2004){Tsvetkov}, {Pavlyuk}, \&
  {Bartunov}}]{2004AstL...30..729T}
{Tsvetkov}, D.~Y., {Pavlyuk}, N.~N., \& {Bartunov}, O.~S. 2004, Astronomy
  Letters, 30, 729

\bibitem[{{Turatto}(2003)}]{2003LNP...598...21T}
{Turatto}, M. 2003, in Lecture Notes in Physics, Berlin Springer Verlag, Vol.
  598, Supernovae and Gamma-Ray Bursters, ed. {K.~Weiler}, 21--36

\bibitem[{{Turatto} {et~al.}(2007){Turatto}, {Benetti}, \&
  {Pastorello}}]{2007AIPC..937..187T}
{Turatto}, M., {Benetti}, S., \& {Pastorello}, A. 2007, in American Institute
  of Physics Conference Series, Vol. 937, Supernova 1987A: 20 Years After:
  Supernovae and Gamma-Ray Bursters, ed. {S.~Immler, K.~Weiler, \& R.~McCray},
  187--197

\bibitem[{{Uomoto} {et~al.}(1999){Uomoto}, {Smee}, {Rockosi}, {Burles}, {Pope},
  {Friedman}, {Brinkmann}, {Gunn}, {Nichol}, \& {SDSS
  Collaboration}}]{1999AAS...195.8701U}
{Uomoto}, A., {Smee}, S., {Rockosi}, C., {et~al.} 1999, in Bulletin of the
  American Astronomical Society, Vol.~31, American Astronomical Society Meeting
  Abstracts, 1501

\bibitem[{{van den Bergh}(1988)}]{1988PASP..100..344V}
{van den Bergh}, S. 1988, \pasp, 100, 344

\bibitem[{{van den Bergh}(1997)}]{1997AJ....113..197V}
{van den Bergh}, S. 1997, \aj, 113, 197

\bibitem[{{van den Bergh} {et~al.}(2002){van den Bergh}, {Li}, \&
  {Filippenko}}]{2002PASP..114..820V}
{van den Bergh}, S., {Li}, W., \& {Filippenko}, A.~V. 2002, \pasp, 114, 820

\bibitem[{{van den Bergh} {et~al.}(2003){van den Bergh}, {Li}, \&
  {Filippenko}}]{2003PASP..115.1280V}
{van den Bergh}, S., {Li}, W., \& {Filippenko}, A.~V. 2003, \pasp, 115, 1280

\bibitem[{{van den Bergh} {et~al.}(2005){van den Bergh}, {Li}, \&
  {Filippenko}}]{2005PASP..117..773V}
{van den Bergh}, S., {Li}, W., \& {Filippenko}, A.~V. 2005, \pasp, 117, 773

\bibitem[{{van den Bergh} \& {Tammann}(1991)}]{1991ARA&A..29..363V}
{van den Bergh}, S. \& {Tammann}, G.~A. 1991, \araa, 29, 363

\bibitem[{{van Dyk} {et~al.}(1996){van Dyk}, {Hamuy}, \&
  {Filippenko}}]{1996AJ....111.2017V}
{van Dyk}, S.~D., {Hamuy}, M., \& {Filippenko}, A.~V. 1996, \aj, 111, 2017

\bibitem[{{van Dyk} {et~al.}(2000){van Dyk}, {Peng}, {King}, {Filippenko},
  {Treffers}, {Li}, \& {Richmond}}]{2000PASP..112.1532V}
{van Dyk}, S.~D., {Peng}, C.~Y., {King}, J.~Y., {et~al.} 2000, \pasp, 112, 1532

\bibitem[{{Wagner} {et~al.}(2004){Wagner}, {Vrba}, {Henden}, {Canzian},
  {Luginbuhl}, {Filippenko}, {Chornock}, {Li}, {Coil}, {Schmidt}, {Smith},
  {Starrfield}, {Klose}, {Tich{\'a}}, {Tich{\'y}}, {Gorosabel}, {Hudec}, \&
  {Simon}}]{2004PASP..116..326W}
{Wagner}, R.~M., {Vrba}, F.~J., {Henden}, A.~A., {et~al.} 2004, \pasp, 116, 326

\bibitem[{{Wang} {et~al.}(2010){Wang}, {Deng}, \& {Wei}}]{2010MNRAS.405.2529W}
{Wang}, J., {Deng}, J.~S., \& {Wei}, J.~Y. 2010, \mnras, 405, 2529

\bibitem[{{Wang} \& {Wei}(2008)}]{2008ApJ...679...86W}
{Wang}, J. \& {Wei}, J.~Y. 2008, \apj, 679, 86

\bibitem[{{Wild} \& {Schildknecht}(1987)}]{1987IAUC.4374....1W}
{Wild}, P. \& {Schildknecht}, T. 1987, \iaucirc, 4374, 1

\bibitem[{{Yahil} {et~al.}(1977){Yahil}, {Tammann}, \&
  {Sandage}}]{1977ApJ...217..903Y}
{Yahil}, A., {Tammann}, G.~A., \& {Sandage}, A. 1977, \apj, 217, 903

\end{thebibliography}
\end{document}